\documentclass[conference]{IEEEtran}
\IEEEoverridecommandlockouts
\usepackage{amsmath,amssymb,amsfonts}
\usepackage{listings}
\usepackage{algorithm}
\usepackage{algorithmicx}
\usepackage[noend]{algpseudocode}
\usepackage{graphicx}
\usepackage{textcomp}
\usepackage{xcolor}
\usepackage{tikz}
\usetikzlibrary{shapes,positioning,calc,arrows,chains}
\usepackage[unicode, hidelinks]{hyperref}
\usepackage[hyphenbreaks]{breakurl}
\usepackage[style=ieee, bibencoding=utf8, citestyle=numeric-comp]{biblatex}
\def\BibTeX{{\rm B\kern-.05em{\sc i\kern-.025em b}\kern-.08em
    T\kern-.1667em\lower.7ex\hbox{E}\kern-.125emX}}
\usepackage{booktabs}
\usepackage{colortbl}
\usepackage{multirow}
\usepackage{flushend}
\usepackage{verbatim}
\usepackage{subfig}
\usepackage[inline]{enumitem}

\algnewcommand{\algorithmicand}{\textbf{ and }}
\algnewcommand{\algorithmicor}{\textbf{ or }}
\algnewcommand{\OR}{\algorithmicor}
\algnewcommand{\AND}{\algorithmicand}

\definecolor{light-gray}{gray}{0.80}

\begin{document}

\title{Sydr-Fuzz: Continuous Hybrid Fuzzing and Dynamic Analysis for Security
Development Lifecycle}

\author{
\IEEEauthorblockN{
  Alexey Vishnyakov\IEEEauthorrefmark{1},
  Daniil Kuts\IEEEauthorrefmark{1},
  Vlada Logunova\IEEEauthorrefmark{1}\IEEEauthorrefmark{3},
  Darya Parygina\IEEEauthorrefmark{1}\IEEEauthorrefmark{2}, \\
  Eli Kobrin\IEEEauthorrefmark{1}\IEEEauthorrefmark{2},
  Georgy Savidov\IEEEauthorrefmark{1}\IEEEauthorrefmark{2} and
  Andrey Fedotov\IEEEauthorrefmark{1}
}
\IEEEauthorblockA{
  \IEEEauthorrefmark{1}Ivannikov Institute for System Programming of the RAS
}
\IEEEauthorblockA{
  \IEEEauthorrefmark{2}Lomonosov Moscow State University
}
\IEEEauthorblockA{
  \IEEEauthorrefmark{3}Moscow Institute of Physics and Technology
}
Moscow, Russia \\
\{vishnya, kutz, vlada, pa\_darochek, kobrineli, avgor46, fedotoff\}@ispras.ru
}

\maketitle

\begin{tikzpicture}[remember picture, overlay]
\node at ($(current page.south) + (0,0.65in)$) {
\begin{minipage}{\textwidth} \footnotesize
 Vishnyakov A., Kuts D., Logunova V., Parygina D., Kobrin E., Savidov G.,
 Fedotov A. Sydr-Fuzz: Continuous Hybrid Fuzzing and Dynamic Analysis for
 Security Development Lifecycle. 2022 Ivannikov ISPRAS Open Conference (ISPRAS),
 IEEE, 2022, pp. 111-123. DOI:
 \href{https://www.doi.org/10.1109/ISPRAS57371.2022.10076861}{10.1109/ISPRAS57371.2022.10076861}.

 \copyright~2022 IEEE. Personal use of this material is permitted. Permission
 from IEEE must be obtained for all other uses, in any current or future media,
 including reprinting/republishing this material for advertising or promotional
 purposes, creating new collective works, for resale or redistribution to
 servers or lists, or reuse of any copyrighted component of this work in other
 works.
\end{minipage}
};
\end{tikzpicture}

\begin{abstract}
Nowadays automated dynamic analysis frameworks for continuous testing are in
high demand to ensure software safety and satisfy the security development
lifecycle~(SDL) requirements. The security bug hunting efficiency of
cutting-edge hybrid fuzzing techniques outperforms widely utilized
coverage-guided fuzzing. We propose an enhanced dynamic analysis pipeline to
leverage productivity of automated bug detection based on hybrid fuzzing. We
implement the proposed pipeline in the continuous fuzzing toolset Sydr-Fuzz
which is powered by hybrid fuzzing orchestrator, integrating our DSE tool Sydr
with libFuzzer and AFL++. Sydr-Fuzz also incorporates security predicate
checkers, crash triaging tool Casr, and utilities for corpus minimization and
coverage gathering. The benchmarking of our hybrid fuzzer against alternative
state-of-the-art solutions demonstrates its superiority over coverage-guided
fuzzers while remaining on the same level with advanced hybrid fuzzers.
Furthermore, we approve the relevance of our approach by discovering 85 new
real-world software flaws within the OSS-Sydr-Fuzz project. Finally, we open
Casr source code to the community to facilitate examination of the existing
crashes.
\end{abstract}

\begin{IEEEkeywords}
dynamic analysis, hybrid fuzzing, continuous fuzzing, crash triage, dynamic
symbolic execution, DSE, error detection, security development lifecycle, SDL,
computer security
\end{IEEEkeywords}

\section{Introduction}

Modern industrial software becomes more and more complicated. It pervades almost
all the spheres of our life from relatively insignificant to critically
important. Introducing the security development lifecycle (SDL)~\cite{howard06,
iso08, gost16} becomes a natural thing in the majority of software development
companies. Hybrid fuzzing~\cite{cha12, pak12, stephens16, yun18, poeplau20,
poeplau21, borzacchiello21, chen22, david21} is one of the widely applied
analysis techniques for searching errors in binary code. Its popularity is
caused by the highly efficient combination of fuzzing and dynamic symbolic
execution (DSE) where the tools work together and share obtained results to help
each other. Launching fuzzing together with DSE outperforms coverage-guided
fuzzing~\cite{vanhauser21}. Fuzzing can quickly increase the code coverage by
overcoming simple constraints but fails at exploring complex code parts. DSE, on
the contrary, copes well with non-trivial branch constraints yet performs slower
analysis. We develop new hybrid fuzzing tool that is based on our symbolic
execution tool Sydr~\cite{vishnyakov20}. We integrate Sydr with two
state-of-the-art fuzzers, AFL++~\cite{fioraldi20} and
libFuzzer~\cite{serebryany16}, and evaluate our tool's efficiency compared to
existing hybrid fuzzers.

Within the modern rapidly progressing industry, it is crucial not only to find
bugs in software but also to report and fix them in adequate time. Integrating a
variety of tools into one tool can increase program analysis efficiency by
orders of magnitude. Based on this idea, we create dynamic analysis toolset
\textit{sydr-fuzz} that unites the hybrid fuzzing, corpus minimization, security
predicates checking, crash triaging, and coverage collection. \textit{Sydr-fuzz}
implements a convenient and productive dynamic analysis pipeline. Running
sequential \textit{sydr-fuzz} pipeline stages allows to maximize the
profit from hybrid fuzzing-based dynamic analysis.

The hybrid fuzzing itself is performed at the beginning pipeline stage. It
results in a corpus of input test cases that provide some new coverage and may
potentially lead to previously unknown errors. As the corpus may contain lots of
seeds that discover the same coverage or errors, it is a good point to prune
redundant seeds from further analysis. Corpus minimization tries to delete as
much test cases as possible while saving the same code coverage, thus keeping
the most profitable seeds.

After the first two steps, we receive the corpus of adequate size containing
test cases set that can be then processed in many ways. \textit{Sydr-fuzz}
suggests three strategies that allow getting various information from the hybrid
fuzzing results. The first one is error detection via applying symbolic security
predicates. This technique is based on dynamic symbolic execution with
additional constraints aimed at detecting four error kinds: null pointer
dereference, division by zero, integer overflow, and out of bounds access. Some
error kinds are related to multiple CWEs. The second strategy is collecting the
code coverage provided by the corpus. And the third one includes crash triaging
with the help of \textit{Casr}~\cite{casr-cluster21} tools. These tools allow to
generate crash reports for errors triggered by the corpus seeds, then
deduplicate and cluster these reports. After this stage, the results are
represented in a form of clusters for potentially different bugs accompanied by
the corresponding test cases.

This paper makes the following contributions:
\begin{itemize}
  \item We design Continuous Hybrid Fuzzing Infrastructure for efficient
    dynamic program analysis. We unite hybrid fuzzing~\cite{cha12, pak12,
    stephens16, yun18, poeplau20, poeplau21, borzacchiello21, chen22,
    david21}, corpus minimization, symbolic security
    predicates~\cite{vishnyakov21}, coverage collection, and crash
    triaging~\cite{casr-cluster21} stages into \textit{sydr-fuzz} and
    propose a dynamic analysis pipeline to maximize its profitability. We present
    OSS-Sydr-Fuzz~\cite{oss-sydr-fuzz} repository inspired by
    OSS-Fuzz~\cite{oss-fuzz} and adapted to hybrid fuzzing with
    \textit{sydr-fuzz}.
  \item We develop new hybrid fuzzer based on DSE tool
    Sydr~\cite{vishnyakov20}. We integrate Sydr with AFL++~\cite{fioraldi20}
    and libFuzzer~\cite{serebryany16} (which is the first integration between
    a DSE-tool and libFuzzer). We combine the power of our dynamic symbolic
    executor Sydr with the state-of-the-art fuzzers. We implement some
    profitable features in Sydr that help reach better evaluation results.
    We introduce symbolic pointers reasoning during hybrid
    fuzzing~\cite{kuts21}.
  \item We evaluate \textit{sydr-fuzz} on Google
    FuzzBench~\cite{fuzzbench} against state-of-the-art coverage-guided and
    hybrid fuzzers. We show that \textit{sydr-fuzz}
    outperforms coverage-guided fuzzers and proves to be comparable to hybrid
    fuzzers, gaining a significant profit from its symbolic executor Sydr~\cite{sydr-results}.
  \item We open source \textit{Casr}~\cite{casr-cluster21} for crash reports
    clustering, deduplication, and severity estimation at
    \href{https://github.com/ispras/casr}{\texttt{https://github.com/ispras/casr}}.
\end{itemize}


\section{Related Work}

\subsection{Hybrid Fuzzing}

Hybrid fuzzing is the state-of-the-art technique for detecting software bugs.
Its power comes from the lightweight and fast fuzzing and accurate symbolic
execution. Fuzzing helps quickly discover new paths, while DSE is responsible
for systematic code exploration.

\subsubsection{QSYM}

The hybrid fuzzer QSYM~\cite{yun18} became one of the first tools that showed
the effectiveness of hybrid fuzzing. Yun et al. implemented a hybrid fuzzer
that is lightweight enough to allow DSE and fuzzer to work in
parallel (while in Driller~\cite{stephens16} DSE is launched for a short period
to help fuzzer when it stops opening new coverage). QSYM utilizes Dynamic Binary
Translation to reduce the number of symbolically emulated instructions, and
refuses to use intermediate representation to eliminate additional overhead. The
authors proposed two optimization techniques to increase analysis efficiency.
Firstly, basic block pruning allows to skip some block emulation if it has been
executed too frequently. Secondly, optimistic solving assumes solving only the
target constraint if the whole path predicate is unsatisfiable. These techniques
do not provide sound analysis but help invert more symbolic branches per time
unit. Another useful technique suggested by QSYM is cache for inverted branches
with two caching modes. The static mode means that every symbolic branch is
inverted only once, while the context mode allows to consider the depth and the
set of the previously executed symbolic branches. QSYM also proposes handling of
symbolic addresses~\cite{cha12} by simply fuzzing them. It searches minimum and
maximum address values with SMT-solver and produces new seed on every
invocation. As for the hybrid fuzzing, QSYM launches the fuzzer in parallel to
the symbolic executor and lets them exchange new test inputs. DSE prefers such
input files that have discovered the new coverage recently and at the same time
have smaller size.

\subsubsection{SymCC, SymQEMU}

SymCC~\cite{poeplau20} developed a new compilation-based symbolic execution
technique. The instrumentation code for concolic execution 
is
inserted into the target application code, hence compiled binary file can be
executed without switching between program code execution and interpreter. The
code for updating symbolic state and handling symbolic computations is generated
only once at compile time. This method also benefits from the ability to apply
all the LLVM IR and CPU optimizations. SymCC bundles the symbolic backend into
the libraries used by the target program. Concreteness checks allow to
significantly reduce the number of symbolically processed computations.

As a continuation of SymCC, the authors presented SymQEMU~\cite{poeplau21} that
proposed applying compilation-based symbolic execution to binary files in the
absence of source code. Such method combines high analysis speed with
architecture-independence. SymQEMU was built on top of QEMU by extending its TCG
component so that symbolic handling code is inserted into the TCG ops IR before
compiling it to the host machine code. The symbolic analysis stops at the
system-call boundary that allows to achieve better performance. SymQEMU also
implements symbolic expression garbage collector for effective memory management.

Both SymCC and SymQEMU support AFL++~\cite{fioraldi20} integration in the way
similar to QSYM.

\subsubsection{FUZZOLIC}

Borzacchiello et al.~\cite{borzacchiello21} suggested an approach similar to SymQEMU.
They greatly increased symbolic execution efficiency due to the
fast analysis built on top of QEMU, and the new
Fuzzy-Sat~\cite{borzacchiello21fuzzysat} solver that utilizes fuzzing for query
solving. FUZZOLIC performs JIT compilation to add instrumentation code at
runtime. It compiles each basic block only once and benefits from inserting
instrumentation code into the target program code. FUZZOLIC is composed by the
tracer and the solver that run in distinct parallel processes, and that is the
first important difference from SymQEMU. The tracer executes the program and
generates symbolic expressions that are sent to the solver responsible for the
queries solving. The processes communicate through the shared memory. Also,
unlike SymQEMU, FUZZOLIC injects symbolic instrumentation into the target code
after QEMU has generated the basic block TCG. It allows to take advantage of TCG
optimizations at the level of the entire basic block. The last significant
difference is that SymQEMU handles only architecture-independent TCG
instructions, while FUZZOLIC can insert symbolic instrumentation for a large
number of architecture-dependent TCG native helpers for the x86 and x86\_64
platforms. The hybrid fuzzing scheme in FUZZOLIC is similar to QSYM. It
takes seeds mutated by AFL++, while the fuzzer takes interesting seeds
provided by the symbolic executor.

\subsubsection{PASTIS}

David et al.~\cite{david21} presented an automated testing infrastructure
combining fuzzing and DSE to validate the alerts received from some static
analysis tool and trigger the bugs if possible. The information provided by SAST
tool is used to add intrinsic functions to the target code. All the code
variants are compiled and sent to the PASTIS broker that performs all the
communication between the testing engines. On the one side,
Honggfuzz~\cite{honggfuzz} represents the fuzzing toolkit, while on the other
side Triton DSE framework~\cite{saudel15} is responsible for symbolic execution.

\subsubsection{SymSan, Jigsaw}

The main insight of SymSan~\cite{chen22} lies in building a concolic executor as
a special form of dynamic data flow analysis. SymSan performs compile-time
symbolic instrumentation of the code in LLVM IR~\cite{llvm}. The use of the
highly-optimized infrastructure from DFSAN~\cite{dfsan} helps reduce the
overhead of storing and retrieving symbolic expressions which form is also
optimized. SymSan proposes using an AST table along with a special AST node
design for storing symbolic expressions. In combination with simple forward
allocation of new nodes, it allows to significantly decrease performance
overhead while handling symbolic expressions. Additionally, SymSan implements
deduplication of the stored AST nodes and simplifying load and store operations.

The authors implemented a hybrid fuzzer based on SymSan and
Angora~\cite{chen18}. They also proposed a novel design to improve the search
throughput and incorporated it into the Jigsaw~\cite{chen22jigsaw} prototype
that is utilized as a solver for the hybrid fuzzer. The approach essence is
evaluating newly generated seeds with JIT-compiled path constraints. Jigsaw
compiles preprocessed AST sub-tasks into LLVM IR functions, uses LLVM’s JIT
engine to compile the IR into a native function, and searches for a satisfying
solution using gradient-guided search.

\subsection{Continuous Fuzzing}

The convergence of secure development and fuzzing is becoming an industry
standard~\cite{iso08, gost16} like unit testing enforcement. Continuous fuzzing
is an approach to organize automated fuzz testing as a routine, e.g. by
incorporating it into CI/CD pipeline. Continuous fuzzing infrastructure can be a
part of organization internal workflow scenario~\cite{onefuzz, 
grizzly}
or appear in a form of fuzzing-as-a-service tool~\cite{serebryany17, oss-sydr-fuzz, fuzzit, cifuzz, warkentin20}. 
The existing solutions include simple fuzz job launchers as well as large
frameworks encompassing enhanced functionality for scalable and
ensemble~\cite{chen19, osterlund21, david21} fuzzing, discovered flaws analysis
and reporting, regression commit bisecting, etc.


\subsubsection{OSS-Fuzz \& ClusterFuzz}

Although automated fuzz-testing is not a
silver bullet, it is quite efficient at discovering bugs and reducing human
analytics required. For instance, ClusterFuzz~\cite{demott11} has already found
over 25,000 bugs in Google-developed products and more than 43,500
bugs~\cite{oss-fuzz-issues} in open-source software included in OSS-Fuzz.
OSS-Fuzz project~\cite{serebryany17} selectively provides open-source community
with access to scalable ClusterFuzz infrastructure capacities running on Google
Cloud Platform~\cite{gcp}. ClusterFuzz is designed to automatically handle any
task within fuzzing lifecycle except for fuzz target writing and bug fixing. This
involves planning and launching fuzzing jobs, collecting statistics, new crashes
deduplication and triage, test case minimization, bisection of commit
introducing regression and bug fix verification. ClusterFuzz supports multiple
coverage-guided fuzzing engines~\cite{serebryany16, fioraldi20, honggfuzz},
black box fuzzing, and a range of sanitizers. As ClusterFuzz does not possess a
build infrastructure, a typical set of a project to participate in OSS-Fuzz
includes a docker image, build configuration, and at least one fuzz target. For
now, over 650 critical and widely-used open-source projects are continuously
fuzzed by OSS-Fuzz.

\subsubsection{OneFuzz}

Another continuous fuzzing framework
OneFuzz~\cite{onefuzz} was open-sourced by Microsoft. Similar to ClusterFuzz,
OneFuzz is currently applied to Microsoft software~(Windows, Edge, etc.) and is
tied to corporate cloud environment Azure~\cite{copeland15}. The supported
fuzzing engines are libFuzzer~\cite{serebryany16}, AFL++~\cite{fioraldi20}, and
Radamsa~\cite{radamsa}. OneFuzz provides an opportunity to benefit from built-in
templates as well as creation of customized fuzzing pipeline is possible.
Discovered flaws are classified by reproduction stability and deduplicated.
Furthermore, OneFuzz enables crash live debugging and fuzzing workflow
monitoring by configurable web hook events.

\subsubsection{Grizzly}

FirefoxCI TaskCluster utilizes Grizzly~\cite{grizzly}~--- a scalable
browser-specific fuzzing framework. Grizzly invokes browser and
fuzzing engine independently, manages data transferring between them during
analysis, and performs test case reduction. The two main interfaces, Target and
Adapter, are responsible for desired browser and fuzzer combination deployment
but the framework is primarily oriented on black box fuzzing techniques.

\subsubsection{Fuzzit}

Incorporated into Gitlab~\cite{gitlab-fuzz} 
service Fuzzit~\cite{fuzzit} was designed for fuzzing integration into project continuous build system. 
It constitutes runners for a collection of coverage-guided fuzzing engines to fuzz different programming
languages~(for instance, in case of C/C++ possible options are libFuzzer and AFL++) and is capable of regression testing.


\subsubsection{syzbot}

Continuous Linux kernel fuzzing system
syzbot~\cite{vyukov18} consistently produces structured reports on detected
kernel crashes. In addition to searching for bugs, the system monitors bug
obsoletion and verifies fixed issues. After conducting patch testing syzbot
validates that corresponding commit reached kernel builds for all tracked
branches to close bug report. Similar-looking crashes can be separately reported
thereupon instead of being merged into existing issue.

\section{Hybrid Fuzzing}

The statistics show that hybrid fuzzing engines, combining the power of the
symbolic execution and the fuzzing, can reach higher coverage than two
simultaneously working fuzzers~\cite{vanhauser21}. Based on this idea, we
implement our hybrid fuzzing tool, combining Sydr~\cite{vishnyakov20} with two
most popular and powerful open-source fuzzers: libFuzzer~\cite{serebryany16} and
AFL++~\cite{fioraldi20}.
There are two main tasks to achieve the goal of making an effective hybrid
fuzzing engine. Firstly, we develop different
heuristics to make Sydr work as effective as it can in combination with fuzzer.
Secondly, we implement Sydr and fuzzers integration, such as inter
communication and seed scheduling for Sydr.

It is necessary to build the target binary in two versions to run
\textit{sydr-fuzz} in
hybrid fuzzing mode. The first one must be built with sanitizers and fuzzing
instrumentation to be used by the corresponding fuzzer. The second binary must
be built without any instrumentation~--- it will be used by Sydr.

We implement various heuristics and features to make Sydr find new interesting
seeds faster and work more effective in combination with fuzzers. For example,
we implement cache in Sydr, that works quite same as QSYM cache, to save time
from trying to invert same symbolic branches. The core idea is that unique hash
is evaluated for every branch and is used as the index in bitmap. Every single
branch corresponds to a byte which is a counter representing how many times the
attempt to invert branch was proceeded. Moreover, we save an execution context
for every branch, which represents what branches were met during execution
before the considered one. If the context changes (i.e. Sydr reaches the branch
with the different execution path), we try to invert the branch. Otherwise we
attempt to invert a single branch when corresponding counter equals 0, 2, 4, 8, and
so on (power of two) until it equals 255, and after that we stop the specific
branch inversion.

Slicing~\cite{vishnyakov20} is another heuristic to fasten the symbolic execution. Every time we
attempt
to invert a branch, we apply the slicing algorithm to its path predicate. The
idea of the algorithm is that we leave only those path constraints in path
predicate which are data dependent on the target branch.
Other path constraints are eliminated, input data for them is
taken from the original program seed.

We implement optimistic and strong optimistic solving~\cite{parygina22}
to avoid formulas under- and overconstraint. Thus, Sydr inverts more
branches and generates more seeds. The core idea is following. When the
original sliced predicate is not satisfiable, we construct the
optimistic predicate consisting of only the target branch constraint. If it is
satisfiable, we construct strong optimistic predicate that is obtained from
original sliced predicate by eliminating some irrelevant path constraints based
on program call stack and branch control dependency analysis. If strong
optimistic predicate matches optimistic predicate or is unsatisfiable, we save
the optimistic seed. If they don't match and strong optimistic predicate is
satisfiable, we save both generated seeds.

We implement function semantics to fasten the symbolic execution for standard
library functions. Instead of stepping into the standard library functions (such
as \texttt{strto*l}, \texttt{*alloc}, \texttt{strcmp}, etc.) and executing their bodies
symbolically, we construct symbolic formulas for their return values, which
helps fasten symbolic execution and avoid overconstraint.

We limit the time that Sydr worker can solve a single query and the
total time that it can spend solving queries to avoid
situations when SMT-solver is stuck solving complicated queries. Moreover, we set
the timeout for the total time that single Sydr process can be executed to avoid
program freeze.

It is also worth mentioning that Sydr have functionality to invert jump tables
(\texttt{switch} statements) and handle symbolic addresses.
There are two ways to handle symbolic addresses
in Sydr: complete symbolic pointers processing and symbolic address fuzzing.
Full support for symbolic pointers~\cite{kuts21} drastically overloads symbolic
engine, leading to worse fuzzing results. However, this mode still allows symbolic
engine to successfully discover new unique coverage. Therefore, the optimal
strategy is to enable this mode periodically instead of using it every Sydr run.
When Sydr runs without symbolic pointers handling enabled, it performs
SMT-fuzzing for all symbolic addresses. It is lightweight but less accurate
method, that iterates over possible symbolic address values using an SMT-solver.

Before starting the hybrid fuzzing process, the initial corpus is automatically
minimized to leave only those seeds that bring any new coverage.

\subsection{Sydr and libFuzzer Integration}

libFuzzer~\cite{serebryany16} is an open-source state-of-the-art coverage-guided
fuzzer, which allows to fuzz libraries and applications effectively. It is
integrated into the Clang C compiler and can be enabled with just setting a
compile flag \texttt{-fsanitize=fuzzer} and adding a fuzzing target into the
project code. We implement Sydr and libFuzzer integration and some interaction
features to make them work better together.

All libFuzzer workers, that are executed simultaneously, store their test cases
in the same corpus directory. So, in libFuzzer integration Sydr takes seeds to
modify and puts generated seeds into the same corpus directory. This allows
libFuzzer to immediately load seeds generated by Sydr into the memory and use
them for further mutations.
We measure Sydr contribution to the hybrid fuzzing process in the following way.
We made a patch~\cite{libfuzzer-patch} to libFuzzer that counts the number of
times it reloaded files that were generated by Sydr, so, we can learn how many
interesting seeds Sydr stores to the corpus directory. All Sydr test cases that
were not reloaded by libFuzzer are regularly removed from corpus directory.

Other important detail of Sydr and libFuzzer interaction is scheduling seeds
for Sydr. We range every file in corpus comparing the following parameters in
the order they are listed:
\begin{enumerate*}
  \item whether the seed discovered new function;
  \item whether the seed brought new coverage;
  \item whether the seed caused libFuzzer features increase;
  \item $t_{creation} / S_{seed}$ value, where
    $t_{creation}$ is the time when the seed was created and $S_{seed}$ is the
    size of the seed.
\end{enumerate*}

Thus, firstly, we compare the fact whether the seed brought new coverage. If
two seeds brought new coverage, we compare whether they caused any features
increase. If both of them brought new coverage and cause features increase, we
prioritize them based on $t_{creation} / S_{seed}$ value:
the seed has higher priority than the other one if it is newer and its size
is smaller~\cite{yun18}.

\subsection{Sydr and AFL++ Integration}

AFL++~\cite{fioraldi20} is a widely used state-of-the-art fuzzer, that has sound interface for
configuring and combining with other fuzzing tools. We run Sydr as a fake
secondary AFL worker to organize hybrid fuzzing with AFL++. That is, the Sydr
worker directory is created in a common output directory for all AFL++ workers.
It contains queue subdirectory so that main AFL worker is able to scan it and
import useful seeds to its own corpus.

\textit{Sydr-fuzz} retrieves seeds for Sydr launching from main AFL worker
queue. These seeds are prioritized according to certain strategy~\cite{yun18}
in order to improve the hybrid fuzzing efficiency. Seeds are scheduled
according to the following criteria (in order of importance):
\begin{enumerate}
  \item \emph{\textbf{New coverage}} Seeds that discover new program coverage
    have the highest priority.
  \item \emph{\textbf{Seed input}} Seeds from initial corpus often increase
    coverage as well, but they are labeled differently by AFL++.
  \item \emph{\textbf{File size}} Smaller seed size is better as it results in
    faster and more efficient symbolic execution.
  \item \emph{\textbf{File name}} Since file name in AFL++ starts with
    \texttt{id}, this ordering allows to select a newer file. Newer seeds are
    more promising to explore new coverage before it's done by fuzzer.
\end{enumerate}

All seeds generated by Sydr are passed through a minimization step before being
added to the Sydr worker queue~\cite{borzacchiello21}. We use a global bitmap to
determine whether seed opens new program coverage, that hasn't been discovered by
symbolic engine yet. The afl-showmap utility is used to collect a bitmap for
every seed that Sydr is launched with. These seed bitmaps are merged into global
bitmap for setting initial coverage. Sydr stores all generated seeds in the
intermediate directory. Periodically, the afl-showmap utility runs on the whole
directory and collects the set of bitmaps, which are sequentially merged into
global bitmap. If global bitmap gets updated during merge, the corresponding
seed is moved to the Sydr worker queue directory. If program crash was detected
during afl-showmap run, the crashing seed is saved in crashes directory instead
of queue. All other seeds are considered uninteresting and removed. Thus, only
files with unique coverage are saved to the \textit{sydr-fuzz} queue.

\textit{Sydr-fuzz} is also capable of running AFL++ in parallel mode. It
launches the specified amount of AFL workers, guided by recommendations in
fuzzer tutorial~\cite{afl-tutorial}. There is always one main AFL instance, the
rest are secondaries. Options such as power schedule, \texttt{MOpt}, and old
queue cycling are mutated between secondary workers. The queue of scheduled
seeds for Sydr is synchronized only with main AFL worker. Seeds generated by
all running Sydr instances are saved in single Sydr worker directory during
\textit{sydr-fuzz} minimization stage.

AFL++ measures assistance from other fuzzing instances by counting files
imported from their queues. The number of files in main worker queue that were
imported from Sydr worker is used to evaluate Sydr contribution to hybrid
fuzzing. This may be done by using AFL++ statistics (in case of single AFL
instance running) or by counting files in main AFL worker queue with
corresponding synchronization tag.

\section{Security Predicates}

We propose security predicates~--- a method for accurate bug detection based on
dynamic symbolic execution~--- as a part of \textit{sydr-fuzz}.
We implement automated
security predicates checking with further verification and
deduplication of seeds generated by security predicates to reveal errors.

\textit{Security predicate} is a boolean predicate that holds true if program
instruction (or function) triggers an error~\cite{vishnyakov21}. Security
predicates implementation is a part of Sydr~\cite{vishnyakov20} that is based on
the following idea. We execute a program symbolically  with the seed that
does not lead to an error. Every time we execute instruction that operates on
symbolic data, we construct corresponding security predicate to check for the
certain error type. Then we conjunct security predicate with the sliced branch
constraints from the path predicate and pass resulting predicate to SMT-solver,
i.e. Bitwuzla~\cite{niemetz20}, to generate a seed that will reproduce the
error. If the predicate is satisfiable, we save the seed and report the error.
There are several types of weaknesses security predicates can detect: null
pointer dereference, division by zero, integer overflow, and out of bounds
access.

Null pointer dereference and division by zero security predicates work in
similar way. In the first case we construct the security predicate to check
whether symbolic address may equal to zero every time we execute the instruction
where the memory access occurs. In the second case we construct the security
predicate to check whether symbolic divisor may equal to zero every time we
execute division instruction such as \texttt{div} or \texttt{idiv} during
symbolic execution.

For out of bounds access error we construct security predicate every time we
execute instruction with memory access. This  security predicate holds true if
the symbolic address can be less than the array lower bound or greater than the
array upper bound. To detect array's bounds we maintain shadow stack and shadow
heap during symbolic execution. We save information about the array bounds
allocated on heap every time we meet the \texttt{*alloc} function call and remove
it when \texttt{free} is called. We change shadow stack according to
call stack, because we consider the upper bound of array allocated on stack as
the function frame beginning address. For arrays allocated on heap we learn
their bounds from shadow heap. For arrays allocated on stack we consider the
upper bound as the current function call site and compute the lower bound
heuristically.

Every time we execute arithmetical instruction during symbolic execution we
construct security predicate for integer overflow error. This predicate holds
true if \texttt{CF} or \texttt{OF} equals to 1 after instruction execution. If
the result is signed, we check only \texttt{OF} flag, and \texttt{CF} otherwise.
The signedness is learnt from previously met conditional instructions, which use
at least one operand same as the analyzed instruction uses. For example,
\texttt{JL} points that the result is signed and we must check only
\texttt{OF} flag. The principal difference between this security predicate and
the others is that we separate the meanings of error source (the arithmetical
instruction where error may occur) and error sink (the place where error can be
used), and check security predicate for the error source only when error sink
that uses this error source is found. We distinguish the following error sinks
types: branch instructions, address dereference instructions, and function
arguments~\cite{wang09}.

We implement automated security predicates checking in \textit{sydr-fuzz}.
Security predicates search for errors only on the path that is defined by the
given seed. Thus, we need to achieve the maximum coverage with the minimum number
of seeds.
Firstly, we run hybrid fuzzing to achieve great coverage. Then we run corpus
minimization to leave the minimum number of files, which achieve the
same coverage as before the minimization. After corpus minimization security
predicates checking starts. Security predicates results often appear to be false
positive, so, we implement automatic verification of security predicates results.
We run target binary built
with sanitizers on seed generated by security predicates. If any sanitizer
reports the error in the place that was reported as error source by security
predicates, we consider this seed as verified. Otherwise, we run
sanitizers-build binary with the seed from corpus, which was used for
security predicates checking, and if seed generated by security predicates
brings new sanitizer warnings, we consider this seed as verified. \textit{Sydr-fuzz}
deduplicates verified security predicates results based on source file, line,
and column of code where error was detected to ease the process of analyzing
security predicates results.

\section{Crash Triaging}

Crash triaging is an important step in dynamic analysis, including fuzzing or
hybrid fuzzing. The number of crashes that a fuzzer produces could be
significant. One can spend a long time to figure out which crashes represent the
same error. We propose an approach that should help developers spend less time
analyzing and fixing various bugs and issues. Our Casr~\cite{casr-cluster21} tools allow one to create
crash reports, deduplicate, and cluster them. The main stages of
\texttt{sydr-fuzz casr} are the following:
\begin{enumerate}
  \item \texttt{casr-san} runs an instrumented binary on all seeds that
    potentially cause crashes and generates crash reports based on sanitizer
    reports (with gdb help if needed).
  \item \texttt{casr-cluster} runs deduplication algorithm on Casr reports
    received from \texttt{casr-san}. Deduplication is based on the stack trace:
    each frame is hashed, then the hash of the entire stack trace is added to the
    hash set. As a result, only unique reports will remain in the hash set
    (details ~\cite{casr-cluster21}), the rest will be removed from the casr
    directory.
  \item \texttt{casr-cluster} starts hierarchical clustering of Casr reports.
    The distance between crashes is calculated based on the stack
    trace~\cite{casr-cluster21}.
  \item \texttt{casr-gdb} generates crash reports for non-instrumented target
    binaries based on the clusters obtained from the third step.
\end{enumerate}

As a result, we get clusters containing potentially different bugs (in the form
of Casr reports). Along with each report there is the corresponding seed that
leads to this crash. Some steps (3, 4) can be skipped by setting the appropriate
options.

Crash report contains information about crash such as OS and package versions,
executed command line, stack trace, open files and network connections, register
state, part of the source code that caused the crash, with the corresponding crash line,
etc. So, the developer does not need to run the gdb and analyze crash
manually, all the necessary information is already in the report.

Also, our Casr tools allow one to estimate crash severity. We divide
them into three broad classes (just like in gdb exploitable~\cite{gdb-exploitable}):
exploitable, probably exploitable, and not exploitable. The classes include
various errors that may occur during the program execution, such as stack
overflow, double free, and others. Crash class is determined based on the stack
trace, disassembly of the code section that caused the exception, the signal
that came to the program, and some other information. Classified crashes help
developers understand which ones should be analyzed and fixed first.

\section{Dynamic Analysis Pipeline}

We present a novel dynamic analysis framework \emph{sydr-fuzz} that comprises a
toolkit for hybrid fuzzing orchestration, corpus minimization, error detection,
crash triaging, and coverage collection. The functionality is divided into
separate tasks implemented by simple CLI that can be effortlessly integrated
into continuous build system. The following pipeline is proposed to leverage
analysis profitable impact.

\subsection{Hybrid Fuzzing (\texttt{sydr-fuzz run})} 

First of all, hybrid fuzzing session is launched. After merging and minimizing
initial corpus directories into a single project directory, the fuzzing process
starts. While the corpus of an individual worker~(Sydr or fuzzer instance) is
evolving, the orchestrator evaluates new seeds and spreads the most beneficial
ones to other workers. The session is halted according to configurable
parameters like quantity of discovered crashes, session timeout, and coverage
growth stagnation timeout. The last one, named \emph{exit-on-time}, sets the
fuzzing time-limit when the coverage increase stops.

\subsection{Corpus Minimization (\texttt{sydr-fuzz cmin})} 

The next step consists in extracting the most useful seeds by corpus
minimization. This reduces examination of seeds that cause equivalent program
behaviour during the next stages.

\subsection{Security Predicates (\texttt{sydr-fuzz security})} 

Then security predicate checkers are applied to minimized corpus directory. In
order to limit time consumption of this step for the large corpus, there is a
fixed number of random seeds to be analyzed. As the project corpus is retained
for the next continuous fuzzing iterations, eventually, an extensible proportion
of thoroughly examined seeds will be accumulated. After conducting results
verification and detected error deduplication, the summary is reported for
unique flaws triggered by checkers.

\subsection{Crash Triaging (\texttt{sydr-fuzz casr})} 

At the moment when crash triggering seeds are gathered, Casr crash triage is
employed to alleviate potential flaws investigation. During this phase crash
reports are automatically created and distilled into clusters classified by
severity of potential bug. Commonly, the number of issues to be manually
analyzed decreases by orders of magnitude. The result report contains a list of
cluster summaries that include number of united crashes, target application
source code line where the crash occurs, crash triggering seed, error type
and severity.

\subsection{Coverage Collecting (\texttt{sydr-fuzz cov-report})} 

Finally, corpus coverage is collected as standard fuzzing metrics. There are
several possible \emph{sydr-fuzz cov-*} options which conform to the
corresponding llvm-cov~\cite{llvm-cov} commands.

\section{Continuous Hybrid Fuzzing Infrastructure}

The continuous fuzzing integration utilizing \emph{sydr-fuzz} framework can be
demonstrated with OSS-Sydr-Fuzz project~\cite{oss-sydr-fuzz}. OSS-Sydr-Fuzz is
inspired by mentioned above OSS-Fuzz~\cite{serebryany17} and is intended to
approve the power of hybrid fuzzing on real software. The corresponding trophy
list~\cite{trophy-list} is presented in Section IX.

The required sources include code of the project to be tested and an additional
OSS-Sydr-Fuzz~\cite{oss-sydr-fuzz} fuzzing repository. The repository contains a range of manually
prepared facilities to configure and run hybrid fuzzing session. The key tasks
to start one are project code building, fuzzing tools deployment, and fuzz target
preparation. Consequently, the standard OSS-Sydr-Fuzz project package involves
build script, Dockerfile instructions, composed fuzz targets, and hybrid fuzzing
configuration file for each fuzz target. Moreover, the repository may provide
extra materials like initial seed corpus, dictionaries, etc.

\begin{figure}[h]
  \centering
  \includegraphics[width=0.48\textwidth]{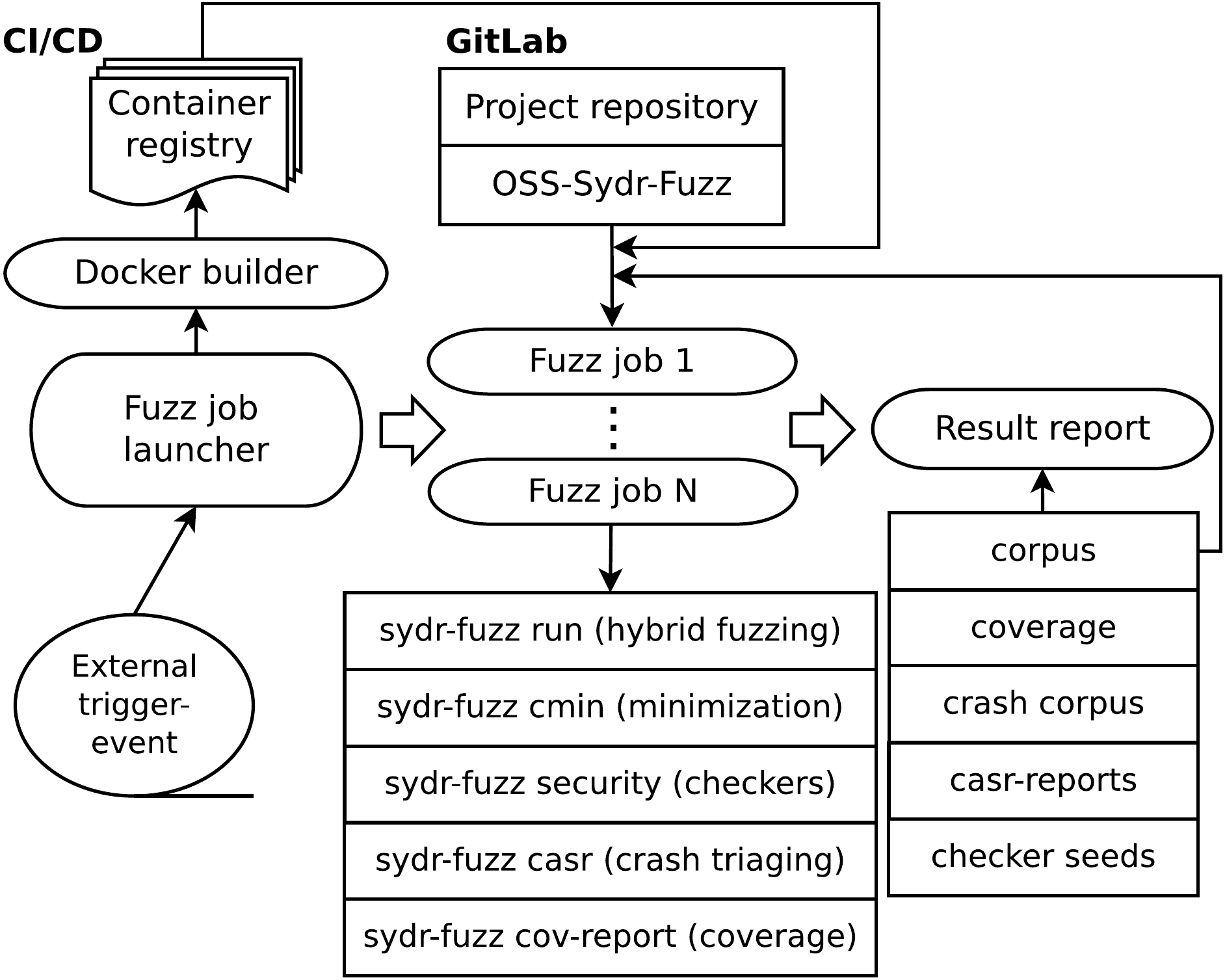}
  \caption{OSS-Sydr-Fuzz CI Architecture.}
  \label{fig:ci-scheme}
\end{figure}

Depicted on the scheme~(Fig.~\ref{fig:ci-scheme}) fuzzing CI workflow is
implemented on GitLab platform~\cite{gitlab}. The system applies dynamic analysis
pipeline introduced in the previous section. The fuzzing is launched when an external
trigger event occurs. Currently, it is run manually for the selected project and
branch. Alternatively, it could be organized as a scheduled routine or triggered
by new commits as well.

Before starting the chosen project analysis, fuzzing job launcher activates
docker builder task. First of all, the builder checks on the up-to-date commit
and requests the container registry whether a docker image from the previous
iteration exists and can be reused. The obtained or rebuilt container is utilized
for all project fuzz targets. Every fuzz target has an individual corpus and a
separate fuzzing job that sequentially launches the analysis pipeline stages.
In addition, there is an option to extend the initial seed corpus with seeds
gained during previous fuzzing sessions. Apart from logs and analysis statistics,
fuzzing job output includes resulting corpus, coverage information, Casr reports,
and error triggering seeds.

\section{Implementation}

\textit{Sydr-fuzz} is a Rust-written tool that configures, launches, and manages
other tools to organize the entire dynamic analysis workflow.
Crash triaging implemented as a set of tools written in Rust. It includes \texttt{casr-san}
binary for crash report generation, \texttt{casr-gdb} binary for generating
detailed crash reports without sanitizers, and \texttt{casr-cluster} for crashes
deduplication and clustering. \texttt{casr-cli} binary represents the generated
crash reports in a human readable format.

\textit{Sydr-fuzz} uses Sydr as symbolic execution engine. Sydr workflow consists of two
processes for symbolic and concrete execution running in parallel. Concrete
executor is build on top of DynamoRIO~\cite{bruening04} framework, and symbolic
executor utilizes Triton~\cite{saudel15} framework. We also extend Triton to
support Bitwuzla~\cite{niemetz20} SMT-solver for constraint solving.

The efficiency of symbolic engine is extremely important when organizing hybrid
fuzzing. We implement several enhancements for Sydr to improve its performance.
One of the main features is asynchronous solving of SMT-queries. It moves branch
inversion routine into a separate thread, that does not interact with symbolic
and concrete execution processes. Whenever new symbolic branch is encountered during
the analysis, the corresponding branch inversion job is pushed to the queue.
One or several solving threads retrieve jobs from this queue and process them.
Asynchronous solving allows to generate new seeds almost immediately after analysis
startup avoiding long interruptions of program analysis caused by comlicated SMT-queries
solving.

Another major improvement is suspending path predicate building if there are too
many branches to invert. Usually branches get discovered by symbolic execution
orders of magnitude faster than SMT-solver processes them. We suspend concrete
executor process if queue of symbolic branches to invert reaches a certain
threshold. Suspending concrete executor allows to keep the size of the path
predicate appropriately to the seed generation rate and save computational
resources. Concrete executor resumes when solving queue is emptied.

By default Sydr creates output directory with a certain structure, that allows
convenient storing of different informational files for inverted symbolic
branches. However, such structure turned out to be ineffective for hybrid
fuzzing organization. Most of informational files are redundant and having
multiple subdirectories only complicates the analysis. So, we implemented a flat
output directory structure in Sydr for hybrid fuzzing mode, that contains only
set of generated files. In this case Sydr generates unique filenames
itself and places new seeds directly to libFuzzer corpus (or to intermediate
directory for AFL++). A flat output directory allows \textit{sydr-fuzz}
to skip traversing Sydr directory and copying seeds, which improves hybrid
fuzzing efficiency.

A hybrid fuzzing implementation differs for libFuzzer and AFL++. AFL++ stores
all information about analysis in special files in its output directory.
Information about generated seeds contained directly in their filenames.
Therefore, \textit{sydr-fuzz} only needs to parse corresponding files and check
proper directories (queue, crashes, hangs) to synchronize tools and track the
state of fuzzing process. On the contrary, libFuzzer only prints all information
about analysis to log. Therefore, \textit{sydr-fuzz} parses libFuzzer logs to
retrieve information required for seeds prioritization and exit-on-time
functionality. Only one libFuzzer log is parsed at a time even when there are
several fuzzing jobs running. Due to the single corpus directory we assume that
all libFuzzer workers are synchronized immediately, so parsing one log file is
sufficient to get actual information. If a bug is found and currently parsed log
is finished, then we select next log file to parse from all active libFuzzer
workers.

Security predicates are implemented as part of Sydr and represented as
SMT-solving job similar to branch inversion job. Security predicates could be
checked during a symbolic execution together with path exploration. However,
security predicate checking is a complex task for SMT-solver, so it is better to
separate it from regular branch inversion tasks in order to keep Sydr
performance for seed generation. We implemented a special mode in Sydr that
allows to run symbolic execution only for security predicates checking.

\section{Evaluation}

We continuously use \textit{sydr-fuzz} to perform program analysis. The
OSS-Sydr-Fuzz repository~\cite{oss-sydr-fuzz} contains the list of tested
projects and their build instructions, configuration setup for the environment
and fuzzing process. During one year of \textit{sydr-fuzz} usage, 85
previously unknown defects were discovered in 22 open source
projects~\cite{trophy-list}. Of these, 13 errors were found by security
predicates implemented in Sydr.

We used Google FuzzBench framework~\cite{fuzzbench} to compare
\textit{sydr-fuzz} with modern state-of-the-art fuzzing tools. The FuzzBench
experiments were deployed on the server with 256Gb RAM and two AMD EPYC 7702
CPUs (64 cores each).
\textit{Sydr-fuzz} was sequentially tested against 4 fuzzers:
libFuzzer~\cite{serebryany16}, AFL++~\cite{fioraldi20}, SymQEMU~\cite{poeplau21},
and FUZZOLIC~\cite{borzacchiello21}. We selected 15 targets from FuzzBench intended for
coverage evaluation. Experiment was configured to perform 10 trials per fuzzer
and target combination, each trial ran on a single CPU core for 23 hours. Due to
the limited server resources we divided each \textit{sydr-fuzz} testing in two
packs of targets. Results of our FuzzBench experiments are
publicly available~\cite{sydr-results}.

During evaluation the following \textit{sydr-fuzz} configuration was used.
Hybrid fuzzing was performed with one instance of fuzzer and one running Sydr
instance at a time. Sydr inverted branches in direct order with one solving
thread. A 10 second limit was used for solving a single SMT-query, and 60
seconds~--- for the total solving time. Every Sydr run was limited to 2 minutes.
A cache was used to prevent Sydr inverting the same branches. Strong optimistic
solutions and symbolic address fuzzing were enabled for Sydr. Every symbolic
address was fuzzed up to 10 different models. Symbolic addresses fuzzing was
stopped for current Sydr run when 1000 such models were generated. Every 25th
launch of Sydr a full symbolic pointers handling mode was enabled instead of
symbolic addresses fuzzing. The path predicate building process was suspended
after 300 branch inversion jobs were scheduled during symbolic execution. The
memory usage for Sydr was limited by 8Gb. When this limit is exceeded, the program
execution is terminated and only branch inversion continues.

\subsection{Sydr-fuzz vs Fuzzers}

Firstly, we compared \textit{sydr-fuzz} with two state-of-the-art fuzzers to
prove the advantages of utilizing symbolic engine. We tested
AFL++~\cite{fioraldi20} and libFuzzer~\cite{serebryany16}
against \textit{sydr-fuzz} configured with corresponding fuzzer. The same
versions of libFuzzer (\href{https://github.com/llvm/llvm-project/commit/de5b16d8ca2d14ff0d9b6be9cf40566bc7eb5a01}{\texttt{de5b16d}}) and AFL++
(\href{https://github.com/AFLplusplus/AFLplusplus/commit/8fc249d210ad49e3dd88d1409877ca64d9884690}{\texttt{8fc249d}}) were used
in \textit{sydr-fuzz} and FuzzBench for evaluation. Also, an identical fuzzer
configuration was used in all tools. Because \textit{sydr-fuzz} has two
instances running (fuzzer and Sydr) on single CPU core in trial, we configured
FuzzBench to launch two workers of libFuzzer (\texttt{-workers=2}) and two
instances of AFL++ (main and secondary nodes).

The results of \textit{sydr-fuzz} and libFuzzer comparison are shown in
Fig.~\ref{fig:libfuzzer_res} in Appendix. \textit{Sydr-fuzz} outperformed
libFuzzer by reached coverage on 9 out of 14 applications. From 5
applications where libFuzzer has better results, only \texttt{sqlite3} has a
strong advantage. For the rest 4 applications the final coverage differs
slightly. Experiment was performed with two FuzzBench launches by 7 benchmarks
each. \textit{Sydr-fuzz} reached more coverage normalized score in both
launches: 98.67\% and 99.63\% for \textit{sydr-fuzz} against 96.51\% and 98.33\%
for libFuzzer respectively.

Fig.~\ref{fig:afl_res} in Appendix shows the results of \textit{sydr-fuzz} vs
AFL++ experiment. \textit{Sydr-fuzz} also showed the best results on 9 out of 14
applications in this experiment. As can be seen from Fig.~\ref{fig:afl_res},
there is a huge margin between \textit{sydr-fuzz} and AFL++ results on the most
of applications in favor of both tools. For two packs of targets
\textit{sydr-fuzz} got a higher average coverage score: 98.75\% and 99.19\% for
\textit{sydr-fuzz} against 94.87\% and 96.70\% for AFL++ respectively.

Thereby, \textit{sydr-fuzz} outperformed both libFuzzer and AFL++ on the most of
evaluated benchmarks and reached a higher total coverage.

\subsection{Sydr-fuzz vs Hybrid Fuzzers}

On the next stage, we evaluated \textit{sydr-fuzz} with modern hybrid fuzzers.
We selected SymQEMU~\cite{poeplau21} and FUZZOLIC~\cite{borzacchiello21}
because these tools also perform AFL++~\cite{fioraldi20} based hybrid fuzzing while symbolically
executing binary code. We made sure that \textit{sydr-fuzz}
uses same AFL++ version (\href{https://github.com/AFLplusplus/AFLplusplus/commit/8fc249d210ad49e3dd88d1409877ca64d9884690}{\texttt{8fc249d}})
and settings as in these tools. The
set of benchmarks slightly differs from those used for fuzzer evaluation due to
inoperability of tested tools with some targets: both tools with \texttt{libxslt}
and \texttt{openssl}, FUZZOLIC with \texttt{woff2}.

The results of \textit{sydr-fuzz} and SymQEMU comparison are shown in
Fig.~\ref{fig:symqemu_res} in Appendix. \textit{Sydr-fuzz} was able to
outperform SymQEMU on 7 out of 13 applications. The results are pretty close on
the most benchmarks. From all benchmarks, where SymQEMU has the better results,
only \texttt{zlib\_uncompress} has a significant advantage over
\textit{sydr-fuzz}. Only 5 trials were launched for \texttt{sqlite3} benchmark
due to SymQEMU instability on this target. For two experiment packs
\textit{sydr-fuzz} reached a higher average coverage: 99.35\% and 99.95\% for
\textit{sydr-fuzz} against 97.03\% and 99.67\% for SymQEMU respectively.

The results of experiments with \textit{sydr-fuzz} and FUZZOLIC are shown in
Fig.~\ref{fig:fuzzolic_res} in Appendix. Only 12 targets were available for
FUZZOLIC evaluation. \textit{Sydr-fuzz} was able to outperform FUZZOLIC on
6 out of 12 benchmarks. Same as SymQEMU, the results are close in this
experiment. FUZZOLIC reached significantly more coverage only on
\texttt{sqlite3} target. \textit{Sydr-fuzz} outperformed FUZZOLIC with a big
advantage on \texttt{libjpeg\_turbo} and \texttt{openthread} targets. The rest
benchmarks have very similar results on 23 hour distance. Nevertheless,
\textit{sydr-fuzz} was able to reach a little more average normalized score by
coverage on the both experiment packs: 99.1\% and 99.84\% for \textit{sydr-fuzz}
vs 99.07\% and 99.81\% for FUZZOLIC respectively.

These results show that \textit{sydr-fuzz} is on the same level with powerful
state-of-the-art hybrid fuzzers and can outperform them in some cases.
Also, \textit{sydr-fuzz} was able to outperform all tested coverage-guided and
hybrid fuzzers on 4 targets: \texttt{freetype2}, \texttt{openthread},
\texttt{libxslt} (SymQEMU and FUZZOLIC couldn't execute \texttt{libxslt}), and
\texttt{woff2} (FUZZOLIC failed to launch). There is also one
target \texttt{sqlite3} on which all other tools have better results than
\textit{sydr-fuzz}. This can be explained by the inefficient work of Sydr
in this particular example, as a result the fuzzer performance decreases.

\begin{table}[htbp]
\caption{Average Number of Useful Seeds Generated by Symbolic Engines}
\begin{center}
\scriptsize
\begin{tabular}{l >{\columncolor[gray]{0.9}}r r r}
\toprule
    \textbf{Application}&\textbf{Sydr}&\textbf{SymQEMU}&\textbf{FUZZOLIC}\\
    freetype2        & \textbf{307.8} & 90.8          & 241.9          \\
    harfbuzz         & \textbf{58.8}  & 34.8          & 21.3           \\
    lcms             & 139.3          & 192.5         & \textbf{203.5} \\
    libpng           & \textbf{30.4}  & 25.7          & 23.9           \\
    libjpeg-turbo    & \textbf{17.5}  & 13.5          & 14.6           \\
    libxml2          & 34.8           & \textbf{41.9} & 26.9           \\
    mbedtls          & 17.1           & 18.0          & \textbf{31.0}  \\
    openthread       & 59.3           & 38.1          & \textbf{72.9}  \\
    re2              & \textbf{2.5}   & 2.0           & 0.1            \\
    sqlite3          & 59.7           & 88.2          & \textbf{96.6}  \\
    vorbis           & 3.1            & \textbf{3.5}  & 2.2            \\
    woff2            & \textbf{24.9}  & 13.0          & ---            \\
    zlib\_uncompress & 1.2            & \textbf{4.7}  & 3.8            \\
\bottomrule
\end{tabular}
\label{tbl:imported}
\end{center}
\end{table}

In addition, we compared the assistance of symbolic engines to the fuzzer. It
can be measured with number of seeds generated by symbolic engine, that were
imported by AFL++. The Table~\ref{tbl:imported} contains an average number of
imported files per 10 trials at the end of analysis. The results show that Sydr
managed to generate more useful seeds than SymQEMU and FUZZOLIC on 6 out of
13 evaluated applications. On other examples the number of imported files is
less but still on the same level, except \texttt{zlib\_uncompress}. These
statistics show that Sydr impact on the fuzzing process is comparable to the
state-of-the-art symbolic engines.

\section{Conclusion}

We have presented Continuous Hybrid Fuzzing Framework \textit{Sydr-Fuzz}
for efficient dynamic program analysis during security development lifecycle.
We have united hybrid fuzzing
orchestration, corpus minimization, error detection, coverage collection, and
crash triaging into a single powerful toolset. We have presented new hybrid
fuzzing integration based on Sydr symbolic executor~\cite{vishnyakov20} and
popular open-source
fuzzers AFL++~\cite{fioraldi20} and libFuzzer~\cite{serebryany16}. We have
created OSS-Sydr-Fuzz~\cite{oss-sydr-fuzz} repository with
open-source software targets for \textit{sydr-fuzz} and discovered 85 new bugs
in 22 projects~\cite{trophy-list}. We have proposed dynamic analysis pipeline
for \textit{sydr-fuzz} to maximize the toolset profitable impact. We open source
Casr tool for crash reports clustering and
deduplication~\cite{casr}.

Our evaluation shows that, on the one hand, \textit{sydr-fuzz} outperforms
modern coverage-guided fuzzers AFL++ and libFuzzer on the majority of estimated
targets, and reaches a higher total coverage. On the other hand,
\textit{sydr-fuzz} proved to be comparable to powerful state-of-the-art hybrid
fuzzers SymQEMU~\cite{poeplau21} and FUZZOLIC~\cite{borzacchiello21}, and can even outperform them in some cases.
The significant profit that \textit{sydr-fuzz} gains from dynamic symbolic
executor Sydr during the fuzzing process, therefore, demonstrates the relevance
of utilizing cutting-edge hybrid fuzzers in dynamic analysis.

\section*{Availability}

The source code for Casr tool is publicly available at
\href{https://github.com/ispras/casr}{\texttt{https://github.com/ispras/casr}}.
OSS-Sydr-Fuzz project can be found at
\href{https://github.com/ispras/oss-sydr-fuzz}{\texttt{https://github.com/ispras/oss-sydr-fuzz}}.
The FuzzBench results are available at
\href{https://sydr-fuzz.github.io/fuzzbench}{\texttt{https://sydr-fuzz.github.io/fuzzbench}}.

\section*{Future Work}

As future directions, we consider the following issues:
\begin{itemize}
    \item Implementing AARCH64 dynamic symbolic execution in Sydr
        and employing our dynamic analysis pipeline for AARCH64 applications.
    \item Supporting fuzzing, corpus minimization, crash triaging, and
        coverage collection for Python code. We plan to integrate
        Atheris~\cite{atheris}~--- coverage-guided fuzzing engine from Google
        based on libFuzzer~--- into \textit{sydr-fuzz}.
    \item Varying AFL\_SYNC\_TIME~\cite{afl-docs-env-vars} environment
        variable value to allow AFL++ observe Sydr-generated test cases more
        frequently.
    \item Parsing debug information, which can help us clarify array boundaries
        for symbolic pointers reasoning and security predicates.
    \item Developing seed scheduler for Sydr based on Katz graph
        centrality~\cite{she22}.
    \item Adding security predicate checkers for integer truncation,
        format string, and command injection errors.
\end{itemize}

\printbibliography

@inproceedings{fioraldi20,
  title={{{AFL++}}: Combining Incremental Steps of Fuzzing Research},
  author={Fioraldi, Andrea and Maier, Dominik and Ei{\ss}feldt, Heiko and Heuse, Marc},
  booktitle={14th USENIX Workshop on Offensive Technologies (WOOT 20)},
  year={2020}
}

@inproceedings{stephens16,
  title={Driller: Augmenting Fuzzing Through Selective Symbolic Execution},
  author={Stephens, Nick and Grosen, John and Salls, Christopher and Dutcher, Andrew and Wang, Ruoyu and Corbetta, Jacopo and Shoshitaishvili, Yan and Kruegel, Christopher and Vigna, Giovanni},
  booktitle={NDSS},
  volume={16},
  number={2016},
  pages={1--16},
  year={2016}
}

@inproceedings{poeplau21,
  title={{{SymQEMU}}: Compilation-based symbolic execution for binaries},
  author={Poeplau, Sebastian and Francillon, Aur{\'e}lien},
  booktitle={Proceedings of the 2021 Network and Distributed System Security Symposium},
  year={2021},
  %doi = {10.14722/ndss.2021.23118},
}

@inproceedings{poeplau20,
  title={Symbolic execution with {{SymCC}}: Don't interpret, compile!},
  author={Poeplau, Sebastian and Francillon, Aur{\'e}lien},
  booktitle={29th USENIX Security Symposium (USENIX Security 20)},
  pages={181--198},
  year={2020}
}

@article{borzacchiello21,
  title={{{FUZZOLIC}}: mixing fuzzing and concolic execution},
  author={Borzacchiello, Luca and Coppa, Emilio and Demetrescu, Camil},
  journal={Computers \& Security},
  volume = {108},
  pages={102368},
  year={2021},
  publisher={Elsevier},
  %doi = {10.1016/j.cose.2021.102368},
}

@inproceedings{borzacchiello21fuzzysat,
  title={Fuzzing symbolic expressions},
  author={Borzacchiello, Luca and Coppa, Emilio and Demetrescu, Camil},
  booktitle={2021 IEEE/ACM 43rd International Conference on Software Engineering (ICSE)},
  pages={711--722},
  year={2021},
  organization={IEEE}
}

@inproceedings{cha12,
 author = {Cha, Sang Kil and Avgerinos, Thanassis and Rebert, Alexandre and Brumley, David},
 title = {Unleashing {{Mayhem}} on Binary Code},
 booktitle = {Proceedings of the 2012 IEEE Symposium on Security and Privacy},
 series = {SP~'12},
 year = {2012},
 pages = {380--394},
 numpages = {15},
 %doi = {10.1109/SP.2012.31},
 publisher = {IEEE},
}

@inproceedings{yun18,
  title={{{QSYM}}: A Practical Concolic Execution Engine Tailored for Hybrid Fuzzing},
  author={Yun, Insu and Lee, Sangho and Xu, Meng and Jang, Yeongjin and Kim, Taesoo},
  booktitle={27th USENIX Security Symposium},
  pages={745--761},
  year={2018},
}

@inproceedings{chen22,
author = {Ju Chen and WookHyun Han and Mingjun Yin and Haochen Zeng and Chengyu Song and Byoungyoung Lee and Heng Yin and Insik Shin},
title = {{SYMSAN}: Time and Space Efficient Concolic Execution via Dynamic Data-flow Analysis},
booktitle = {31st USENIX Security Symposium (USENIX Security 22)},
year = {2022},
% isbn = {978-1-939133-31-1},
% address = {Boston, MA},
pages = {2531--2548},
% url = {https://www.usenix.org/conference/usenixsecurity22/presentation/chen-ju},
publisher = {USENIX Association},
% month = aug,
}

@inproceedings{chen22jigsaw,
  title={{{JIGSAW}}: Efficient and Scalable Path Constraints Fuzzing},
  author={Chen, Ju and Wang, Jinghan and Song, Chengyu and Yin, Heng},
  booktitle={2022 IEEE Symposium on Security and Privacy (SP)},
  pages={1531--1531},
  year={2022},
  organization={IEEE}
}

@inproceedings{chen18,
  title={Angora: Efficient fuzzing by principled search},
  author={Chen, Peng and Chen, Hao},
  booktitle={2018 IEEE Symposium on Security and Privacy (SP)},
  pages={711--725},
  year={2018},
  organization={IEEE}
}

@inproceedings{casr-cluster21,
  title = {{{Casr-Cluster}}: Crash Clustering for Linux Applications},
  author = {Savidov, Georgy and Fedotov, Andrey},
  booktitle = {2021 Ivannikov ISPRAS Open Conference (ISPRAS)},
  pages = {47--51},
  year = {2021},
  publisher = {IEEE},
}

@article{david21,
  title={From source code to crash test-cases through software testing automation},
  author={David, Robin and Salwan, Jonathan and Bourroux, Justin},
  journal={Proc. of the 28th C\&ESAR},
  pages={27},
  year={2021}
}

@inproceedings{vishnyakov20,
  title = {Sydr: Cutting Edge Dynamic Symbolic Execution},
  author = {Vishnyakov, Alexey and Fedotov, Andrey and Kuts, Daniil and Novikov,
            Alexander and Parygina, Darya and Kobrin, Eli and Logunova, Vlada
            and Belecky, Pavel and Kurmangaleev, Shamil},
  booktitle = {2020 Ivannikov ISPRAS Open Conference (ISPRAS)},
  pages = {46--54},
  year = {2020},
  publisher = {IEEE},
  %doi = {10.1109/ISPRAS51486.2020.00014},
}

@inproceedings{vishnyakov21,
  title = {Symbolic Security Predicates: Hunt Program Weaknesses},
  author = {Vishnyakov, Alexey and Logunova, Vlada and Kobrin, Eli and Kuts,
            Daniil and Parygina, Darya and Fedotov, Andrey},
  booktitle = {2021 Ivannikov ISPRAS Open Conference},
  pages = {76--85},
  year = {2021},
  publisher = {IEEE},
  %doi = {10.1109/ISPRAS53967.2021.00016},
}

@inproceedings{kuts21,
  title={Towards Symbolic Pointers Reasoning in Dynamic Symbolic Execution},
  author={Kuts, Daniil},
  booktitle={2021 Ivannikov Memorial Workshop (IVMEM)},
  year={2021},
  organization={IEEE},
  pages={42--49},
  %doi={10.1109/IVMEM53963.2021.00014},
}

@inproceedings{saudel15,
  author    = {Saudel, Florent and Salwan, Jonathan},
  title     = {{{Triton}}: A Dynamic Symbolic Execution Framework},
  booktitle = {Symposium sur la s{\'{e}}curit{\'{e}} des technologies de l'information
               et des communications},
  series    = {SSTIC},
  pages     = {31--54},
  year      = {2015}
}

@phdthesis{bruening04,
  title={Efficient, Transparent, and Comprehensive Runtime Code Manipulation},
  author={Bruening, Derek},
  year={2004},
  school={Massachusetts Institute of Technology, Department of Electrical
          Engineering and Computer Science},
}

@article{niemetz20,
  author    = {Aina Niemetz and
               Mathias Preiner},
  title     = {Bitwuzla at the {{SMT-COMP}} 2020},
  journal   = {CoRR},
  volume    = {abs/2006.01621},
  year      = {2020},
  url       = {https://arxiv.org/abs/2006.01621},
  archivePrefix = {arXiv},
  eprint    = {2006.01621},
}

@book{howard06,
  title={The security development lifecycle},
  author={Howard, Michael and Lipner, Steve},
  volume={8},
  year={2006},
  publisher={Microsoft Press Redmond},
  url={http://msdn.microsoft.com/en-us/library/ms995349.aspx},
}

@mastersthesis{pak12,
  title={Hybrid Fuzz Testing: Discovering Software Bugs via Fuzzing and Symbolic
         Execution},
  school = {School of Computer Science Carnegie Mellon University},
  author={Pak, Brian S},
  year={2012},
}

@book{iso08,
  title={{{ISO/IEC}} 15408-3:2008: Information technology -- Security techniques --
         Evaluation criteria for IT security -- Part 3: Security assurance
         components},
  year={2008},
  %publisher={ISO Geneva, Switzerland},
  url={https://www.iso.org/standard/46413.html},
}

@book{gost16,
  title={{{GOST R}} 56939-2016: Information protection. Secure software development.
         General requirements},
  publisher={National Standard of Russian Federation},
  year={2016},
  url={http://protect.gost.ru/document.aspx?control=7&id=203548},
}

@inproceedings{serebryany16,
  title={Continuous Fuzzing with {{libFuzzer}} and {{AddressSanitizer}}},
  author={Serebryany, Kosta},
  booktitle={2016 IEEE Cybersecurity Development (SecDev)},
  pages={157},
  year={2016},
  organization={IEEE},
  %doi={10.1109/SecDev.2016.043},
}

@inproceedings{llvm,
    author    = {Chris Lattner and Vikram Adve},
    title     = {{{LLVM}}: A Compilation Framework for Lifelong Program Analysis \& Transformation},
    booktitle = {Proceedings of the 2004 International Symposium on Code Generation and Optimization (CGO'04)},
    year      = {2004},
    volume={4},
    pages={75},
}

@misc{llvm-cov,
  title = {The {{LLVM}} Compiler Infrastructure. llvm-cov - emit coverage information},
  url = {https://llvm.org/docs/CommandGuide/llvm-cov.html}
}

@inproceedings{demott11,
  title={{Towards an automatic exploit pipeline}},
  author={DeMott, Jared D and Enbody, Richard J and Punch, William F},
  booktitle={2011 International Conference for Internet Technology and Secured Transactions},
  pages={323--329},
  year={2011},
  organization={IEEE}
}

@misc{vyukov18,
  title = {{{syzbot}}: Automated Kernel Testing},
  author = {Dmitry Vyukov},
  publisher = {Linux Plumbers Conference, Vancouver},
  year = 2018,
  url = {https://lpc.events/event/2/contributions/237/attachments/61/71/syzbot_automated_kernel_testing.pdf}
}

@misc{warkentin20,
  title = {Getting Started Using {{Mayhem}} with Continuous Integration},
  author = {Sheldon Warkentin},
  publisher = {BrightTALK media-company},
  year = 2020,
  url = {https://www.brighttalk.com/webcast/17668/439580}
}

@article{copeland15,
  title={Microsoft azure},
  author={Copeland, Marshall and Soh, Julian and Puca, Anthony and Manning, Mike and Gollob, David},
  journal={New York, NY, USA:: Apress},
  pages={3--26},
  year={2015},
  publisher={Springer}
}

@inproceedings{chen19,
  title={{{EnFuzz}}: Ensemble Fuzzing with Seed Synchronization among Diverse Fuzzers},
  author={Chen, Yuanliang and Jiang, Yu and Ma, Fuchen and Liang, Jie and Wang, Mingzhe and Zhou, Chijin and Jiao, Xun and Su, Zhuo},
  booktitle={28th USENIX Security Symposium (USENIX Security 19)},
  pages={1967--1983},
  year={2019}
}

@inproceedings{osterlund21,
  title={Collabfuzz: A framework for collaborative fuzzing},
  author={{\"O}sterlund, Sebastian and Geretto, Elia and Jemmett, Andrea and G{\"u}ler, Emre and G{\"o}rz, Philipp and Holz, Thorsten and Giuffrida, Cristiano and Bos, Herbert},
  booktitle={Proceedings of the 14th European Workshop on Systems Security},
  pages={1--7},
  year={2021}
}

@misc{gitlab-fuzz,
  title = {Coverage-guided fuzz testing in {{GitLab}}},
  url = {https://docs.gitlab.com/ee/user/application_security/coverage_fuzzing/}
}

@misc{libfuzzer-patch,
  title = {Patch for {{libFuzzer}}: Print reloaded file paths},
  url = {https://reviews.llvm.org/D100303}
}

@misc{afl-tutorial,
  title = {{{AFL++}} tutorial},
  url = {https://github.com/AFLplusplus/AFLplusplus/blob/stable/docs/fuzzing_in_depth.md}
}

@misc{gdb-exploitable,
  title = {{{GDB}} 'exploitable' plugin},
  url = {https://github.com/jfoote/exploitable}
}

@conference{serebryany17,
    author = {Kostya Serebryany},
    title = {{{OSS-Fuzz}} - {{Google{\textquoteright}s}} continuous fuzzing service for open source software},
    year = {2017},
    publisher = {USENIX Association},
}

@misc{oss-fuzz-issues,
  title = {{{OSS-Fuzz}} Issue Report Tracker},
  url = {https://bugs.chromium.org/p/oss-fuzz/issues/list?q=-status%3AWontFix%2CDuplicate%20-component%3AInfra&can=1}
}

@misc{sydr-results,
  title = {{{FuzzBench}} results for {{Sydr-Fuzz}}},
  url = {https://sydr-fuzz.github.io/fuzzbench}
}

@misc{honggfuzz,
  author = {Swiecki, R. and Gröbert, F.},
  title = {Honggfuzz},
  url = {https://github.com/google/honggfuzz}
}

@inproceedings{parygina22,
  %doi = {10.48550/ARXIV.2209.03710},
  %url = {https://arxiv.org/abs/2209.03710},
  author = {Parygina, Darya and Vishnyakov, Alexey and Fedotov, Andrey},
  title = {Strong Optimistic Solving for Dynamic Symbolic Execution},
  booktitle = {Ivannikov Memorial Workshop (IVMEM)},
  year = {2022},
  publisher = {IEEE},
}

@misc{oss-fuzz,
  title = {{{OSS-Fuzz}}: Continuous Fuzzing for Open Source Software},
  url = {https://github.com/google/oss-fuzz}
}

@misc{onefuzz,
  title = {{{OneFuzz}}: A self-hosted Fuzzing-As-A-Service platform},
  url = {https://github.com/microsoft/onefuzz}
}

@misc{grizzly,
  title = {Grizzly Browser Fuzzing Framework},
  url = {https://github.com/MozillaSecurity/grizzly}
}

@misc{fuzzit,
  title = {Fuzzit},
  url = {https://github.com/fuzzitdev/fuzzit}
}

@misc{cifuzz,
  title = {{{cifuzz}}: fuzz tests as easy as unit tests},
  url = {https://github.com/CodeIntelligenceTesting/cifuzz}
}

@misc{oss-sydr-fuzz,
  title = {{{OSS-Sydr-Fuzz}}: Hybrid Fuzzing for Open Source Software},
  url = {https://github.com/ispras/oss-sydr-fuzz}
}

@misc{casr,
  title = {{{CASR}}: Crash Analysis and Severity Report},
  url = {https://github.com/ispras/casr}
}

@misc{trophy-list,
  title = {{{Sydr-Fuzz}} trophy list},
  url = {https://github.com/ispras/oss-sydr-fuzz/blob/master/TROPHIES.md}
}

@inproceedings{fuzzbench,
  title={{{FuzzBench}}: An Open Fuzzer Benchmarking Platform and Service},
  author={Metzman, Jonathan and Szekeres, L{\'a}szl{\'o} and Simon, Laurent and Sprabery, Read and Arya, Abhishek},
  booktitle={Proceedings of the 29th ACM Joint Meeting on European Software Engineering Conference and Symposium on the Foundations of Software Engineering},
  pages={1393--1403},
  year={2021}
}

@misc{gcp,
  title = {Google Cloud Platform},
  url = {https://github.com/GoogleCloudPlatform}
}

@misc{radamsa,
  title = {A Crash Course to {{Radamsa}}},
  url = {https://gitlab.com/akihe/radamsa}
}

@misc{dfsan,
  url = {https://clang.llvm.org/docs/DataFlowSanitizerDesign.html},
  title = {{{DataFlowSanitizer}} Design Document},
  year = {2018}
}

@misc{vanhauser21,
  url = {https://www.fuzzbench.com/reports/experimental/2021-07-03-symbolic/index.html},
  title = {{{FuzzBench}} Symbolic Report},
  year = {2021}
}

@inproceedings{wang09,
  title={{{IntScope}}: Automatically Detecting Integer Overflow Vulnerability in X86 Binary Using Symbolic Execution},
  author={Wang, Tielei and Wei, Tao and Lin, Zhiqiang and Zou, Wei},
  booktitle={NDSS},
  year={2009},
}

@inproceedings{she22,
  title={Effective Seed Scheduling for Fuzzing with Graph Centrality Analysis},
  author={She, Dongdong and Shah, Abhishek and Jana, Suman},
  booktitle={2022 IEEE Symposium on Security and Privacy (SP)},
  pages={2194--2211},
  year={2022},
  organization={IEEE}
}

@misc{atheris,
  title = {{{Atheris}}: A Coverage-Guided, Native {{Python}} Fuzzer},
  url = {https://github.com/google/atheris}
}

@misc{afl-docs-env-vars,
  title = {{{AFL++}} Environment Variables Documentation},
  url = {https://github.com/AFLplusplus/AFLplusplus/blob/stable/docs/env_variables.md}
}

@misc{gitlab,
  title = {{{GitLab}}: The One DevOps Platform},
  url = {https://about.gitlab.com/}
}

\clearpage
\appendix

\begin{figure}[h]%
    \centering
    \subfloat[\centering freetype2]{{\includegraphics[width=4cm]{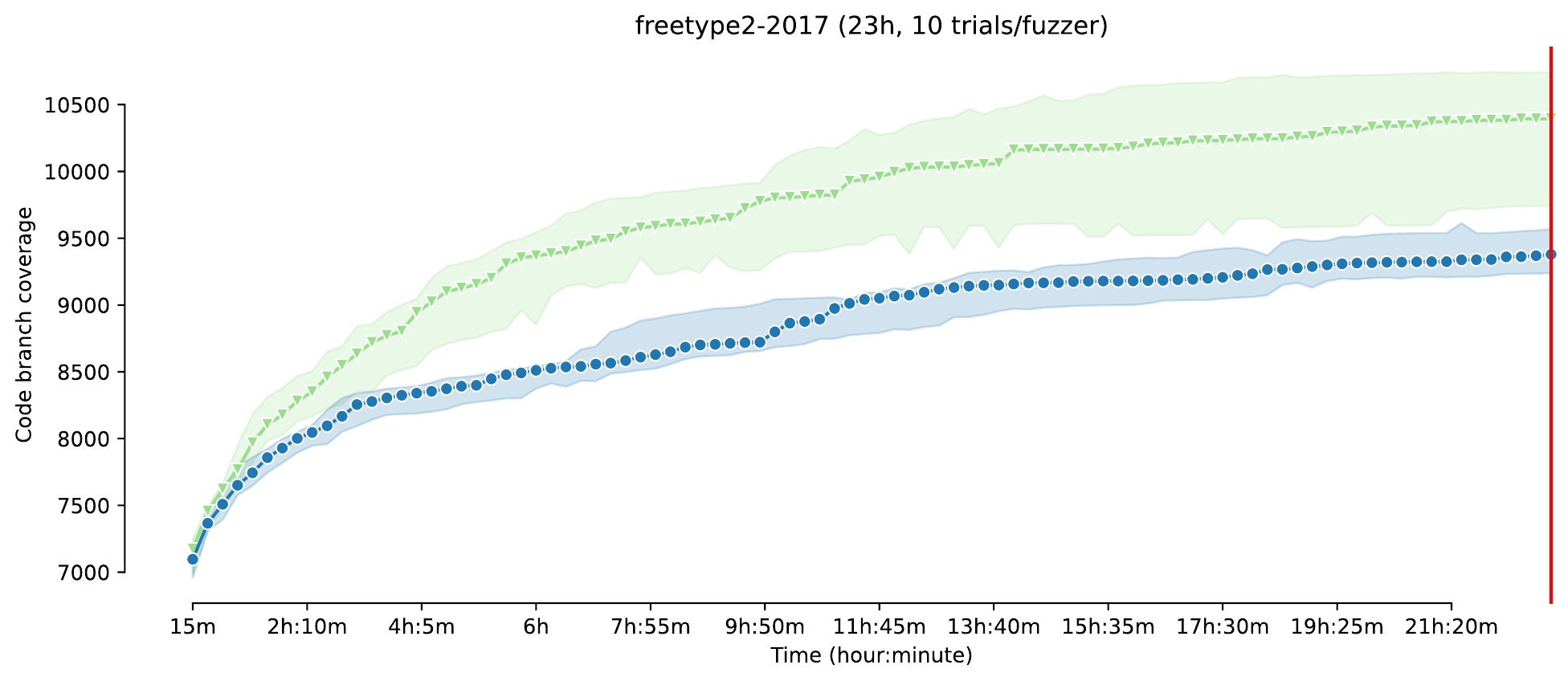} }}%
    \quad
    \subfloat[\centering harfbuzz]{{\includegraphics[width=4cm]{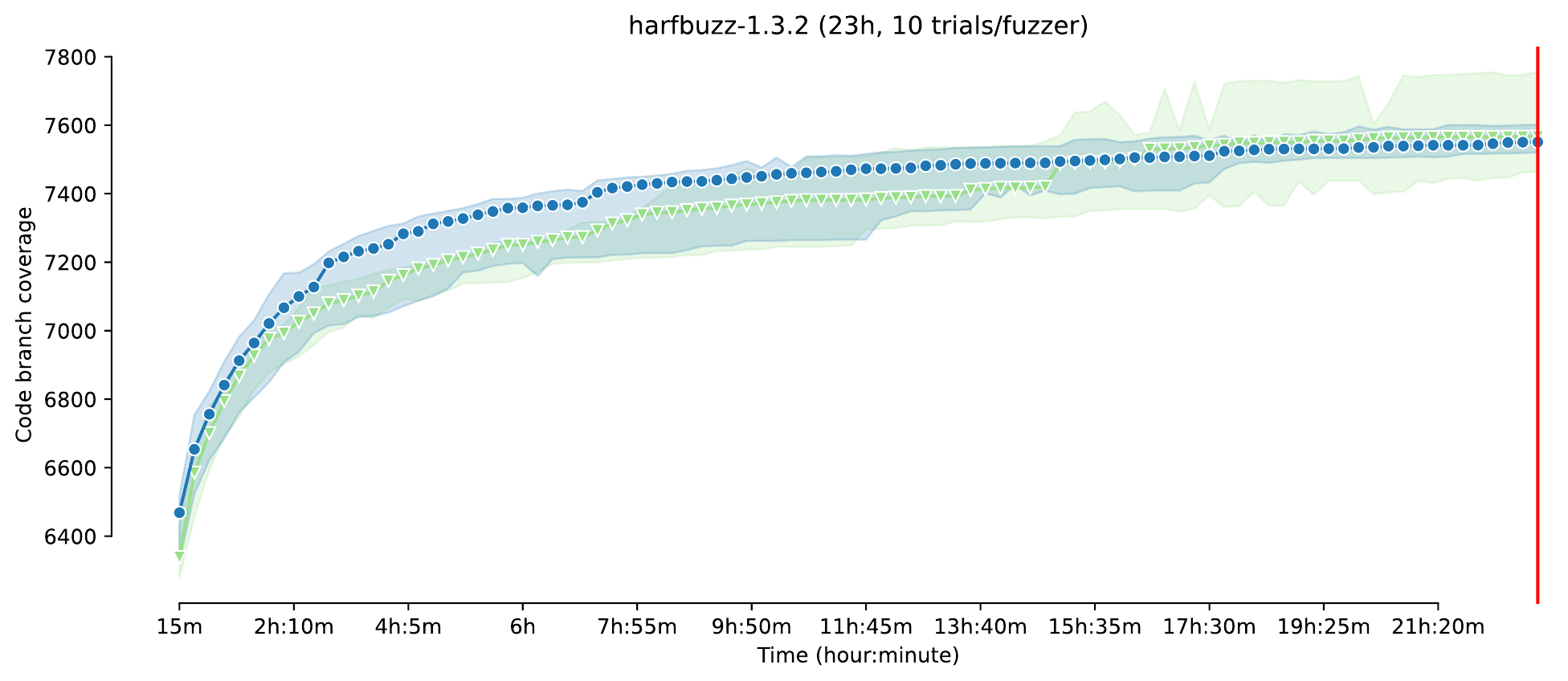} }}%
    \quad
    \subfloat[\centering lcms]{{\includegraphics[width=4cm]{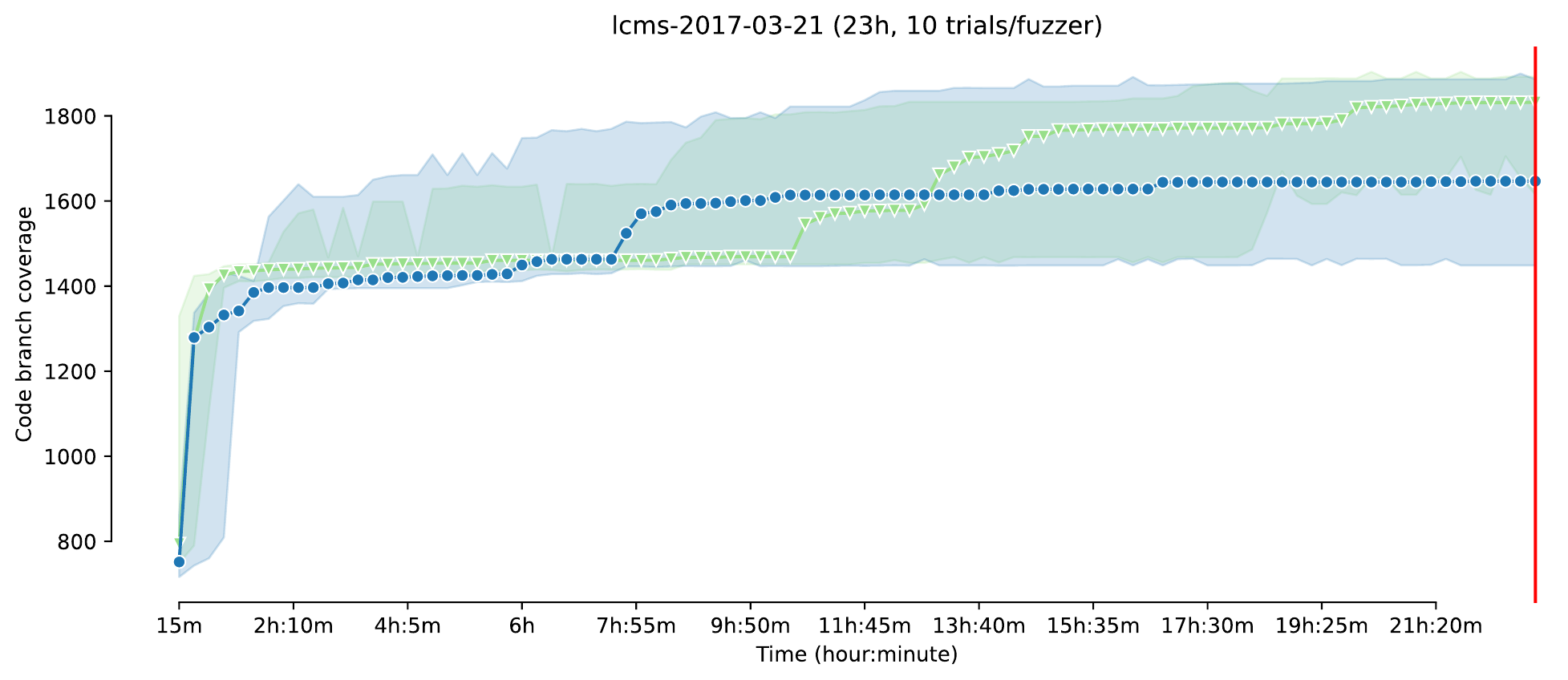} }}%
    \quad
    \subfloat[\centering libjpeg\_turbo]{{\includegraphics[width=4cm]{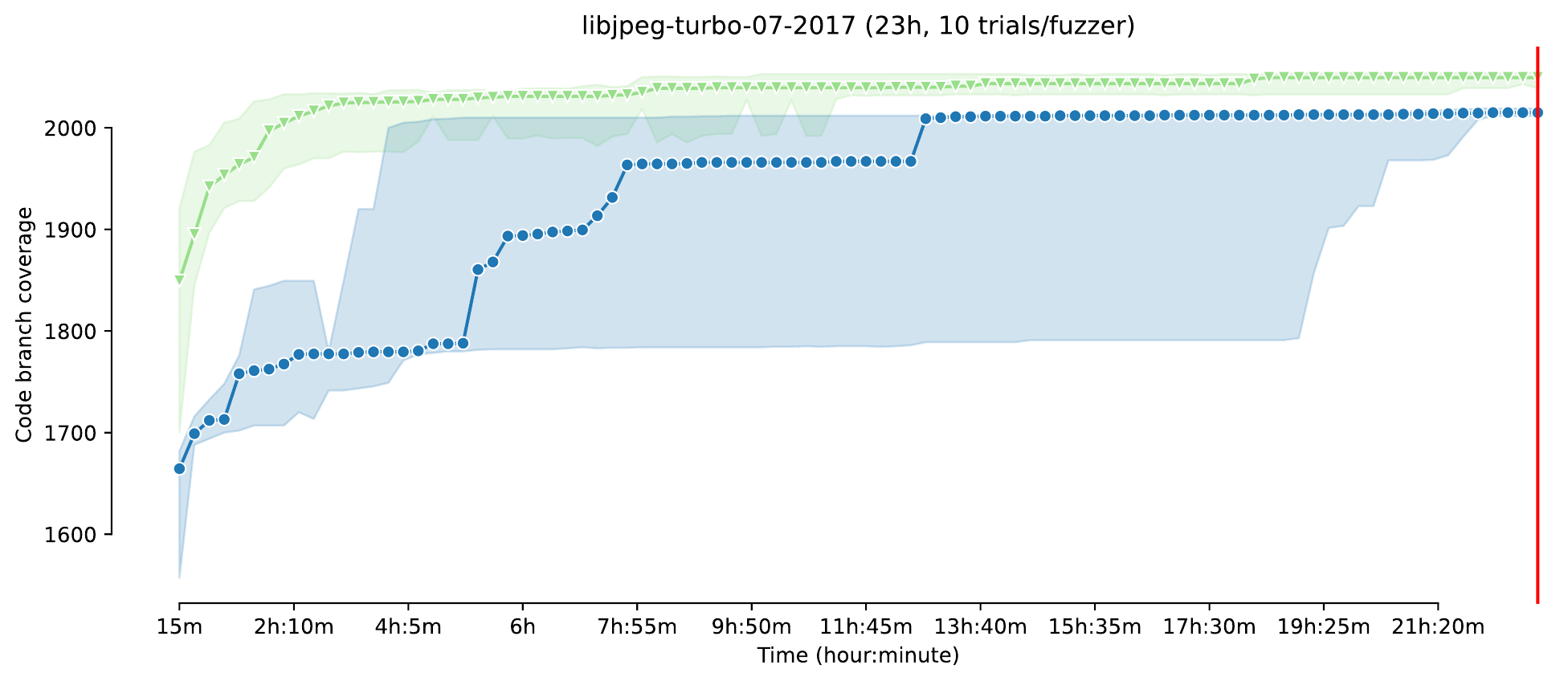} }}%
    \quad
    \subfloat[\centering libpng]{{\includegraphics[width=4cm]{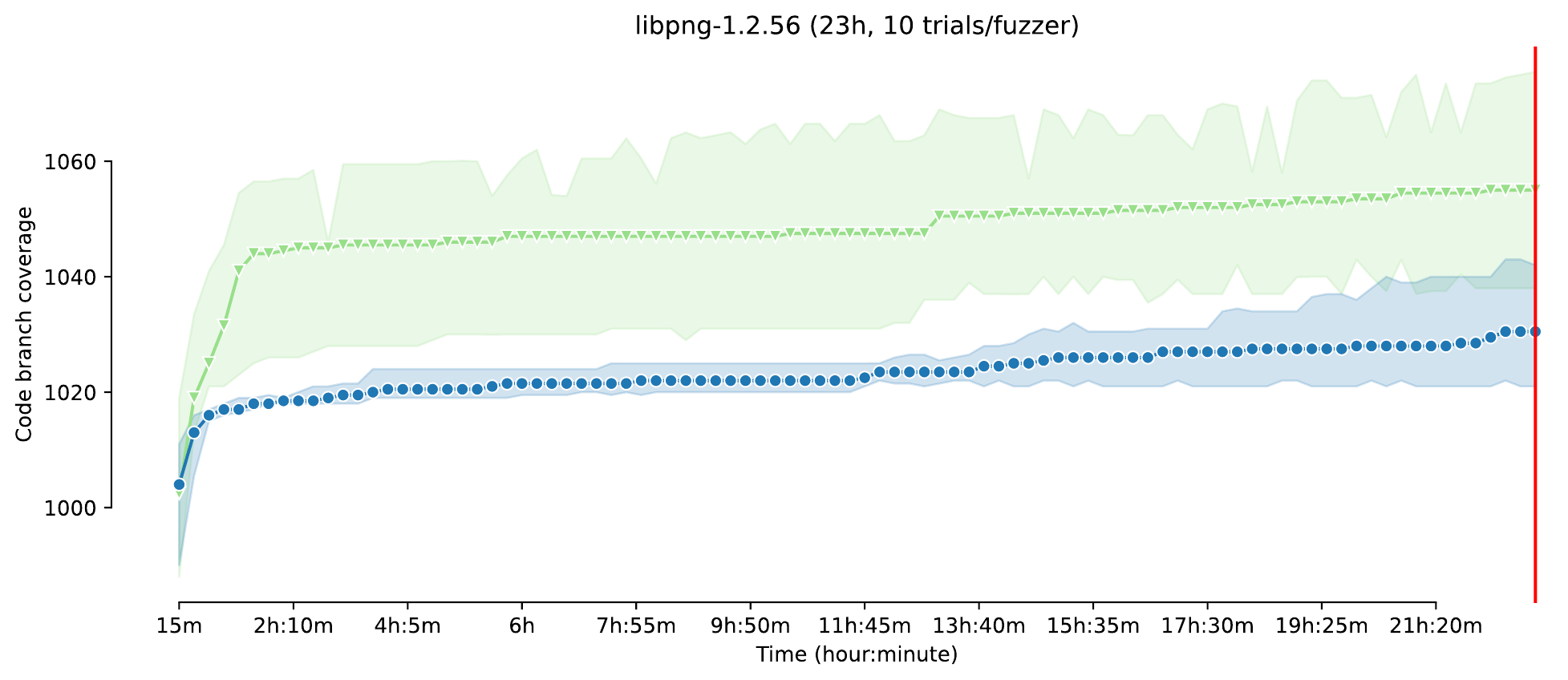} }}%
    \quad
    \subfloat[\centering libxml2]{{\includegraphics[width=4cm]{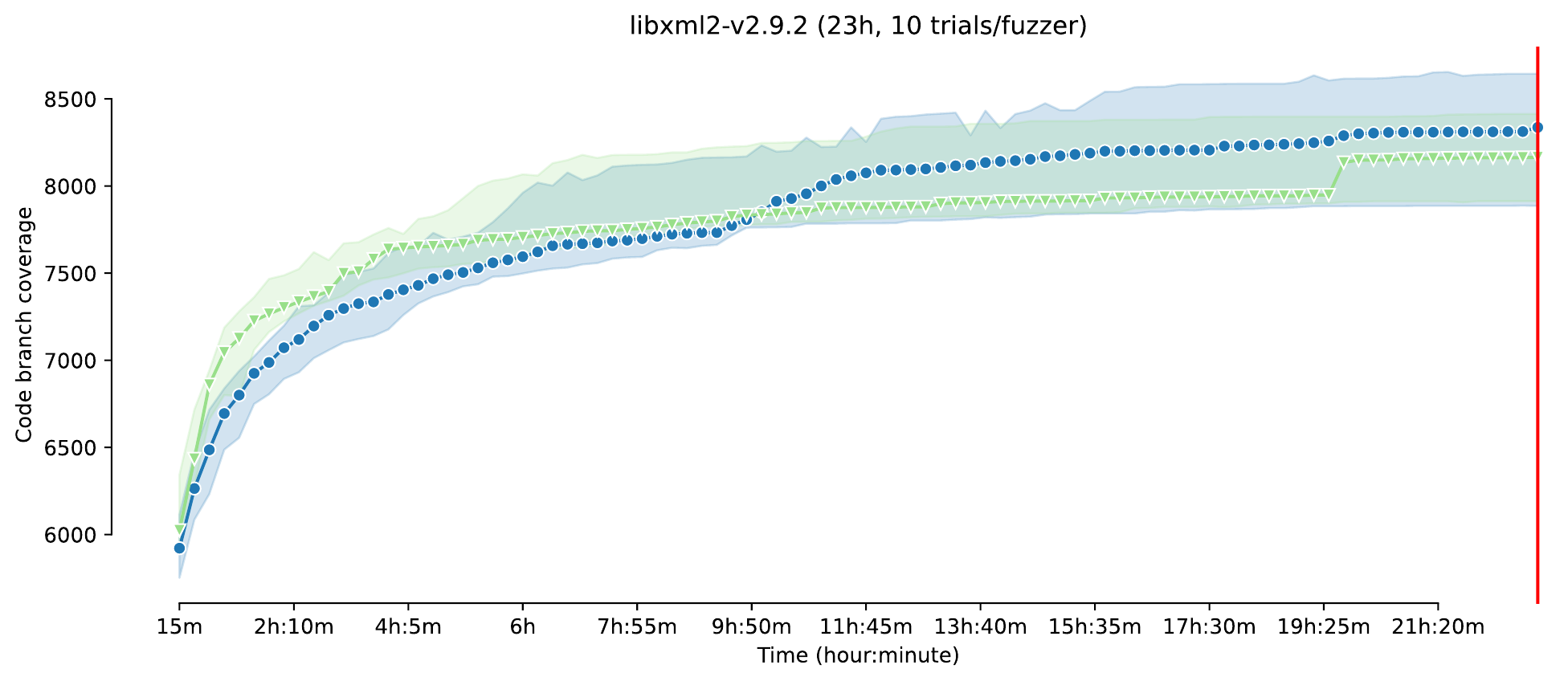} }}%
    \quad
    \subfloat[\centering libxslt]{{\includegraphics[width=4cm]{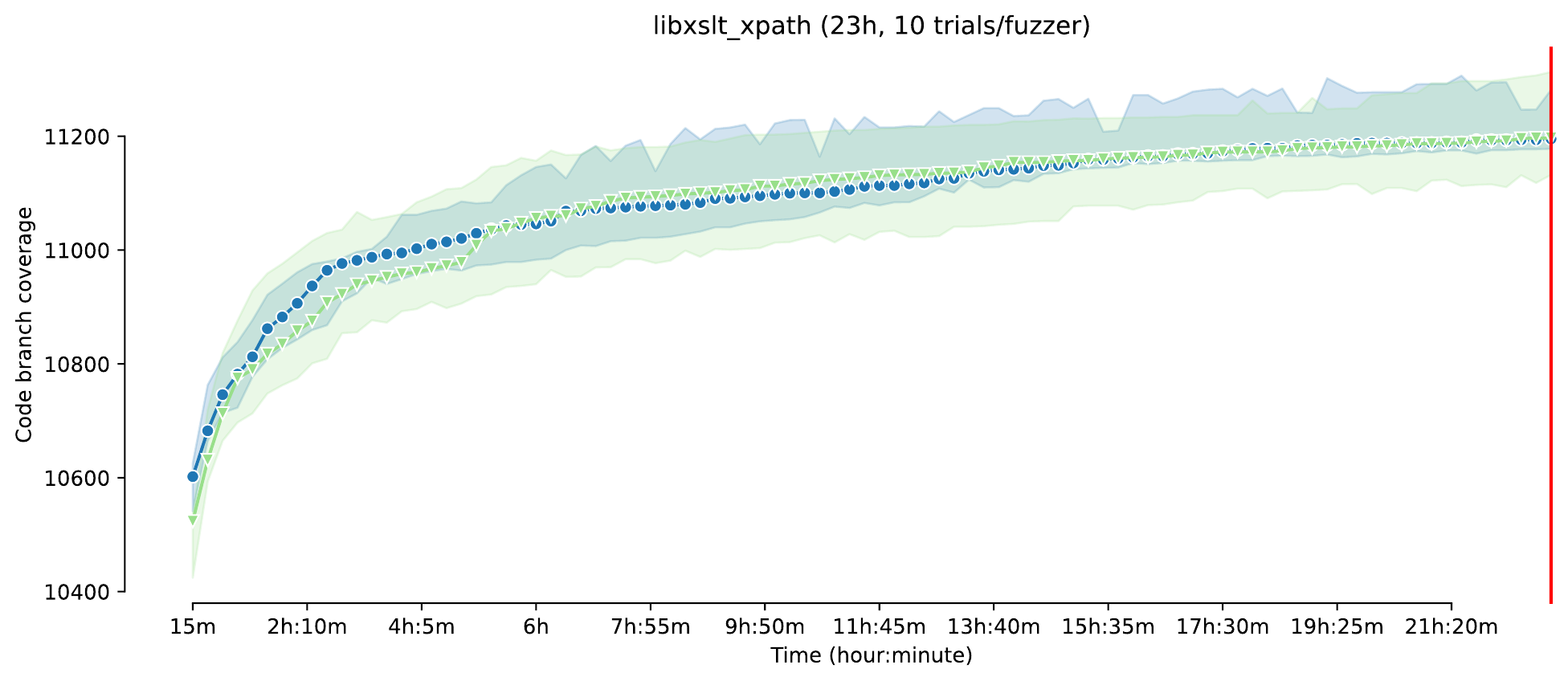} }}%
    \quad
    \subfloat[\centering mbedtls]{{\includegraphics[width=4cm]{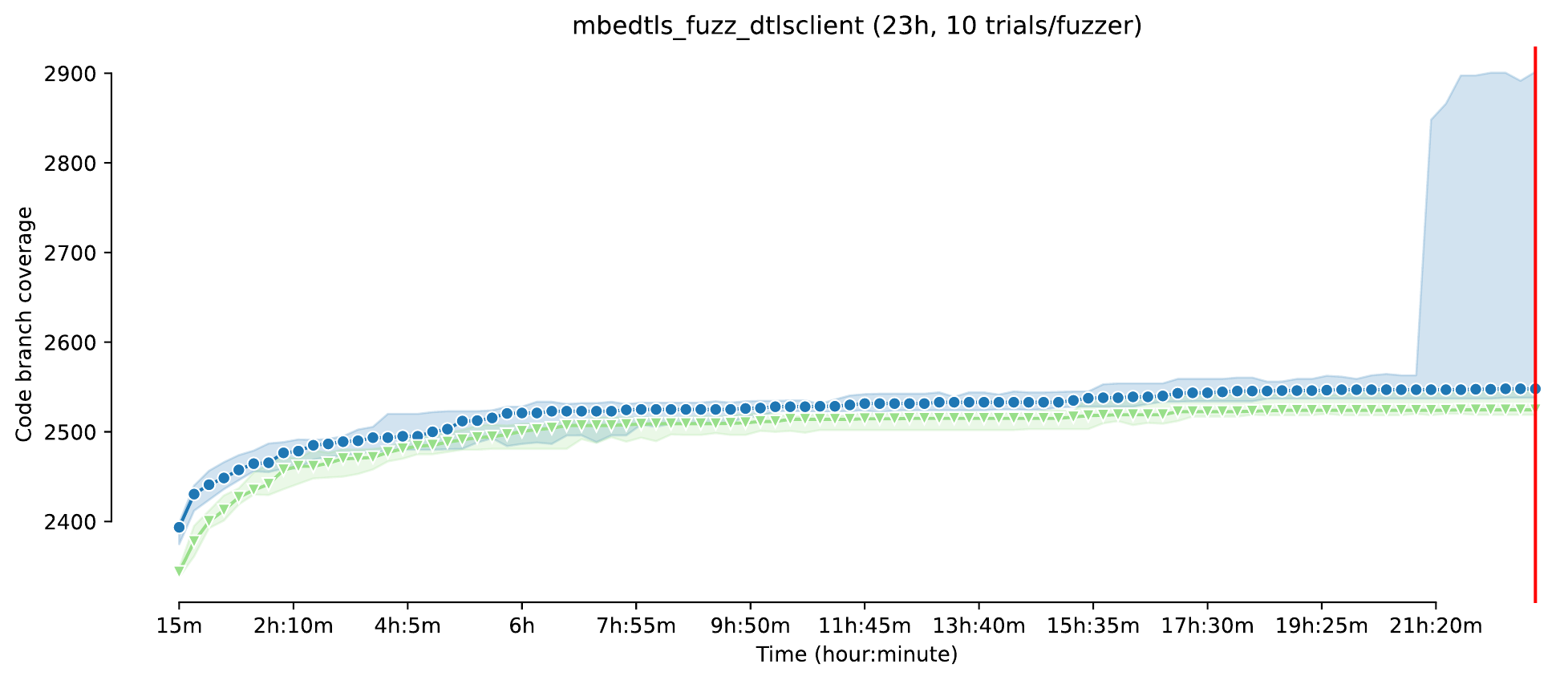} }}%
    \quad
    \subfloat[\centering openssl]{{\includegraphics[width=4cm]{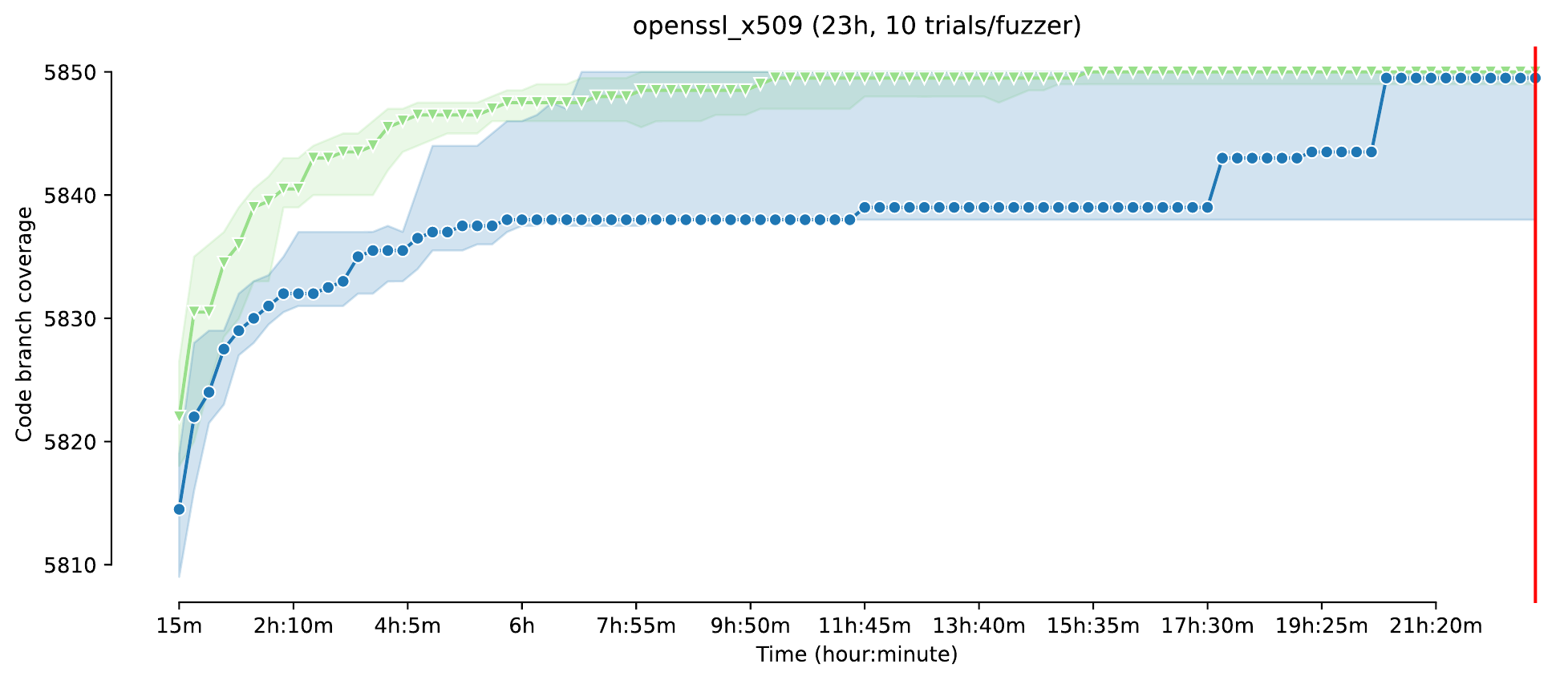} }}%
    \quad
    \subfloat[\centering openthread]{{\includegraphics[width=4cm]{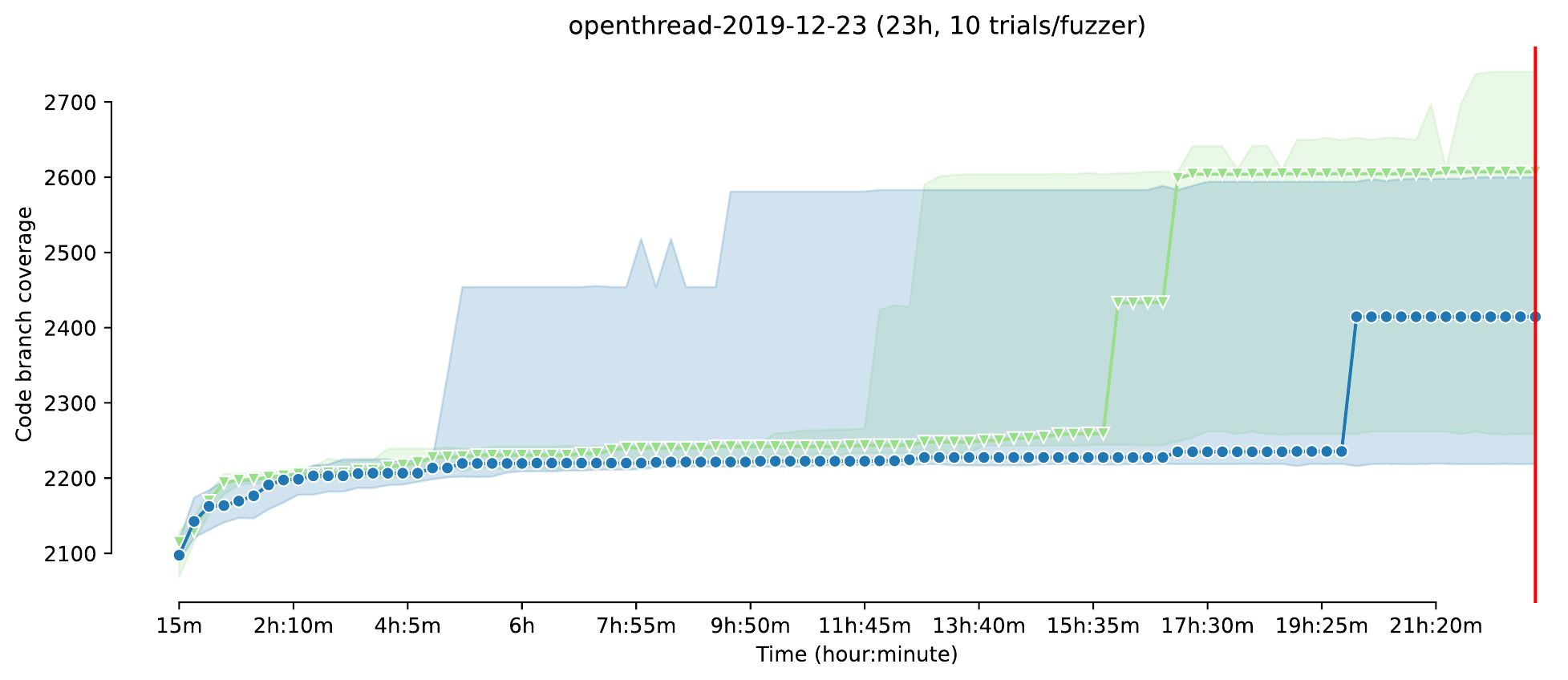} }}%
    \quad
    \subfloat[\centering re2]{{\includegraphics[width=4cm]{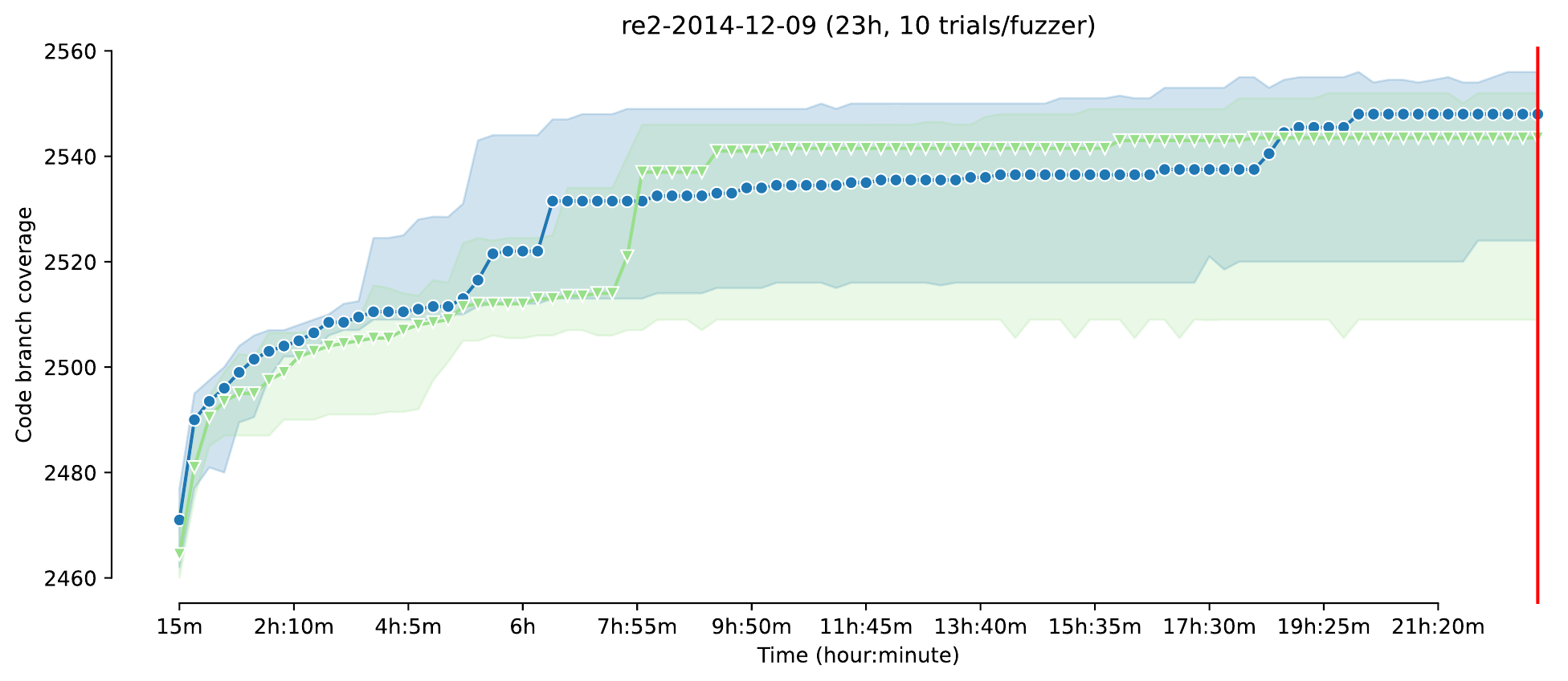} }}%
    \quad
    \subfloat[\centering sqlite3]{{\includegraphics[width=4cm]{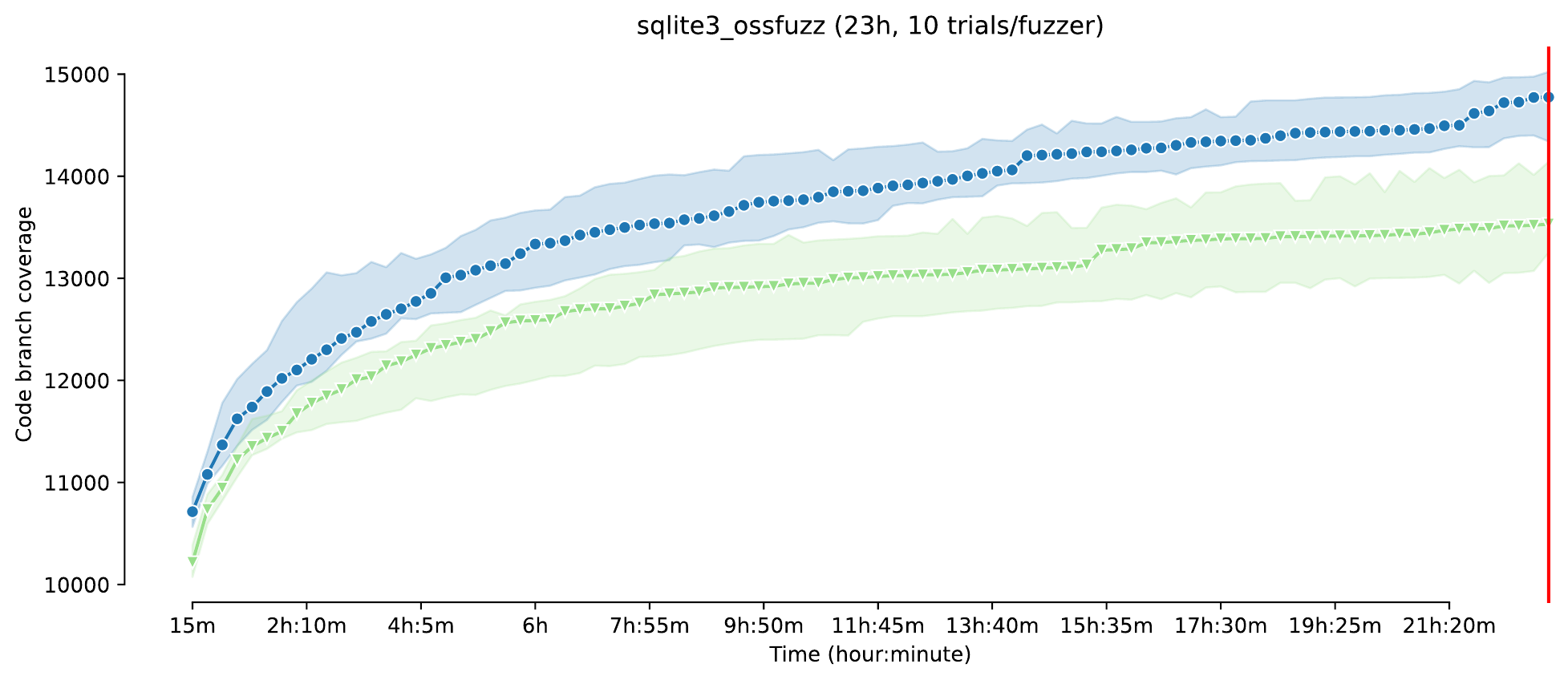} }}%
    \quad
    \subfloat[\centering vorbis]{{\includegraphics[width=4cm]{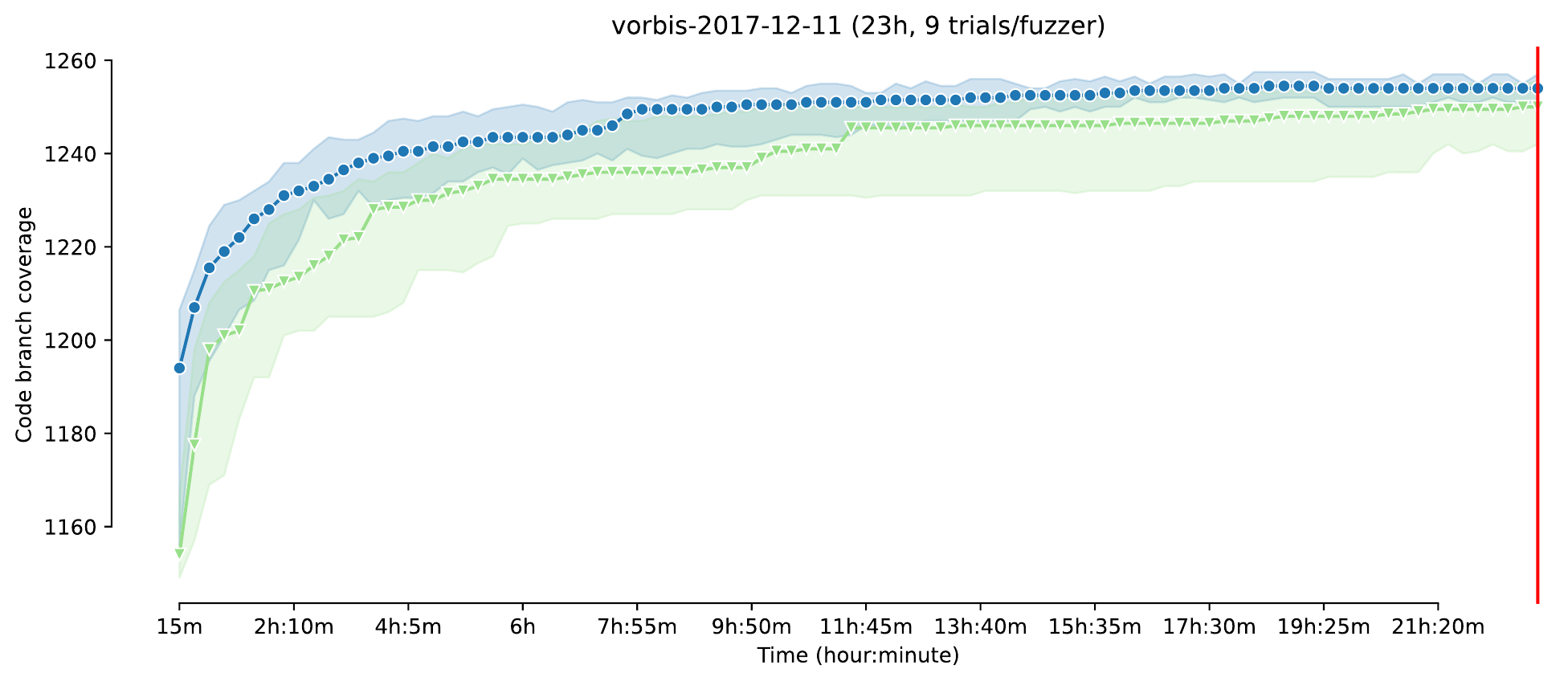} }}%
    \quad
    \subfloat[\centering woff2]{{\includegraphics[width=4cm]{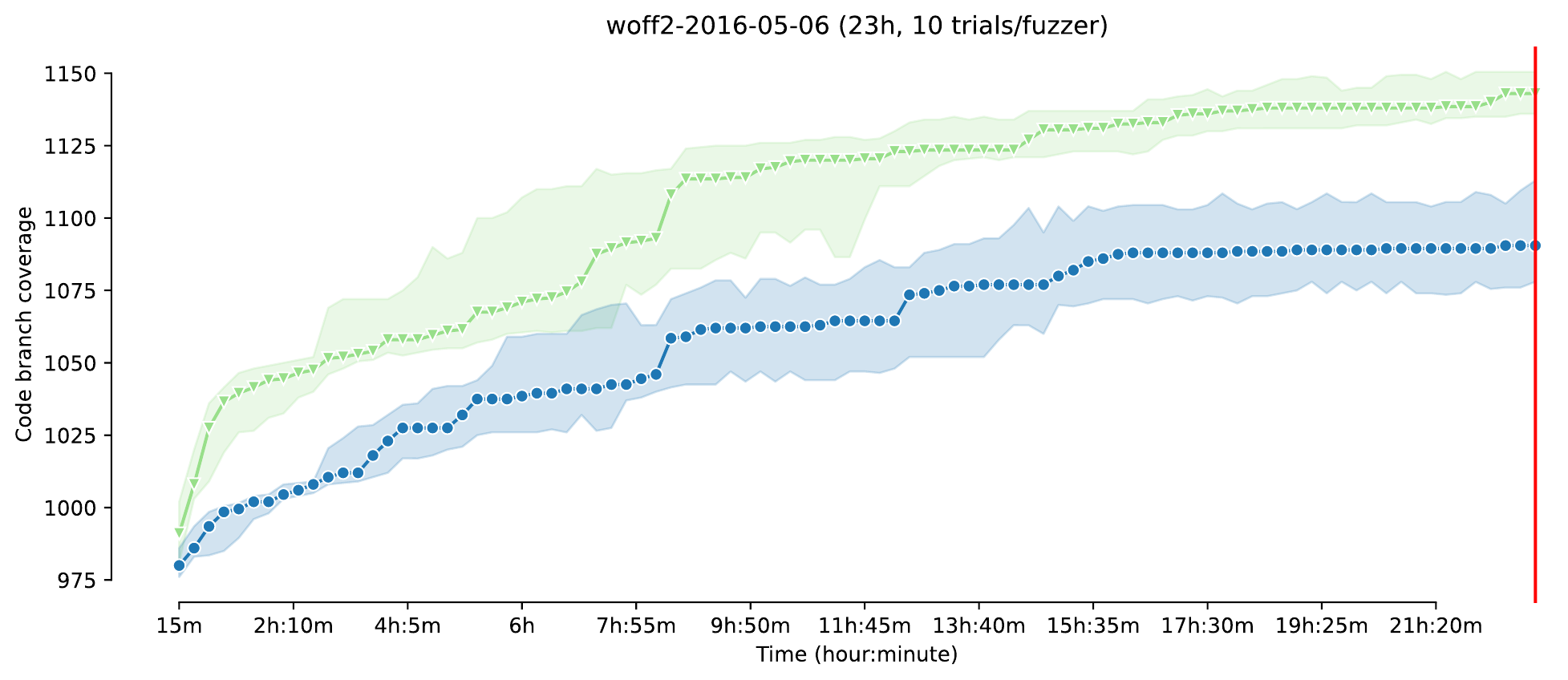} }}%
    \quad
    \subfloat{{\includegraphics[width=4cm]{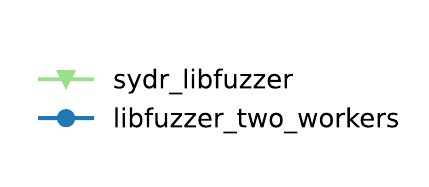} }}%
    \caption{Sydr-Fuzz vs 2xlibFuzzer (23h).}%
    \label{fig:libfuzzer_res}%
\end{figure}

\begin{figure}[h]%
    \centering
    \subfloat[\centering freetype2]{{\includegraphics[width=4cm]{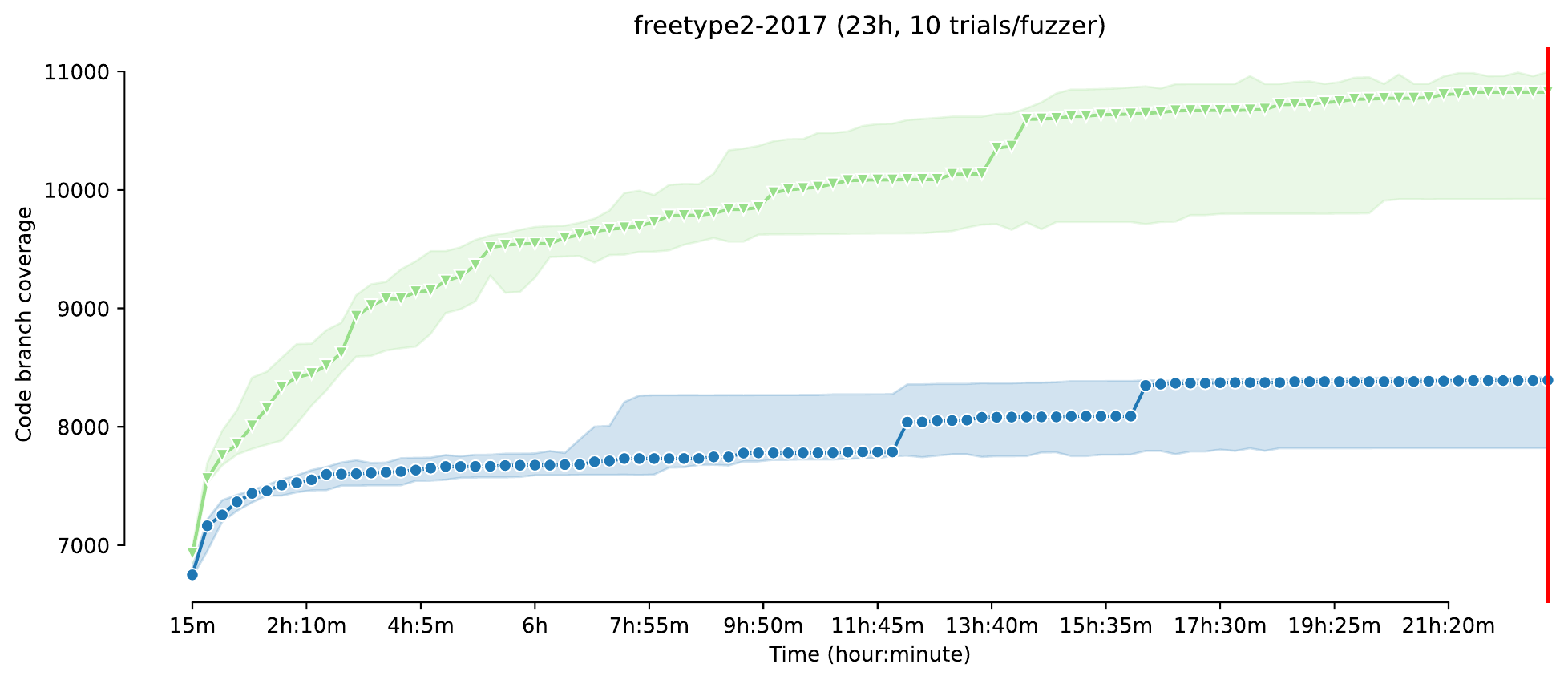} }}%
    \quad
    \subfloat[\centering harfbuzz]{{\includegraphics[width=4cm]{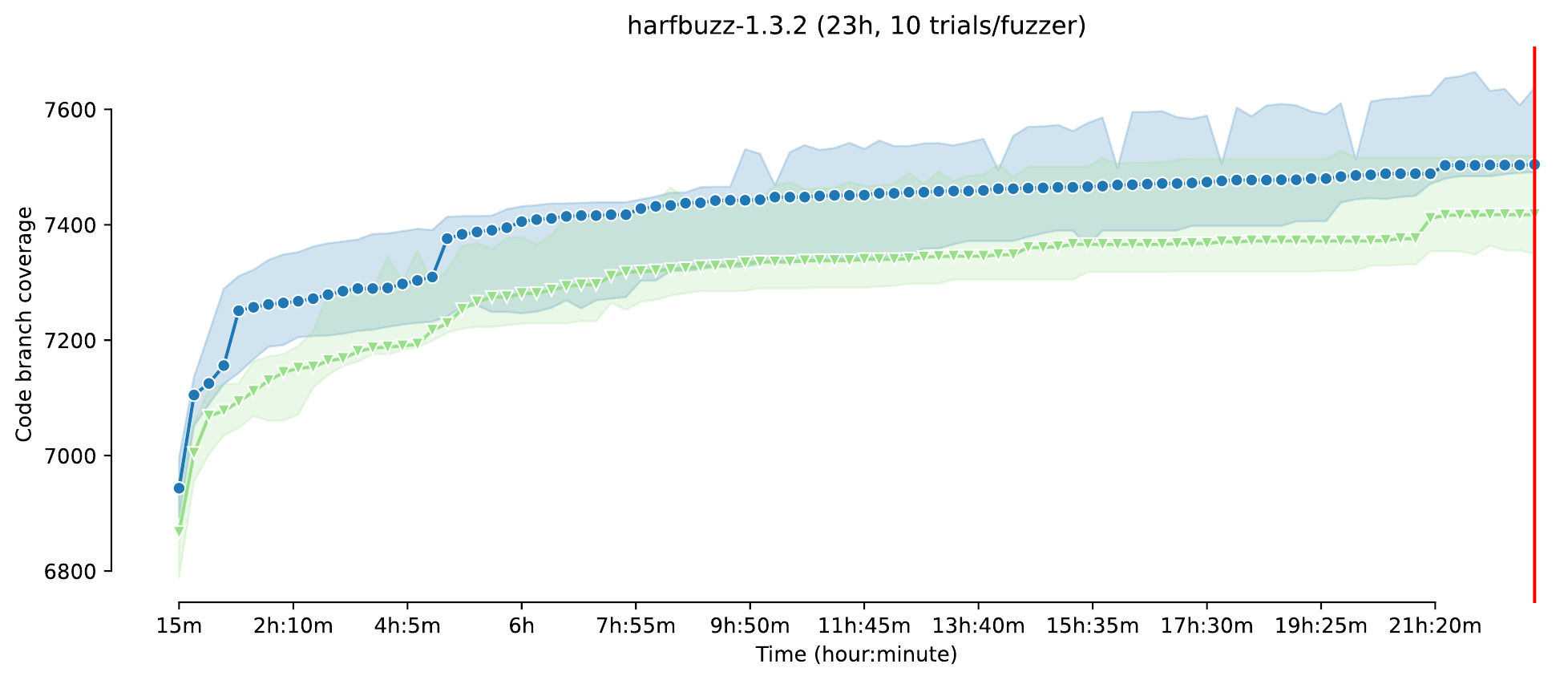} }}%
    \quad
    \subfloat[\centering lcms]{{\includegraphics[width=4cm]{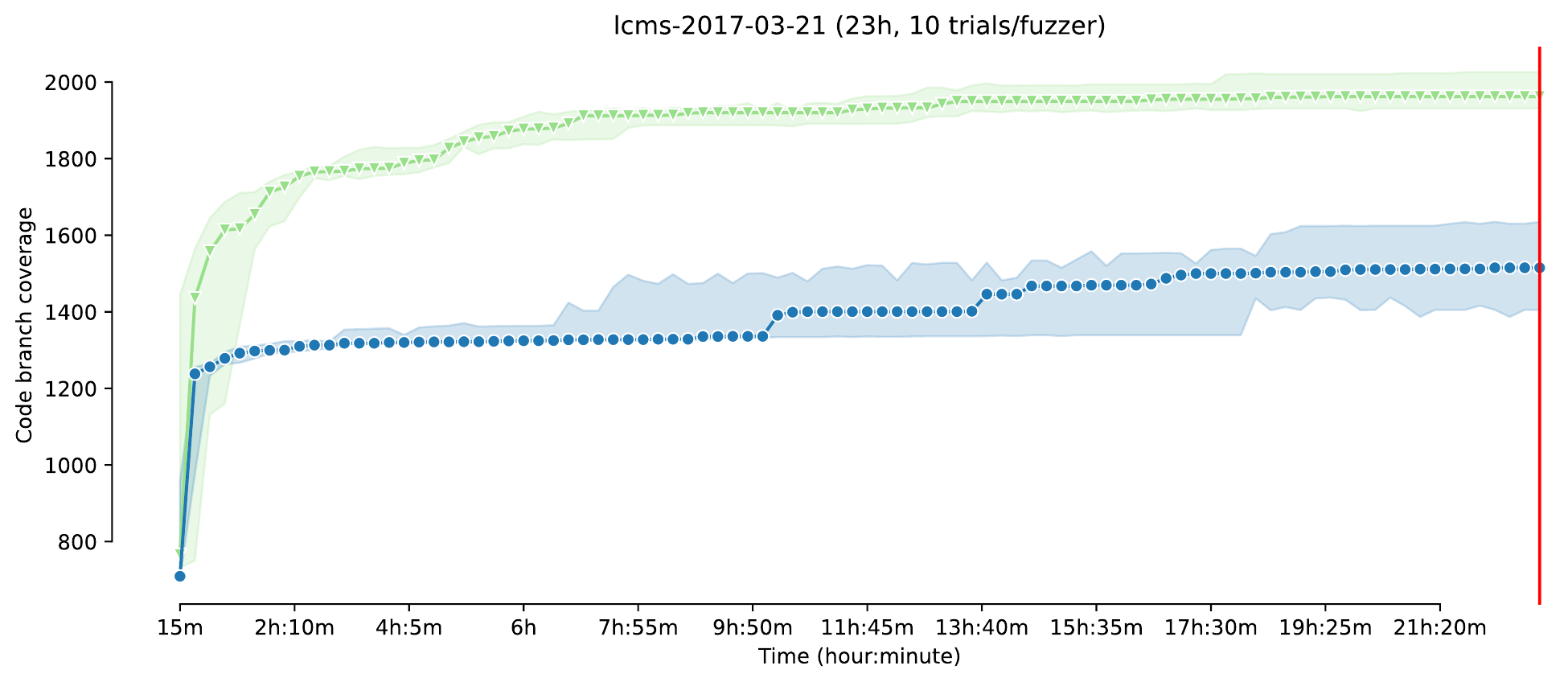} }}%
    \quad
    \subfloat[\centering libjpeg\_turbo]{{\includegraphics[width=4cm]{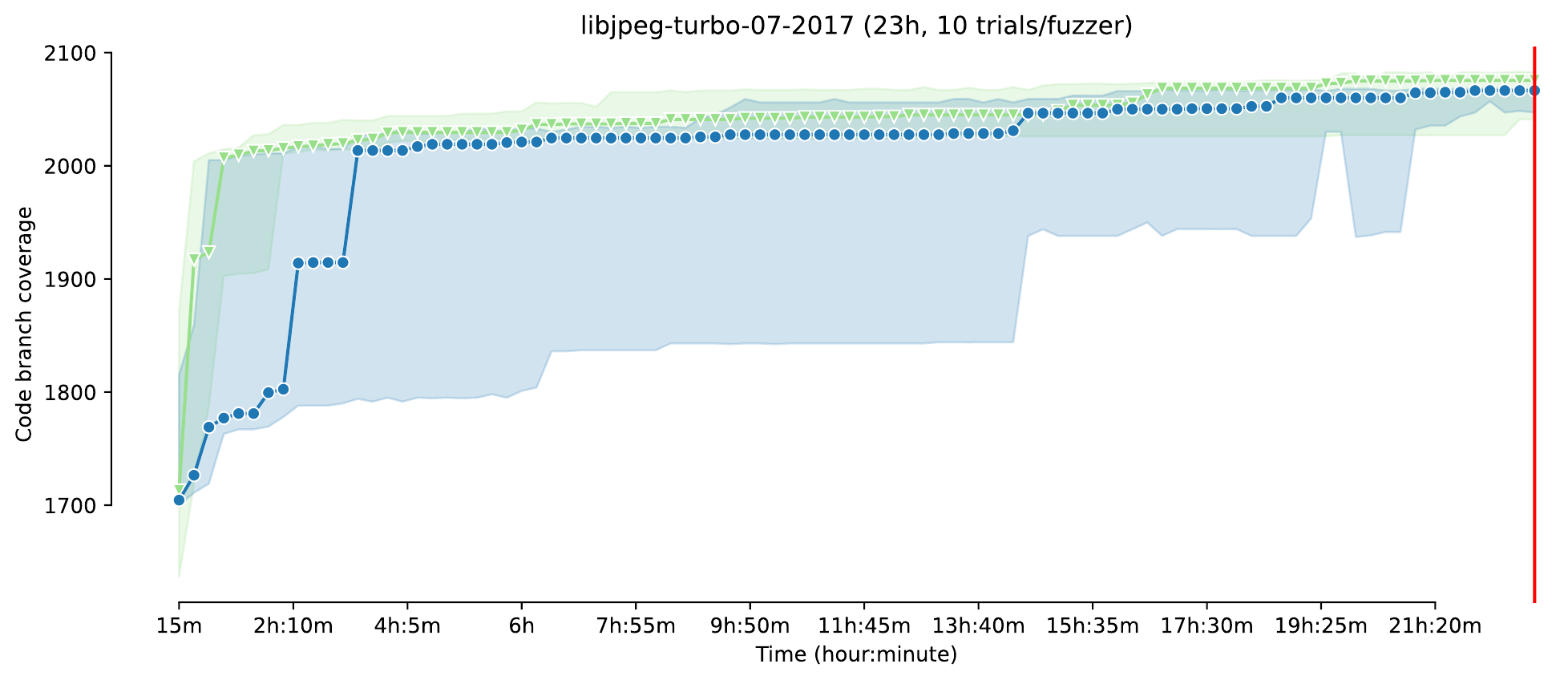} }}%
    \quad
    \subfloat[\centering libpng]{{\includegraphics[width=4cm]{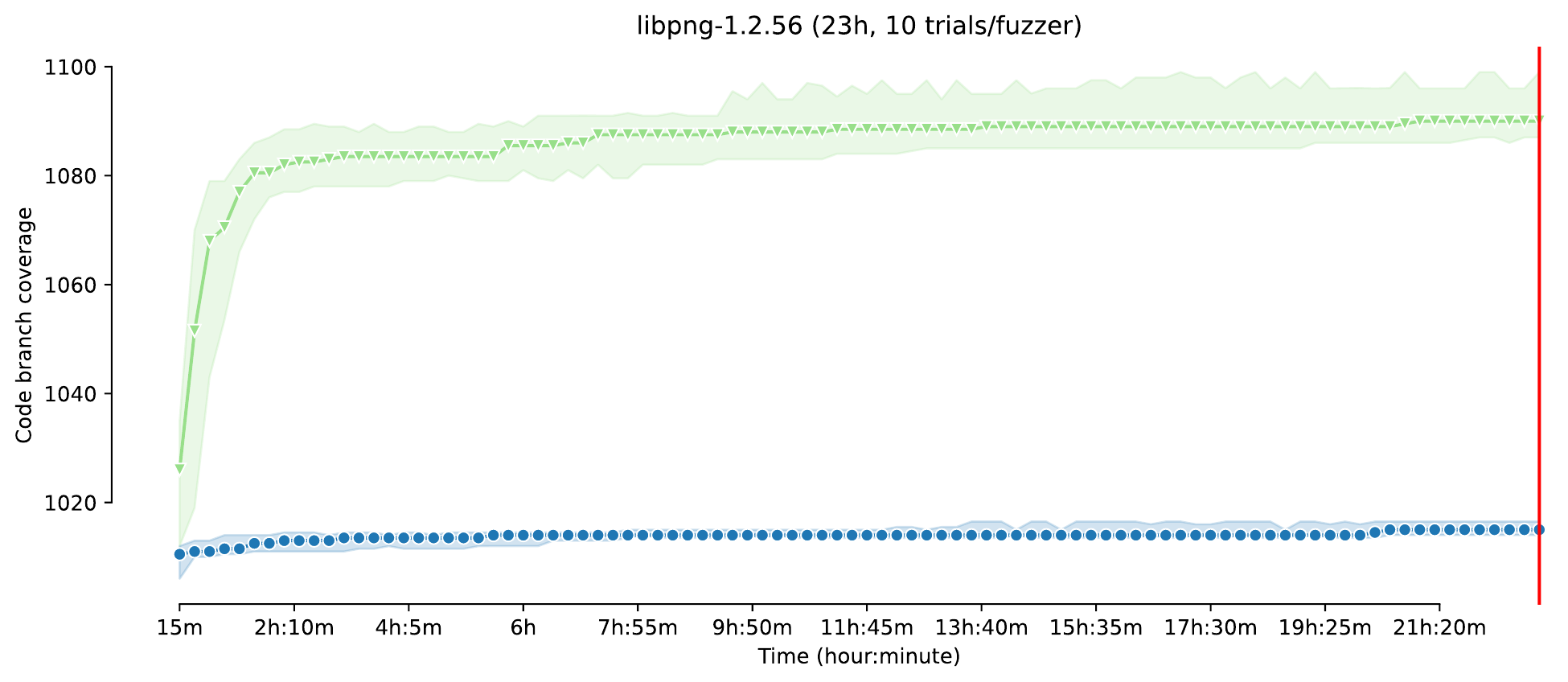} }}%
    \quad
    \subfloat[\centering libxml2]{{\includegraphics[width=4cm]{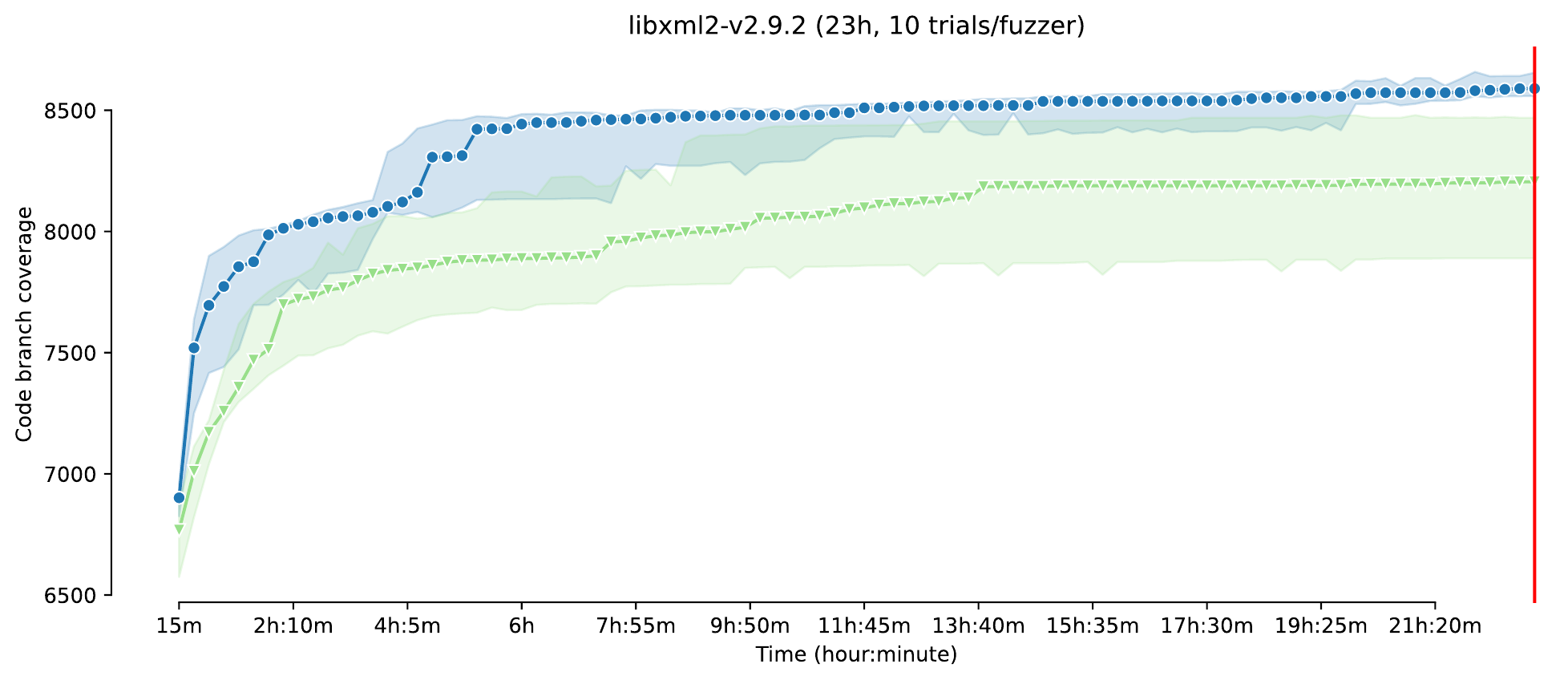} }}%
    \quad
    \subfloat[\centering libxslt]{{\includegraphics[width=4cm]{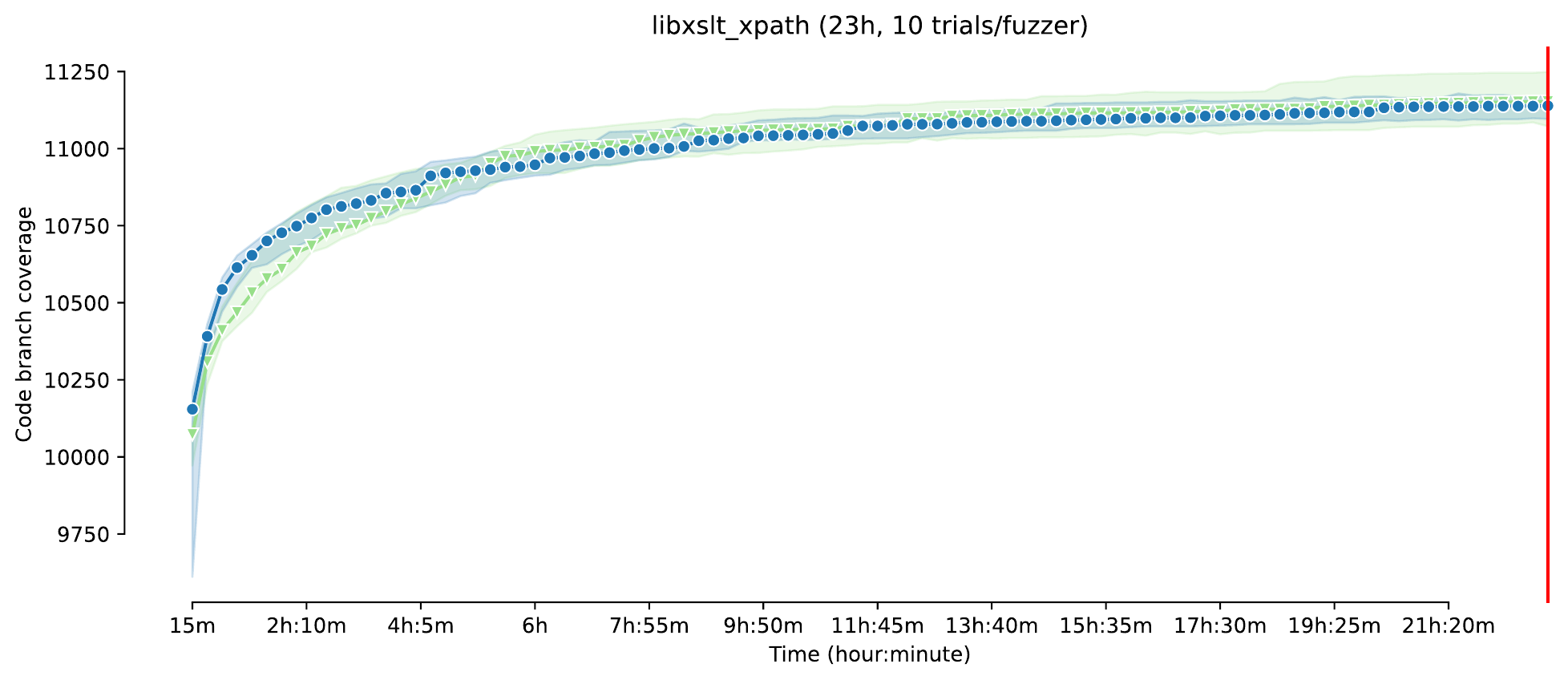} }}%
    \quad
    \subfloat[\centering mbedtls]{{\includegraphics[width=4cm]{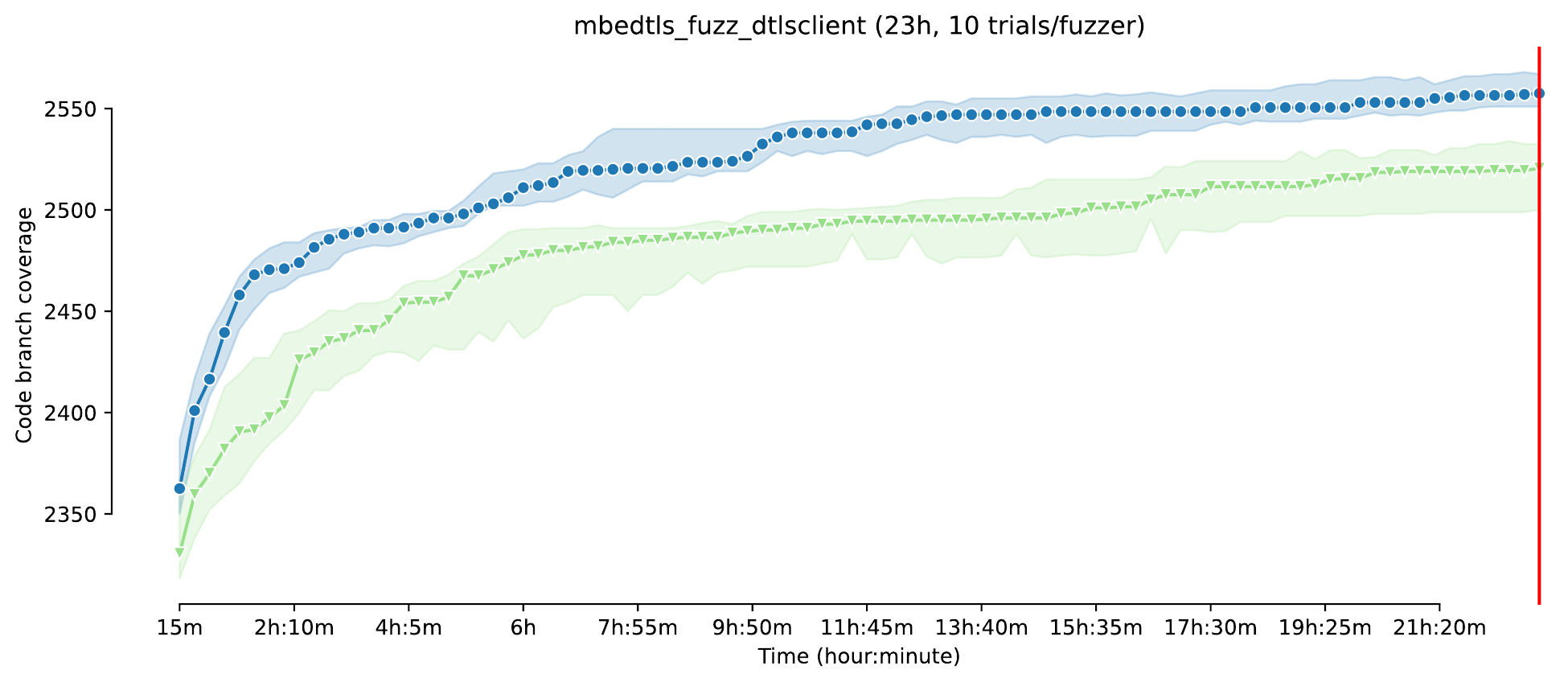} }}%
    \quad
    \subfloat[\centering openssl]{{\includegraphics[width=4cm]{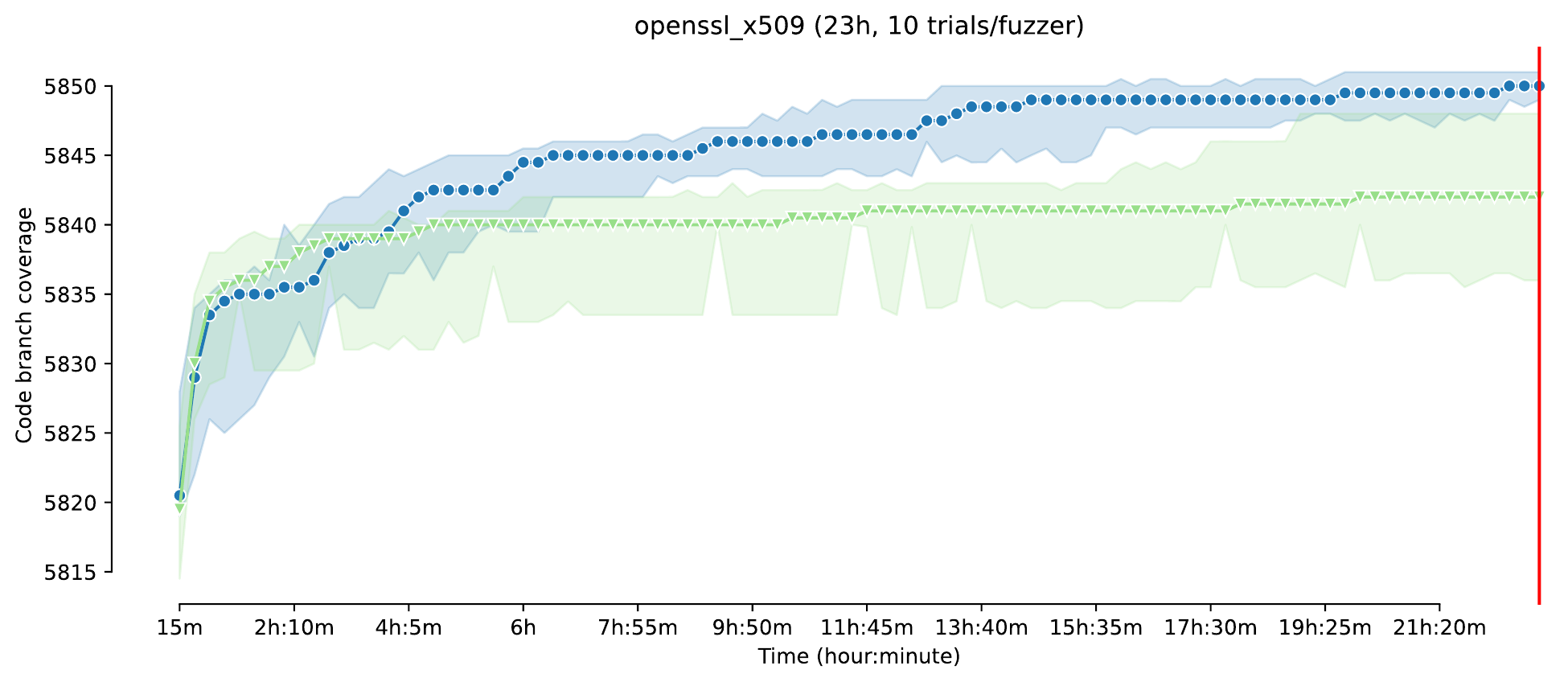} }}%
    \quad
    \subfloat[\centering openthread]{{\includegraphics[width=4cm]{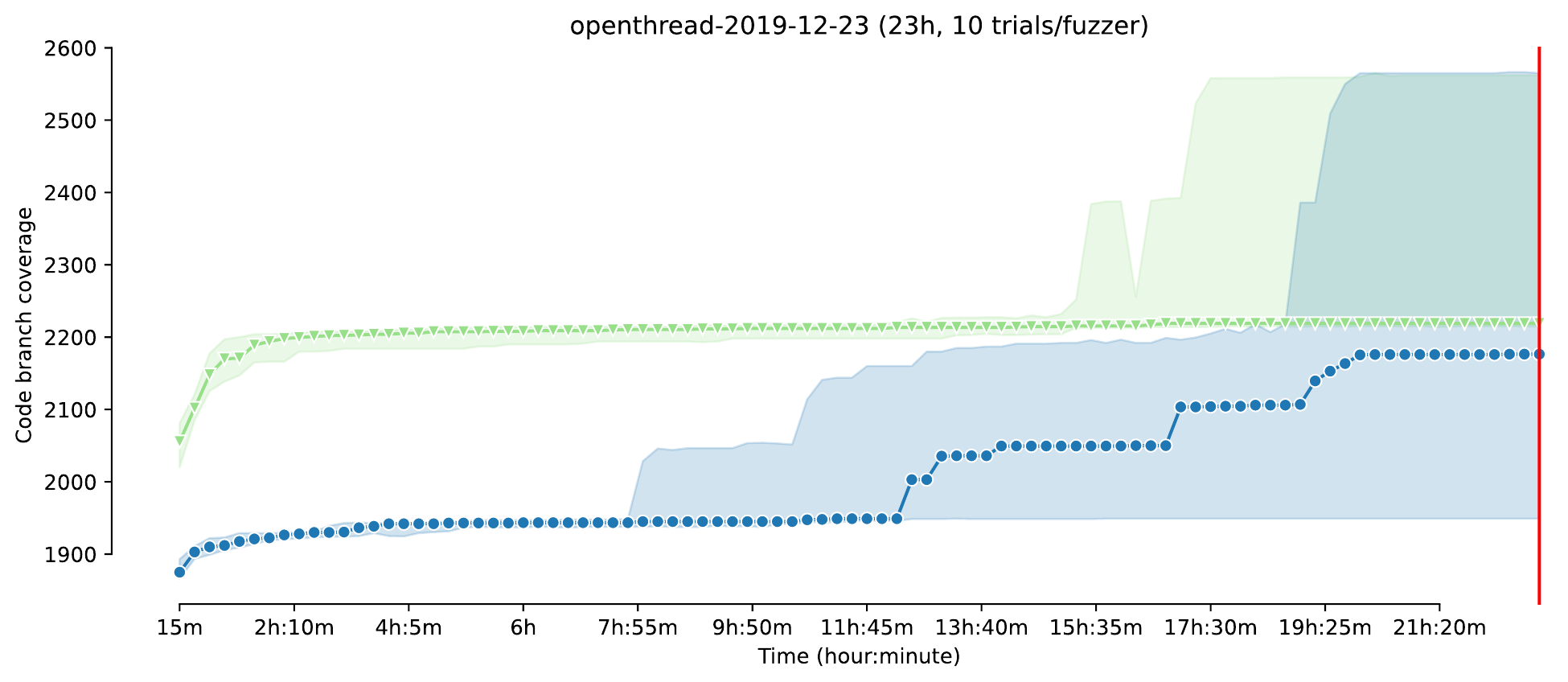} }}%
    \quad
    \subfloat[\centering re2]{{\includegraphics[width=4cm]{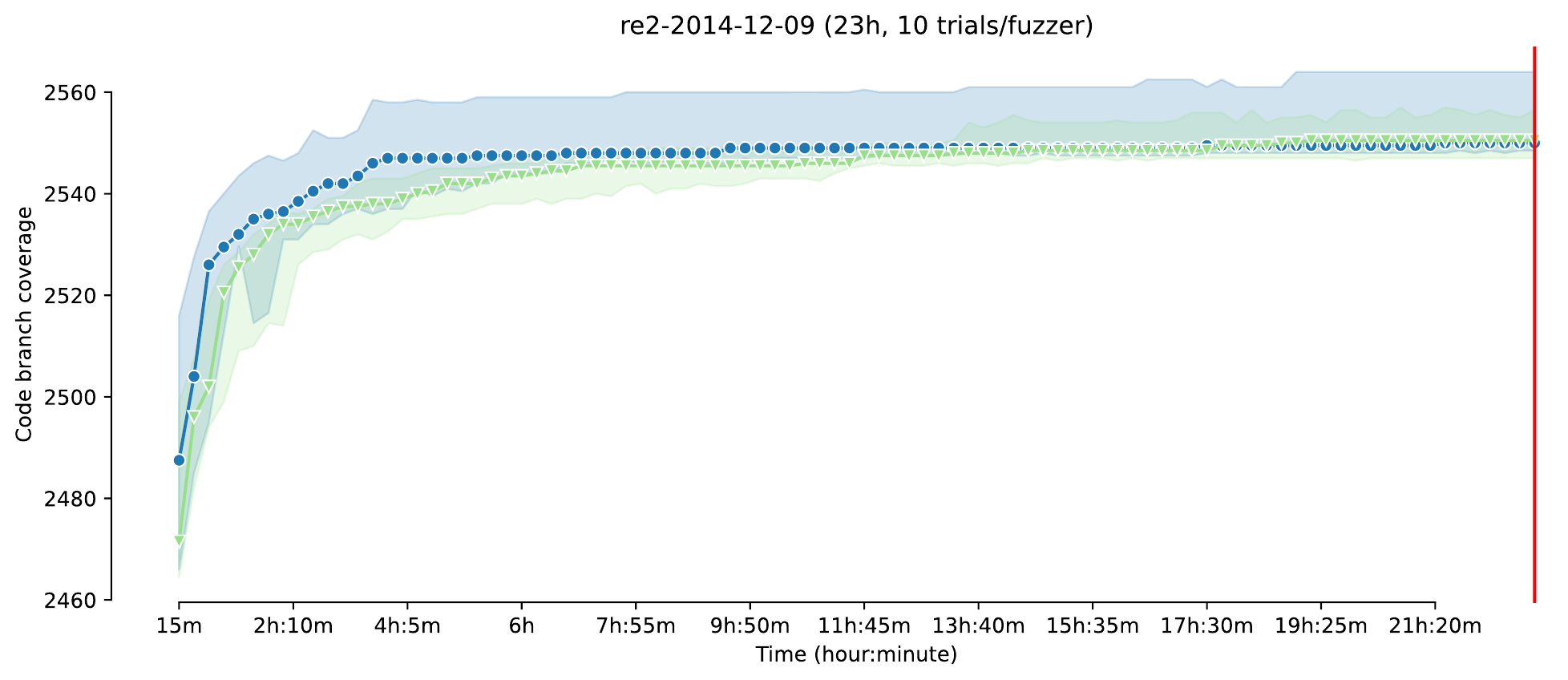} }}%
    \quad
    \subfloat[\centering sqlite3]{{\includegraphics[width=4cm]{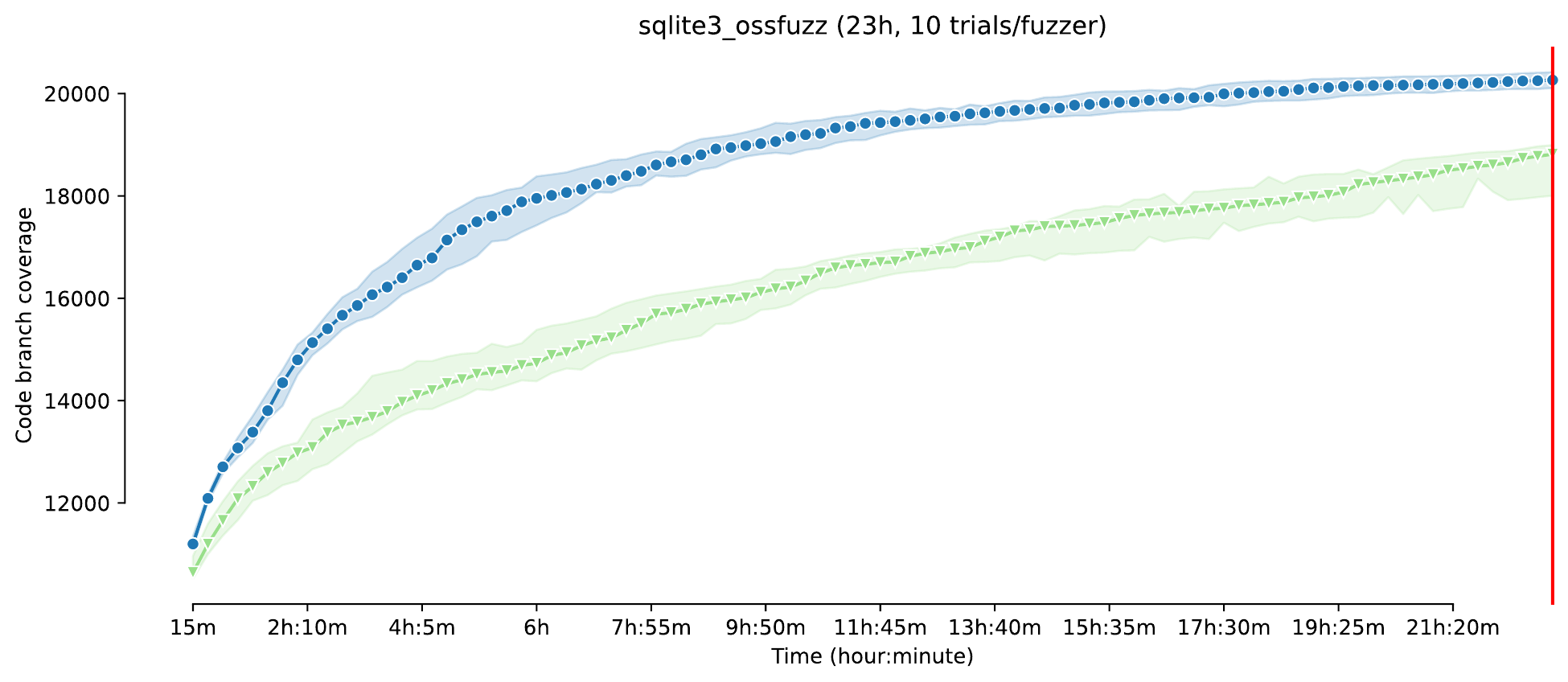} }}%
    \quad
    \subfloat[\centering vorbis]{{\includegraphics[width=4cm]{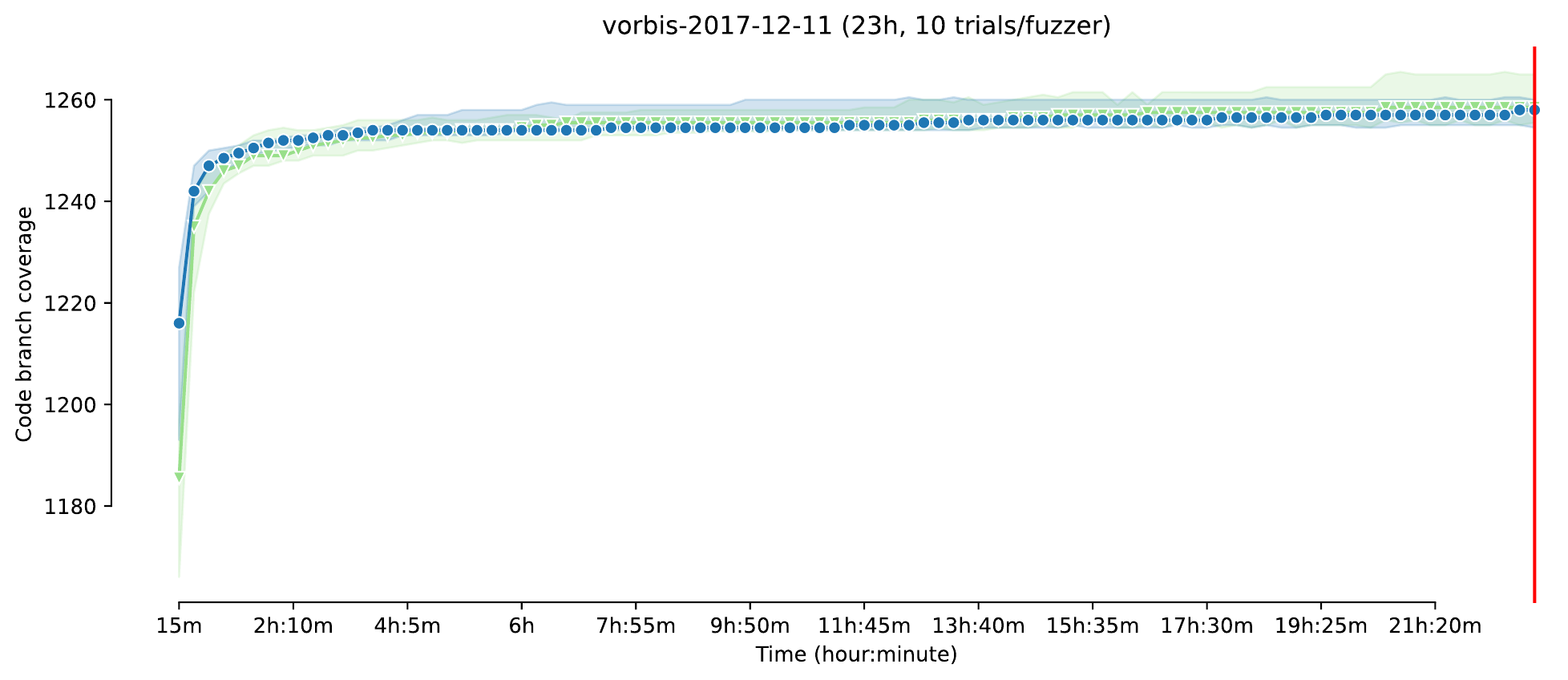} }}%
    \quad
    \subfloat[\centering woff2]{{\includegraphics[width=4cm]{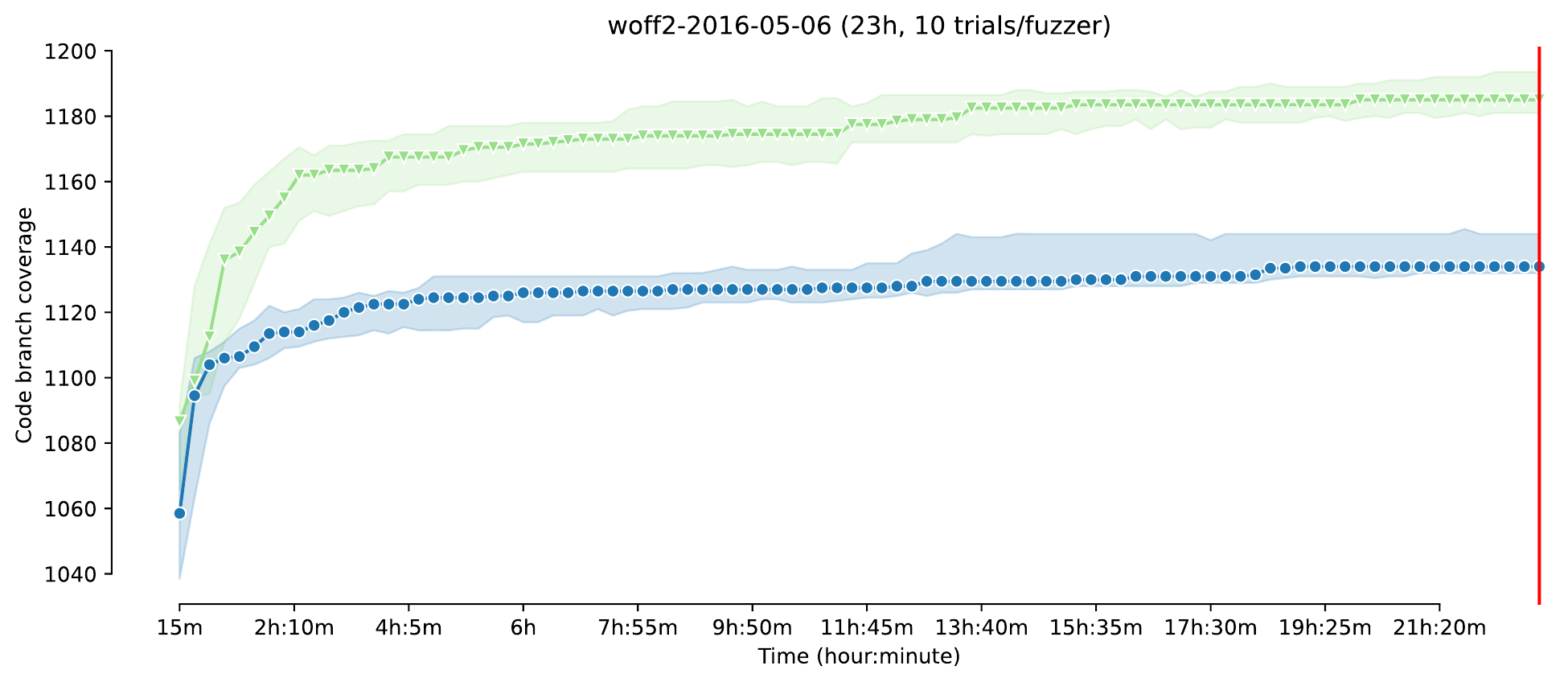} }}%
    \quad
    \subfloat{{\includegraphics[width=4cm]{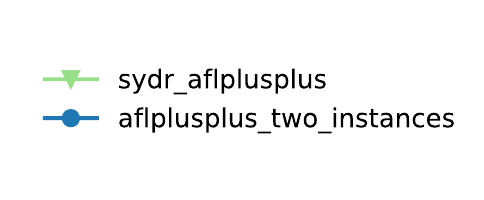} }}%
    \caption{Sydr-Fuzz vs 2xAFL++ (23h).}%
    \label{fig:afl_res}%
\end{figure}

\begin{figure}[h]%
    \centering
    \subfloat[\centering freetype2]{{\includegraphics[width=4cm]{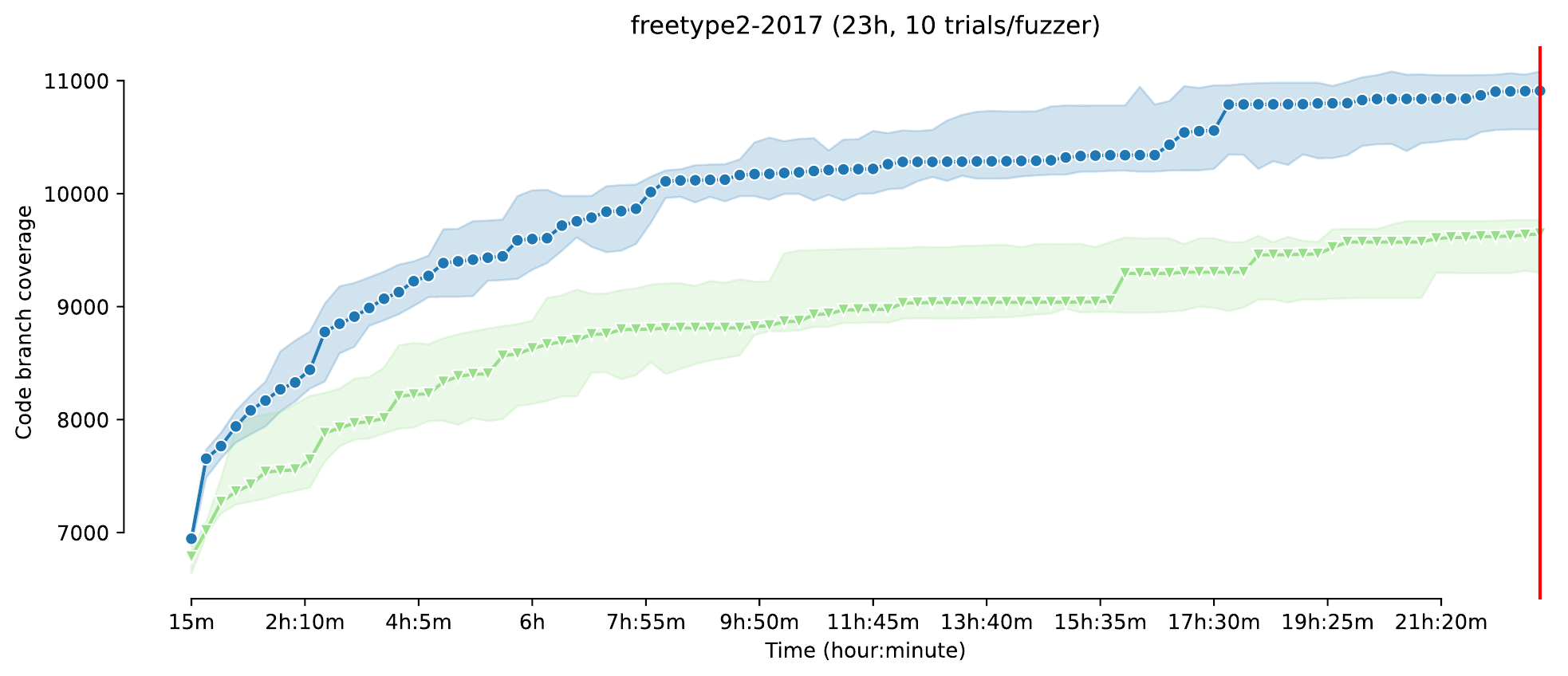} }}%
    \quad
    \subfloat[\centering harfbuzz]{{\includegraphics[width=4cm]{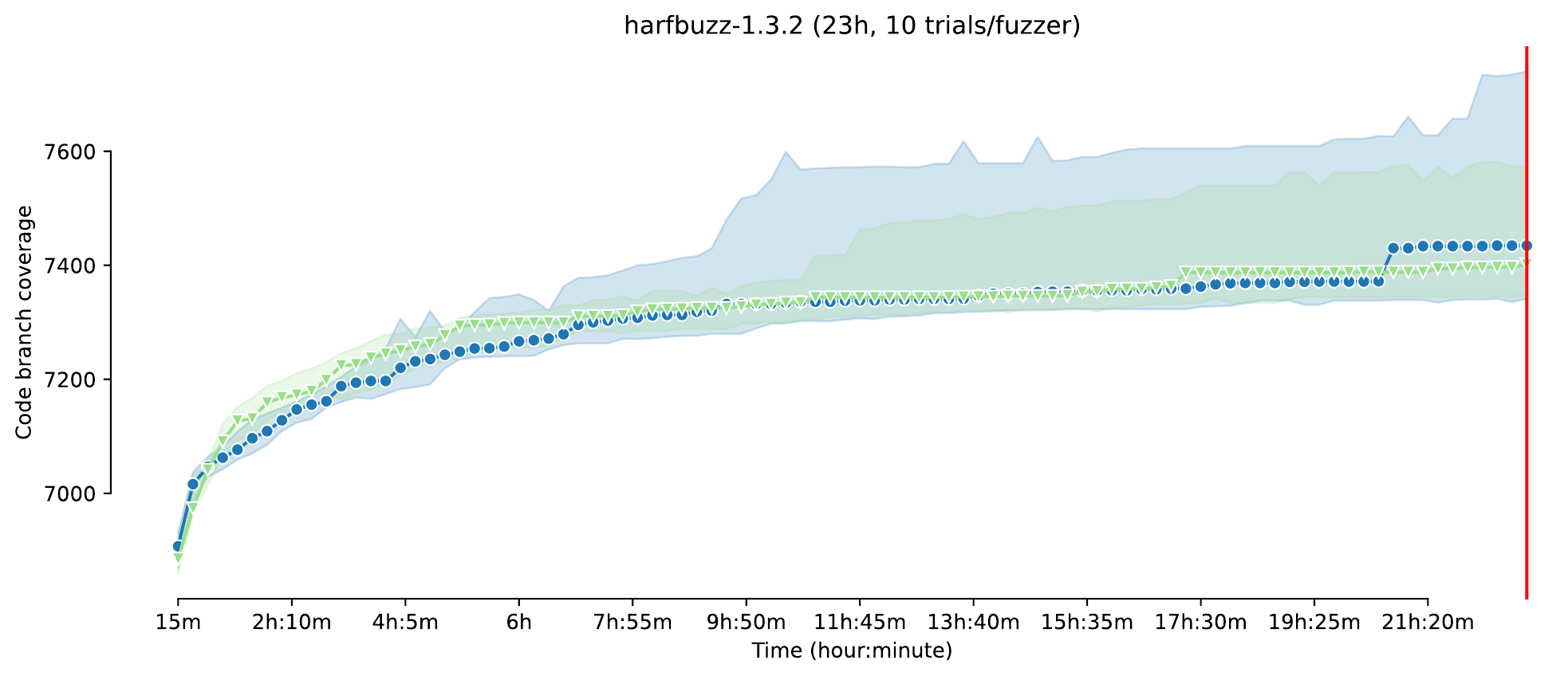} }}%
    \quad
    \subfloat[\centering lcms]{{\includegraphics[width=4cm]{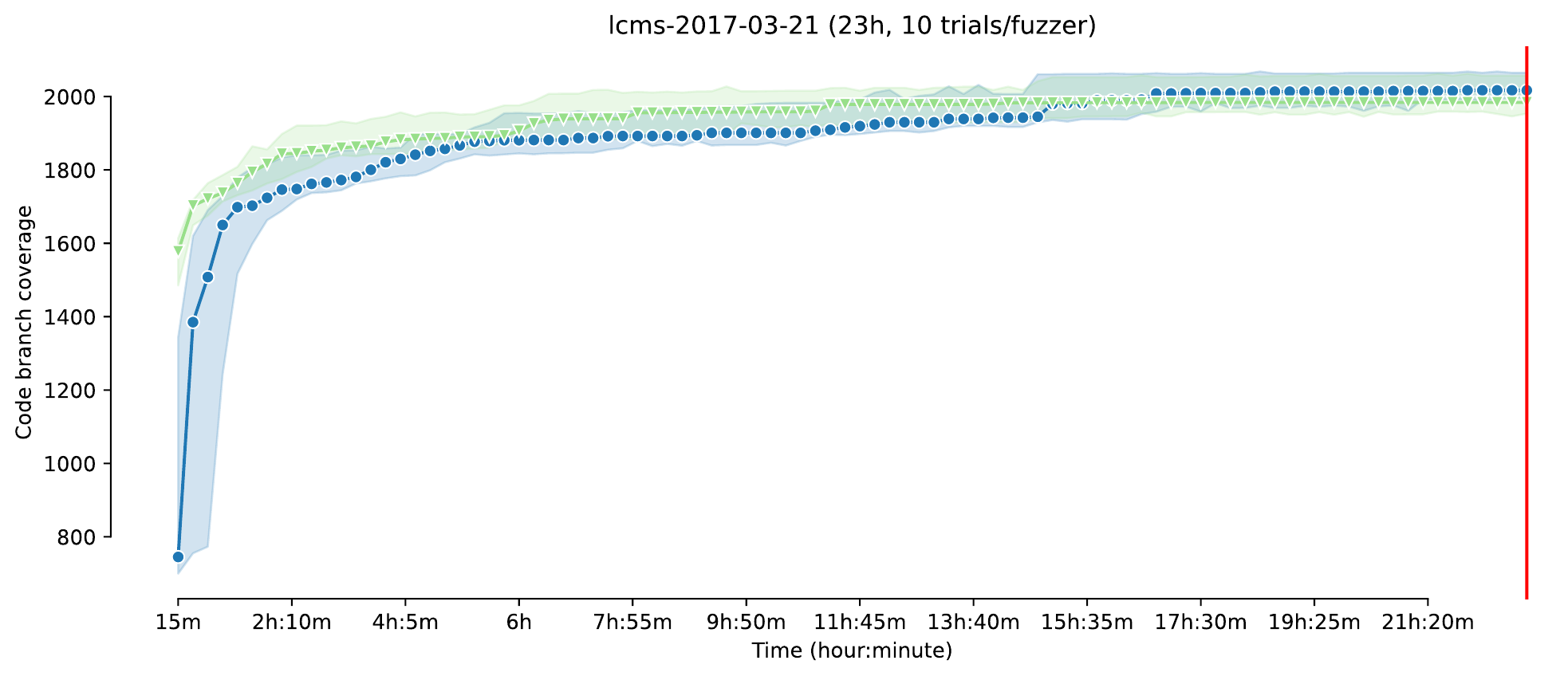} }}%
    \quad
    \subfloat[\centering libjpeg\_turbo]{{\includegraphics[width=4cm]{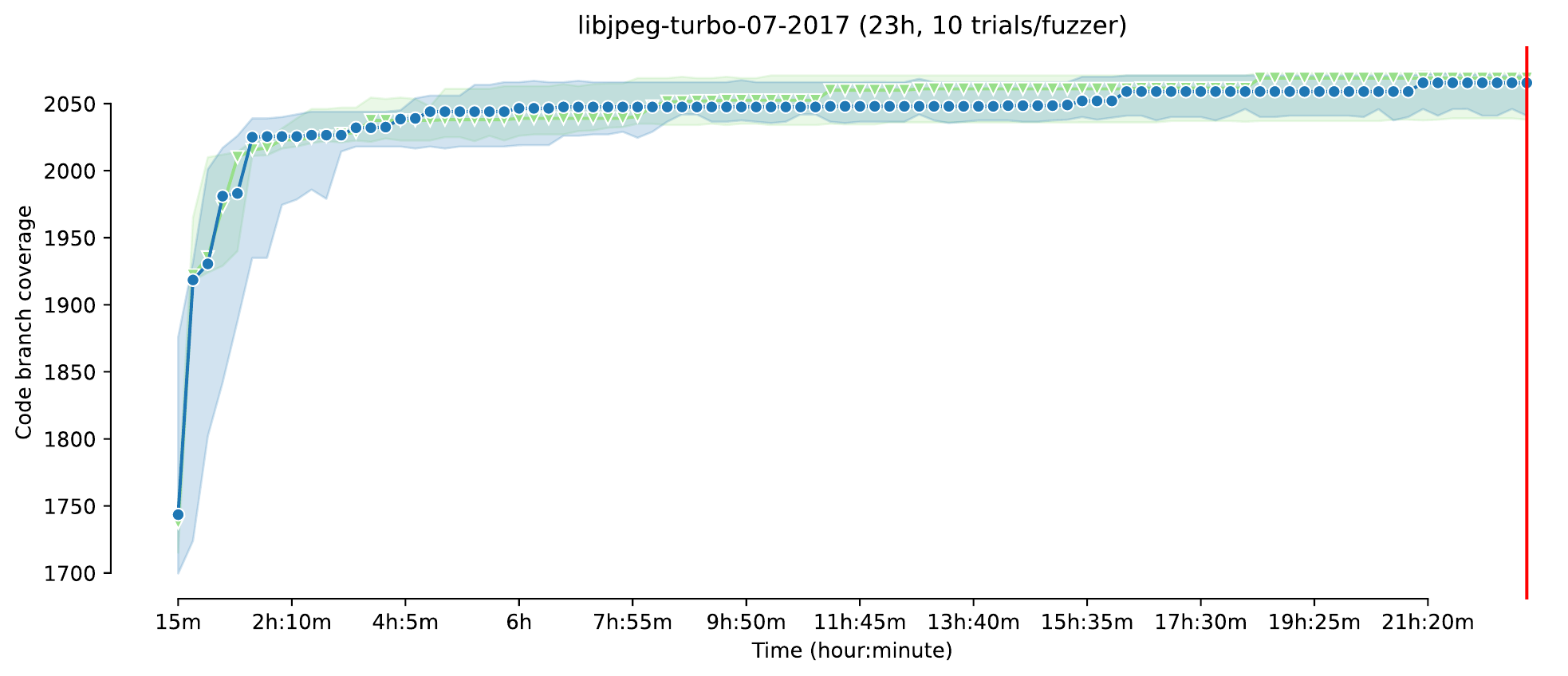} }}%
    \quad
    \subfloat[\centering libpng]{{\includegraphics[width=4cm]{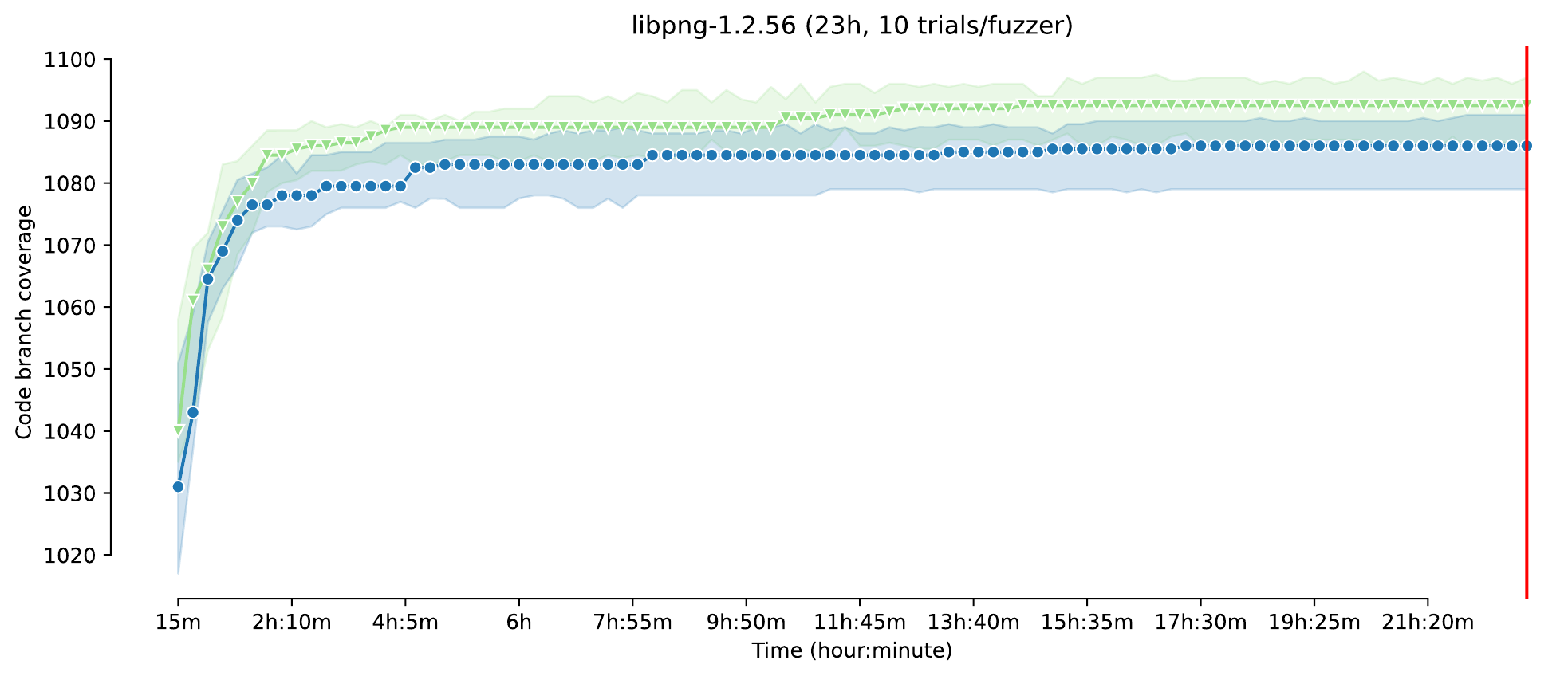} }}%
    \quad
    \subfloat[\centering libxml2]{{\includegraphics[width=4cm]{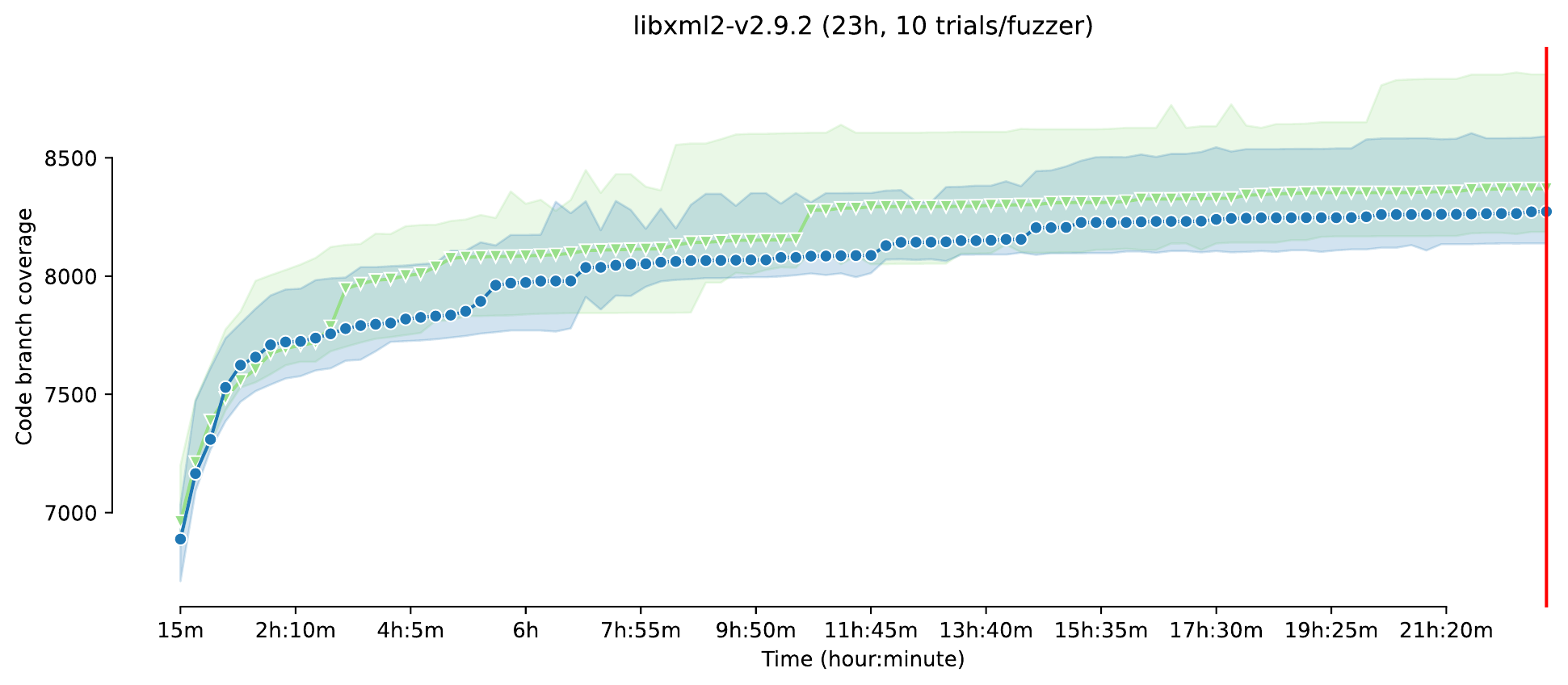} }}%
    \quad
    \subfloat[\centering mbedtls]{{\includegraphics[width=4cm]{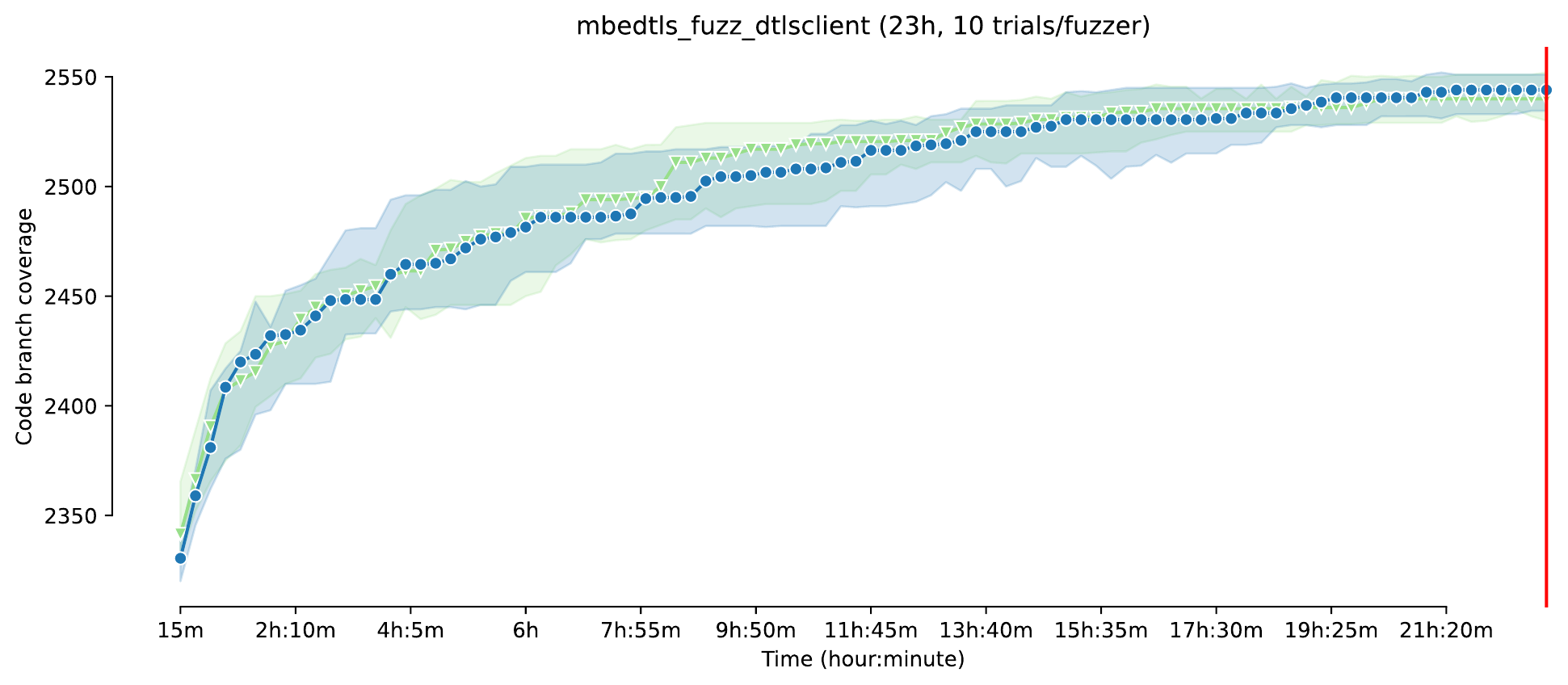} }}%
    \quad
    \subfloat[\centering openthread]{{\includegraphics[width=4cm]{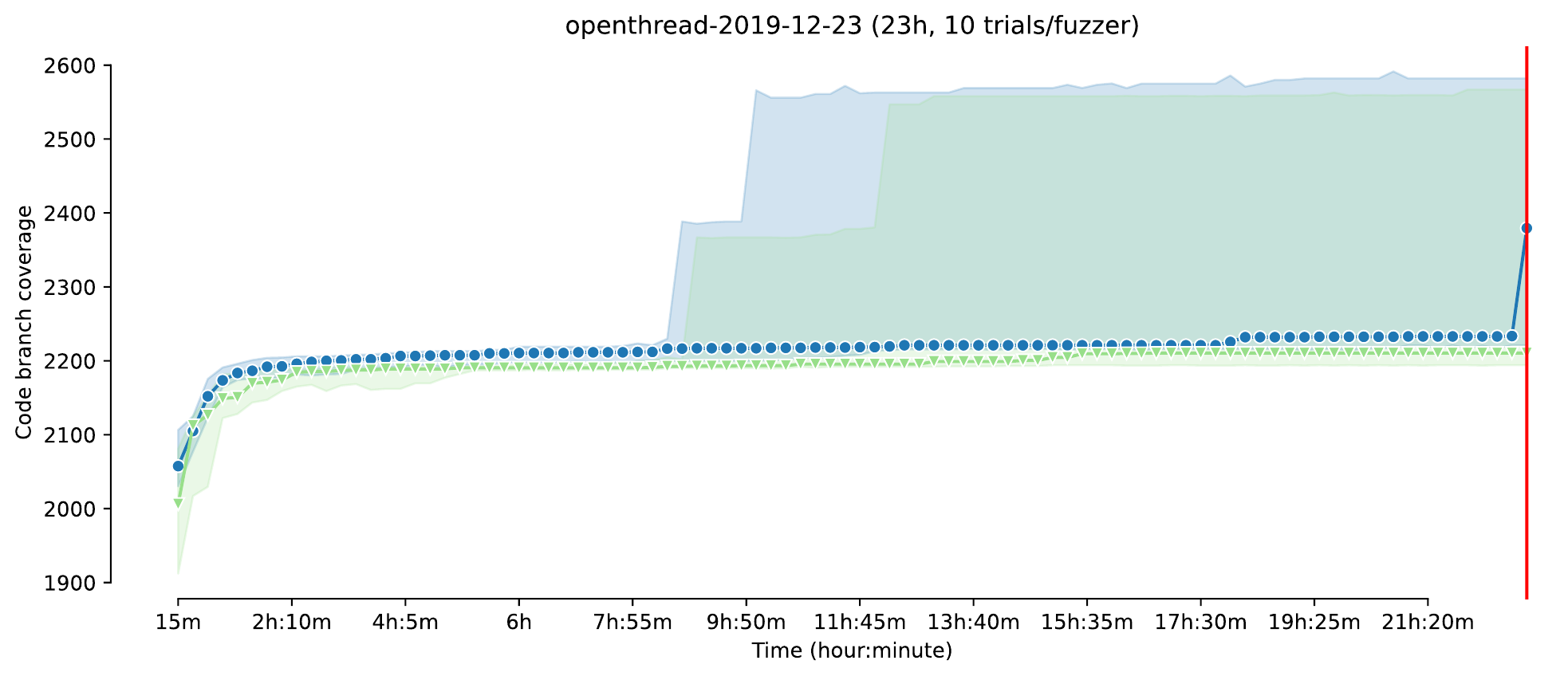} }}%
    \quad
    \subfloat[\centering re2]{{\includegraphics[width=4cm]{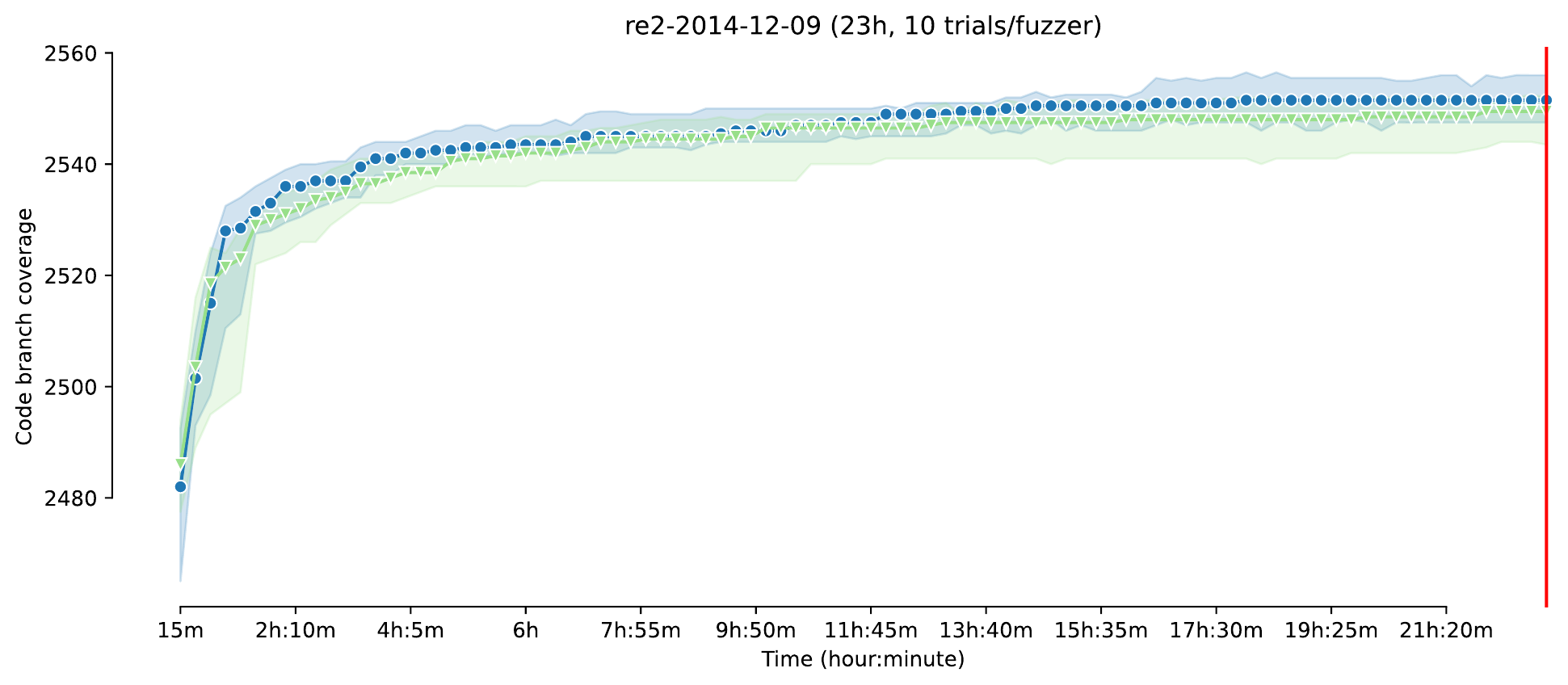} }}%
    \quad
    \subfloat[\centering sqlite3]{{\includegraphics[width=4cm]{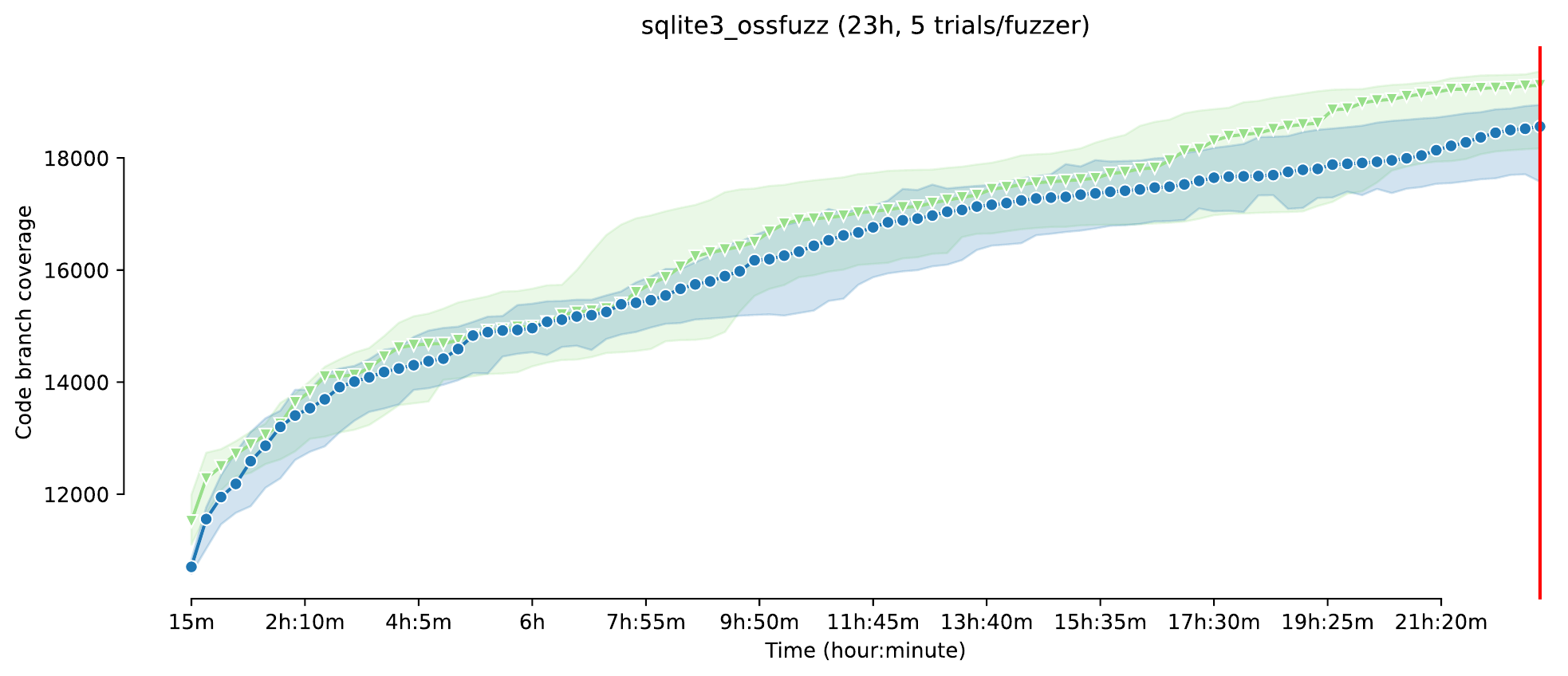} }}%
    \quad
    \subfloat[\centering vorbis]{{\includegraphics[width=4cm]{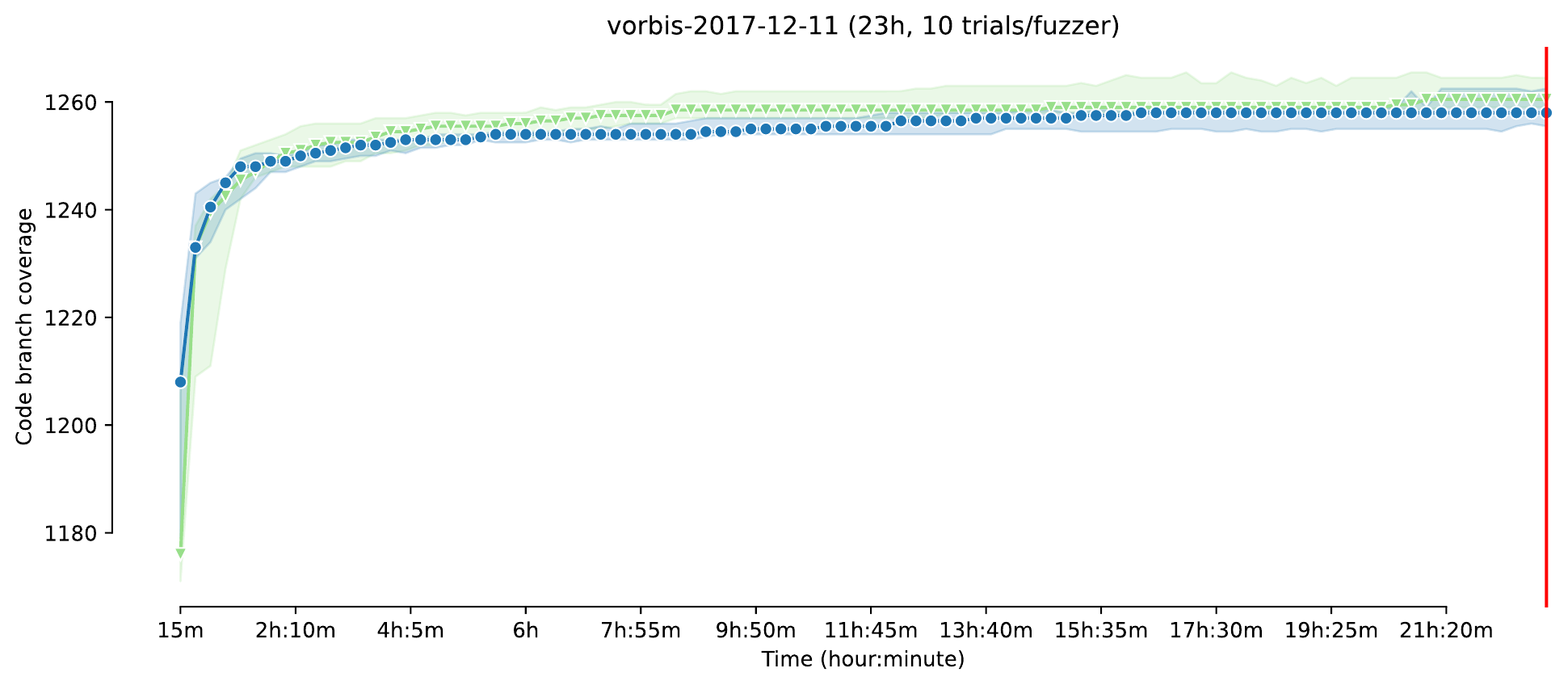} }}%
    \quad
    \subfloat[\centering woff2]{{\includegraphics[width=4cm]{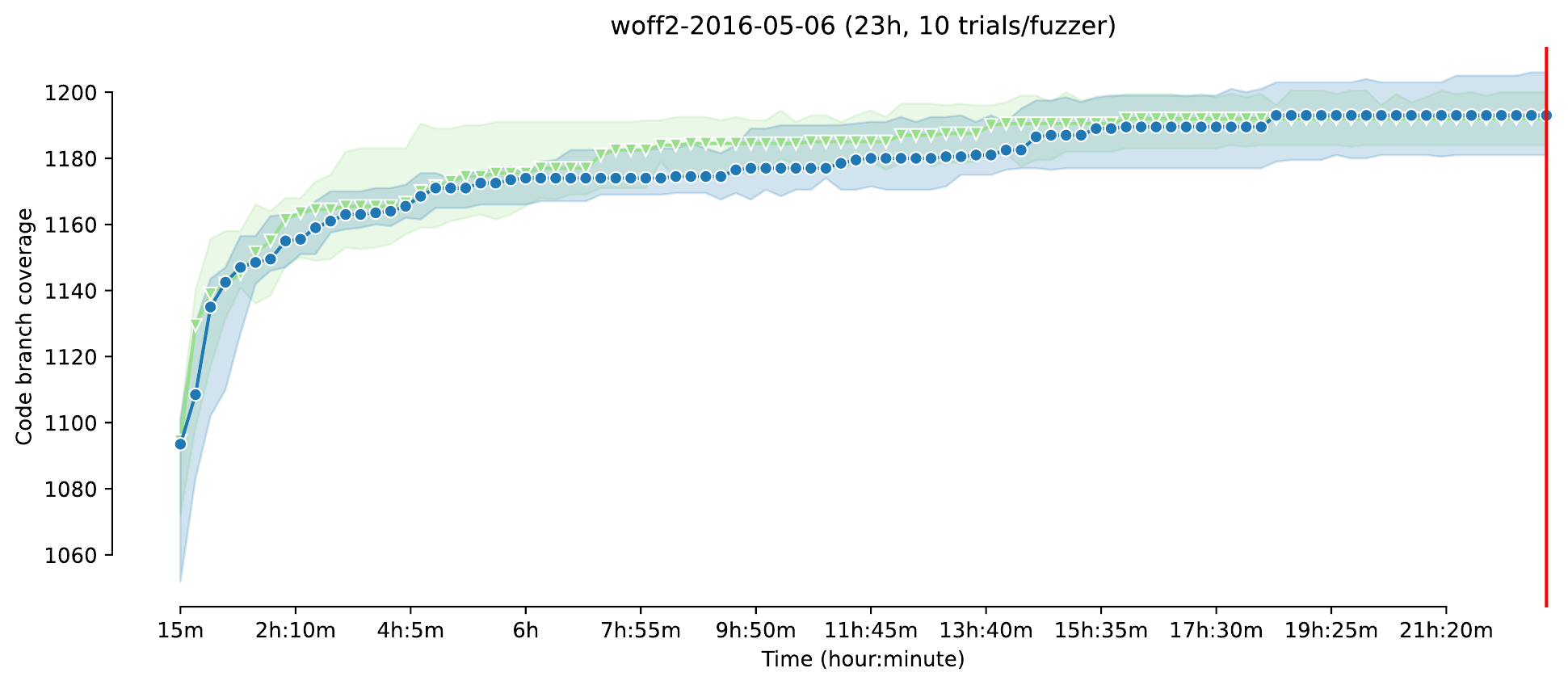} }}%
    \quad
    \subfloat[\centering zlib\_uncompress]{{\includegraphics[width=4cm]{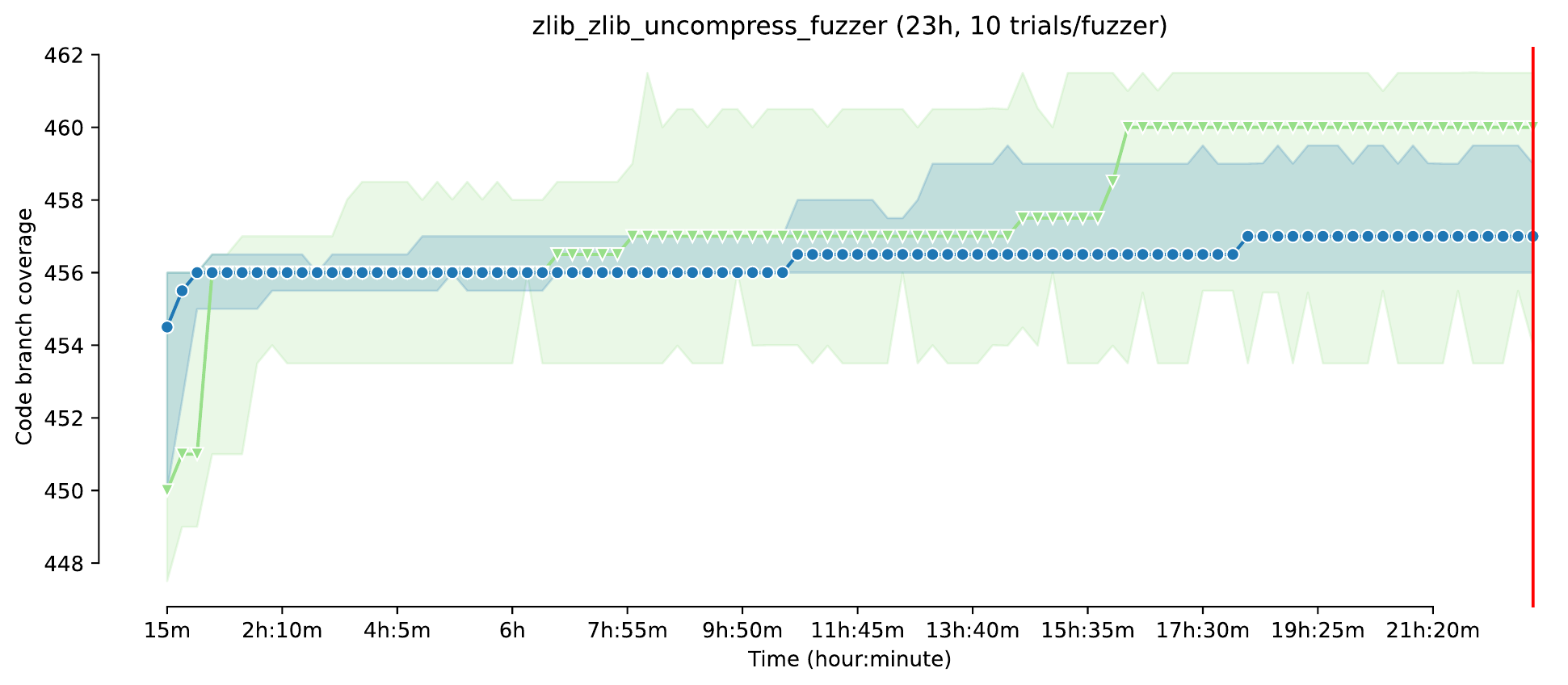} }}%
    \quad
    \subfloat{{\includegraphics[width=4cm]{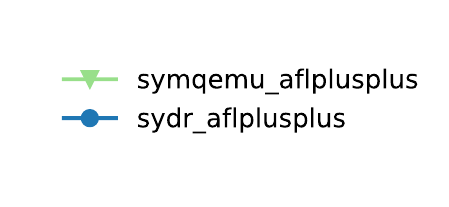} }}%
    \caption{Sydr-Fuzz vs SymQEMU (23h).}%
    \label{fig:symqemu_res}%
\end{figure}

\begin{figure}[h]%
    \centering
    \subfloat[\centering freetype2]{{\includegraphics[width=4cm]{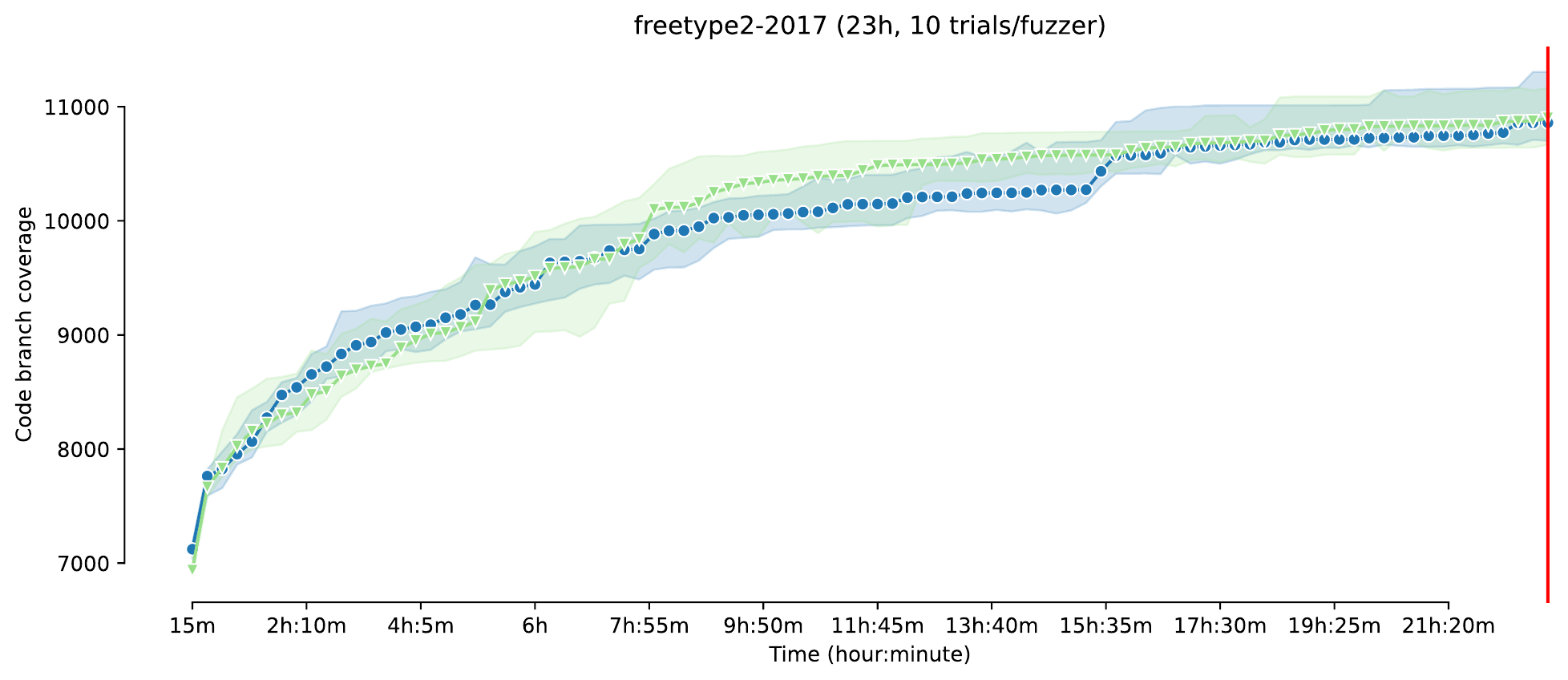} }}%
    \quad
    \subfloat[\centering harfbuzz]{{\includegraphics[width=4cm]{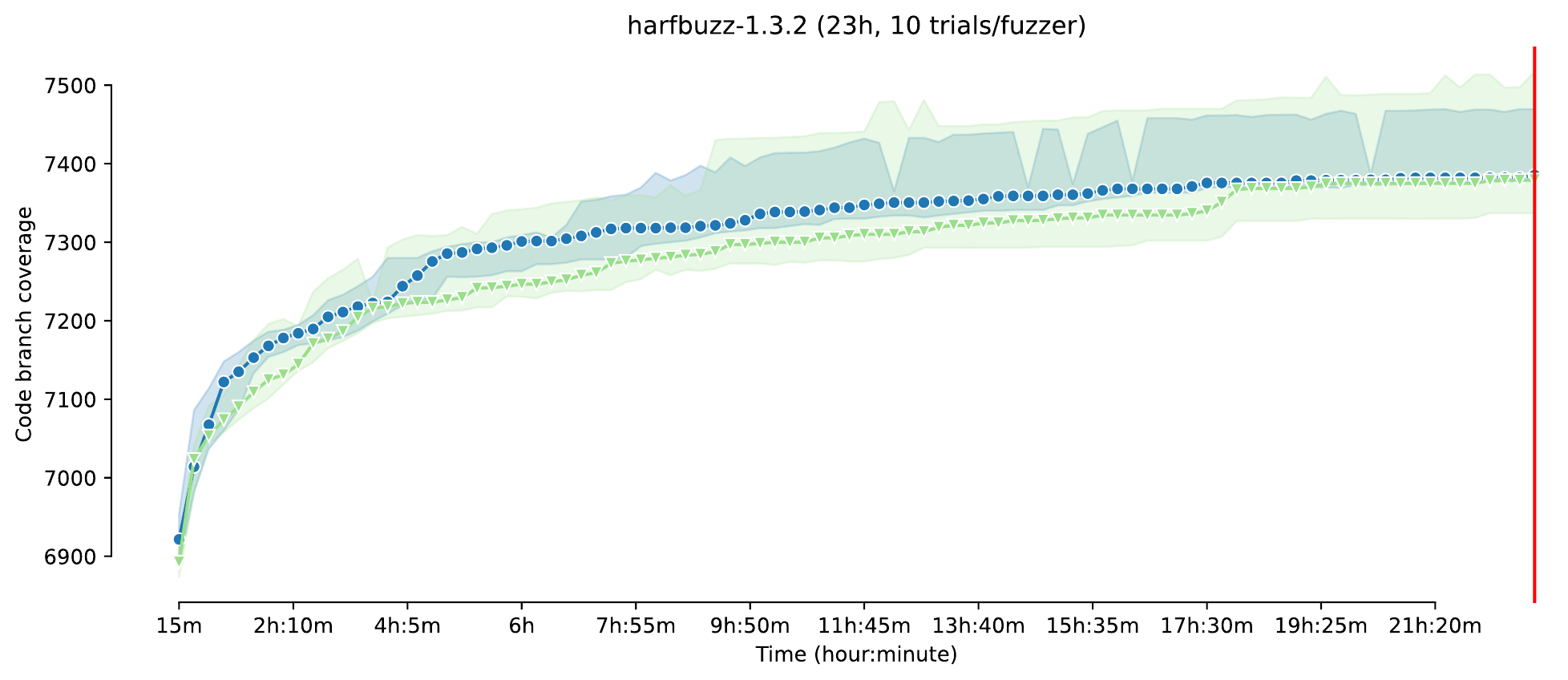} }}%
    \quad
    \subfloat[\centering lcms]{{\includegraphics[width=4cm]{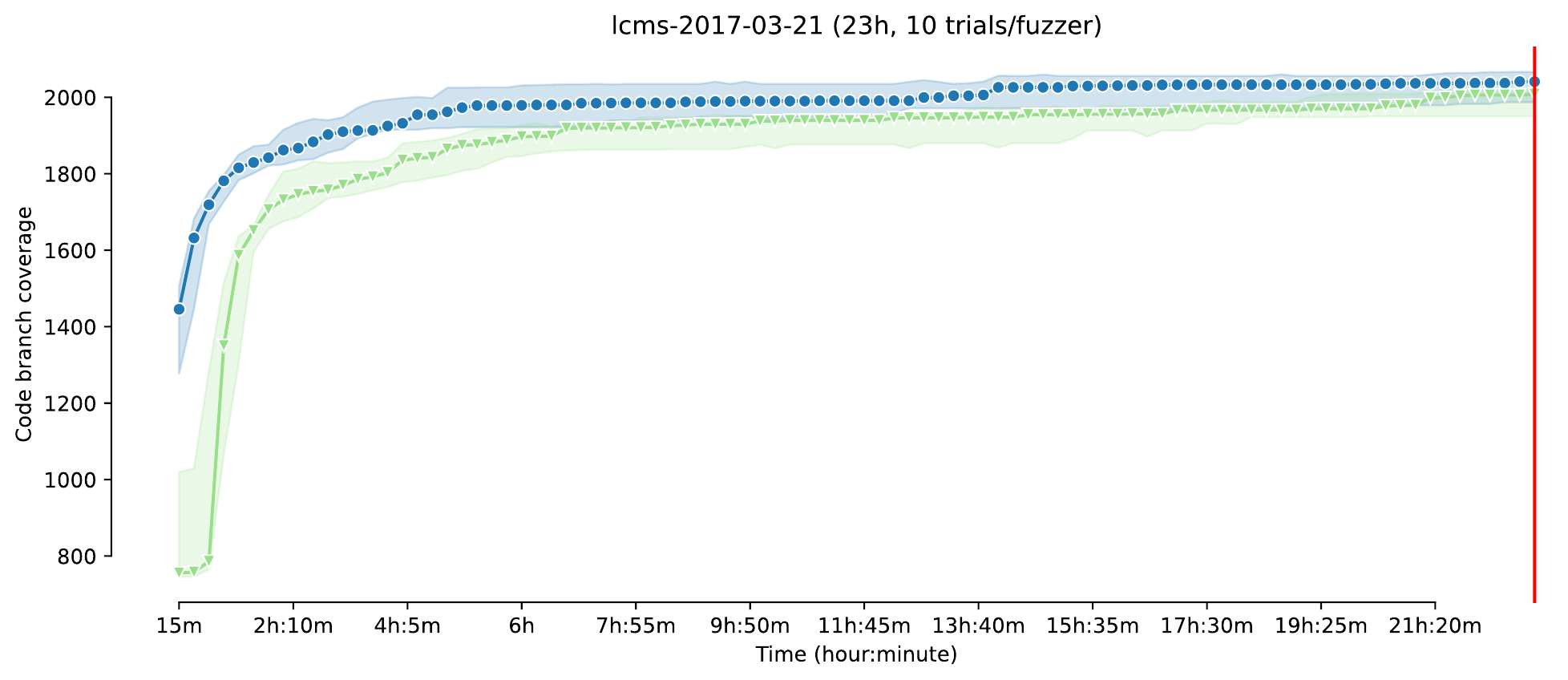} }}%
    \quad
    \subfloat[\centering libjpeg\_turbo]{{\includegraphics[width=4cm]{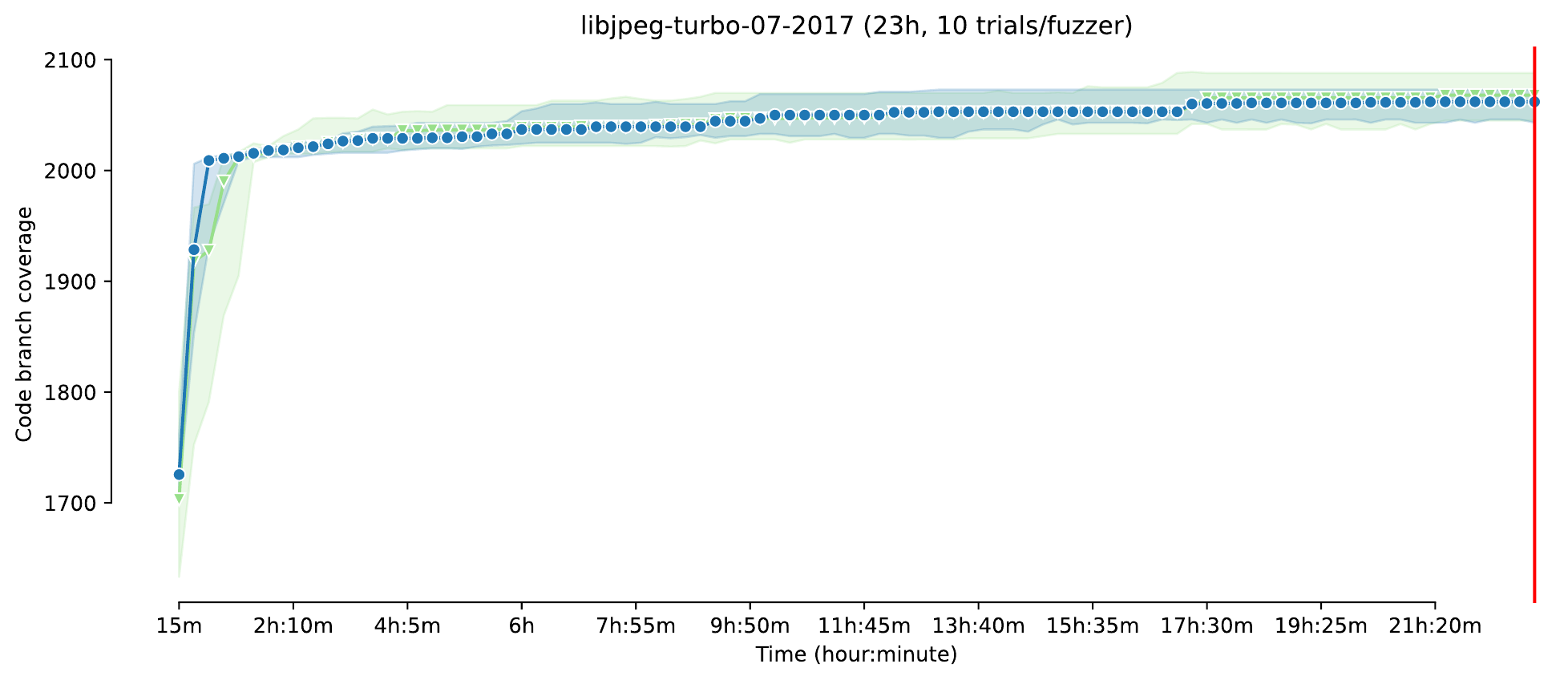} }}%
    \quad
    \subfloat[\centering libpng]{{\includegraphics[width=4cm]{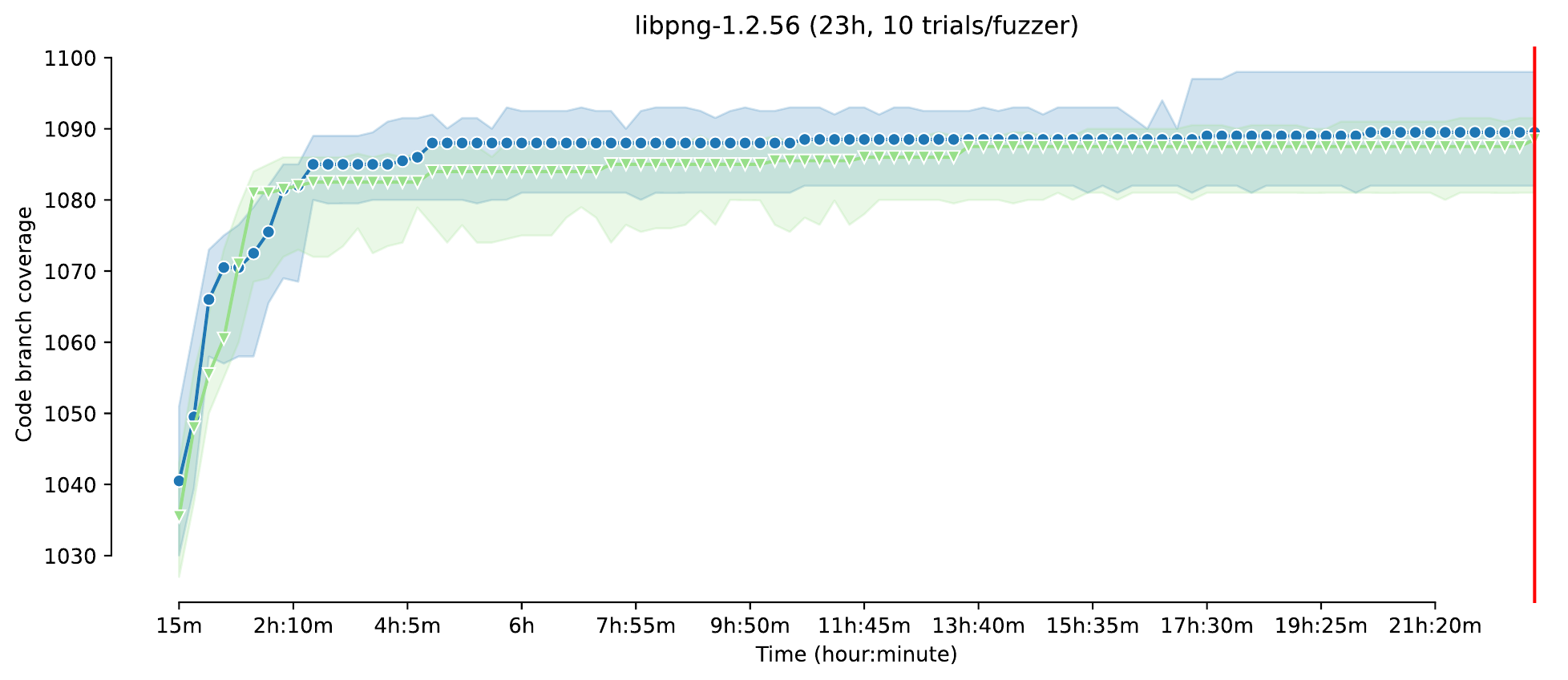} }}%
    \quad
    \subfloat[\centering libxml2]{{\includegraphics[width=4cm]{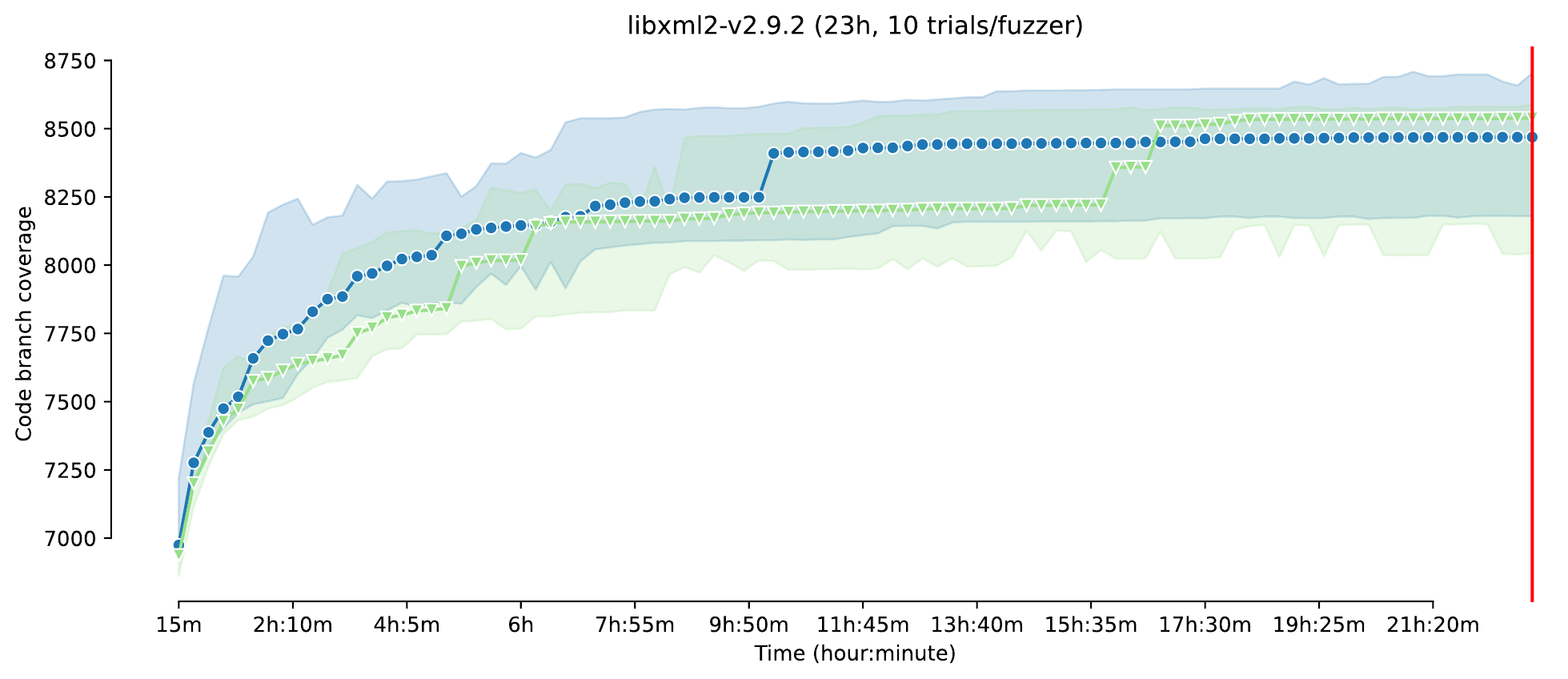} }}%
    \quad
    \subfloat[\centering mbedtls]{{\includegraphics[width=4cm]{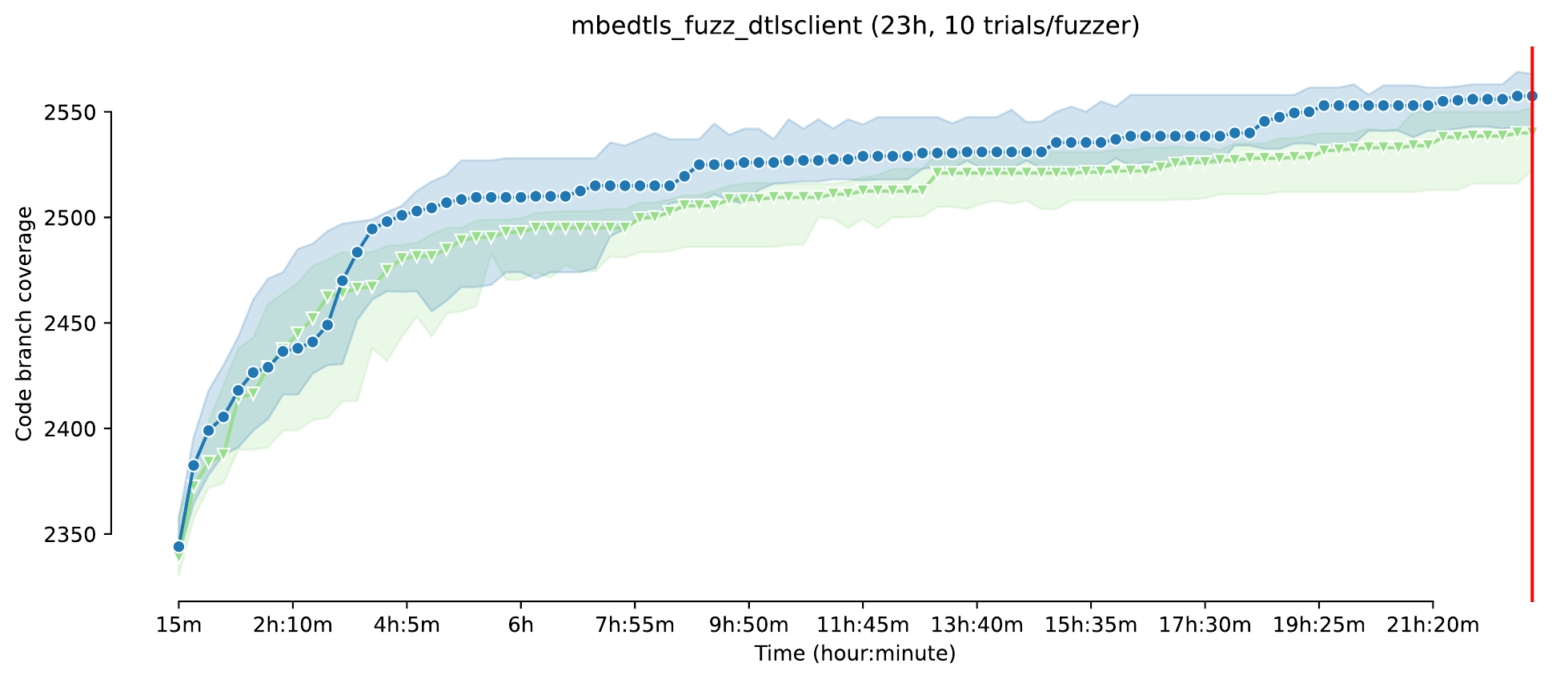} }}%
    \quad
    \subfloat[\centering openthread]{{\includegraphics[width=4cm]{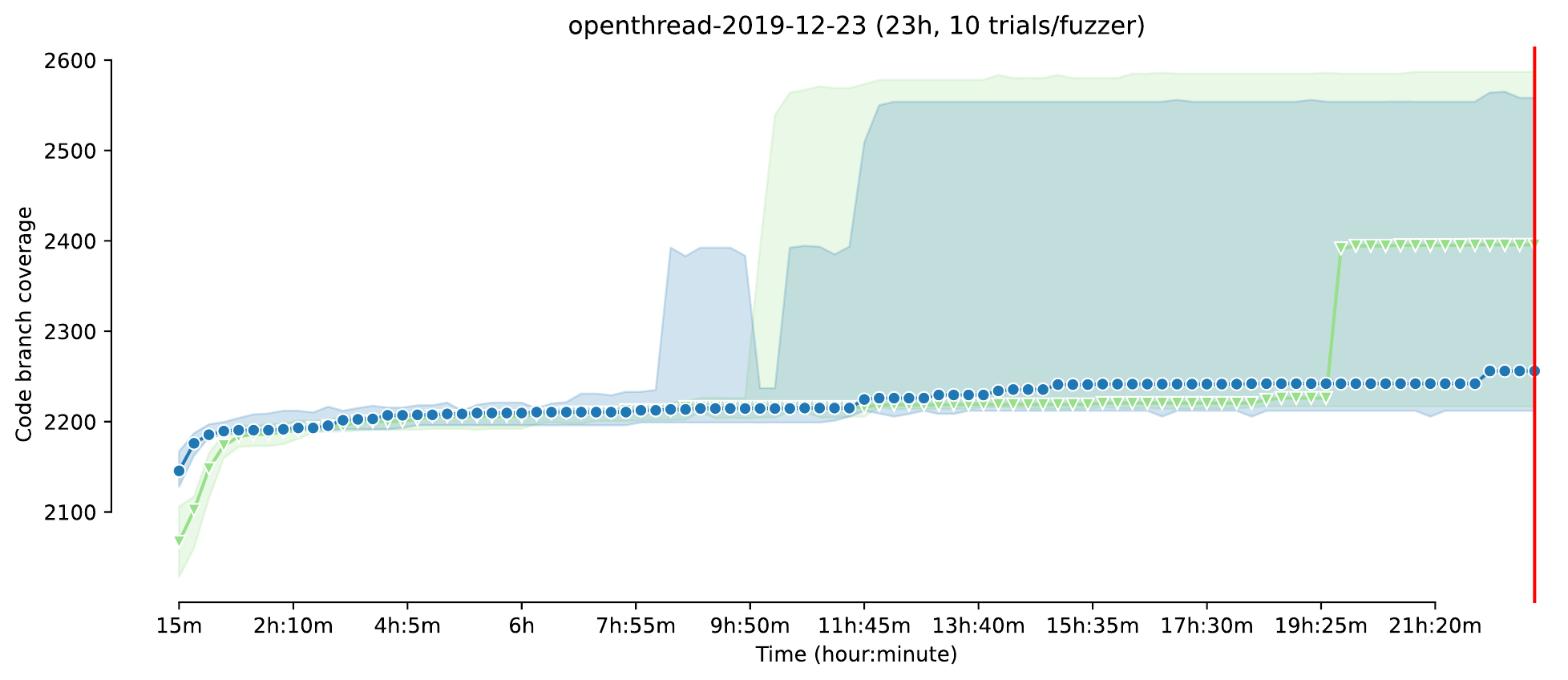} }}%
    \quad
    \subfloat[\centering re2]{{\includegraphics[width=4cm]{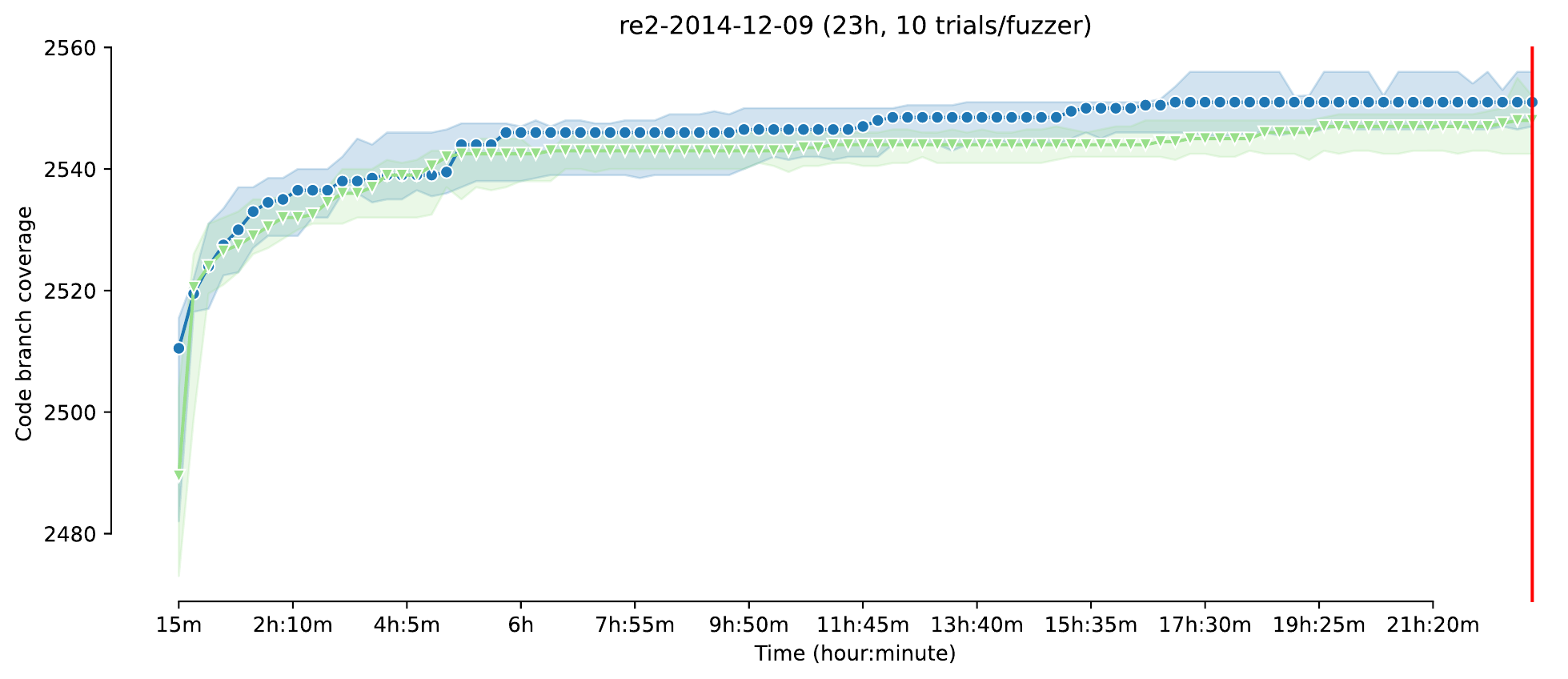} }}%
    \quad
    \subfloat[\centering sqlite3]{{\includegraphics[width=4cm]{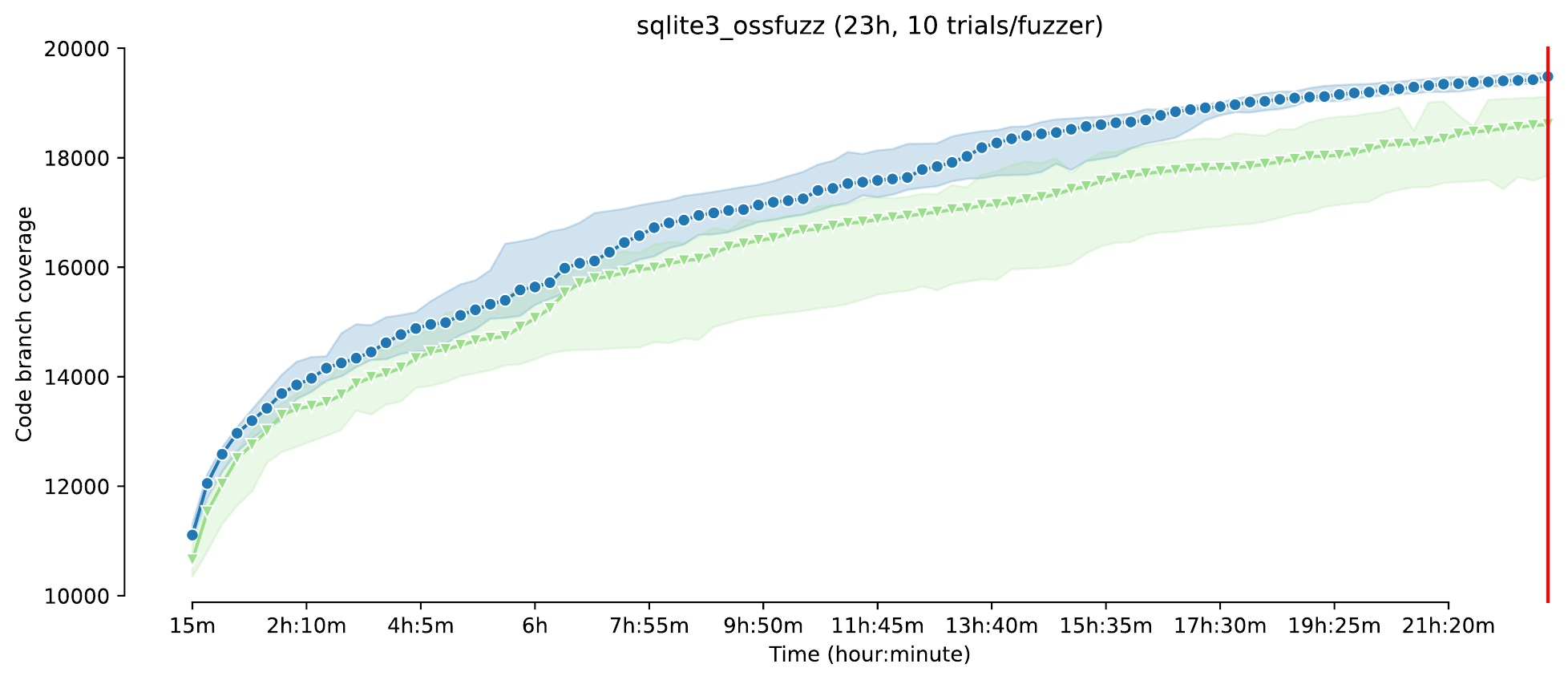} }}%
    \quad
    \subfloat[\centering vorbis]{{\includegraphics[width=4cm]{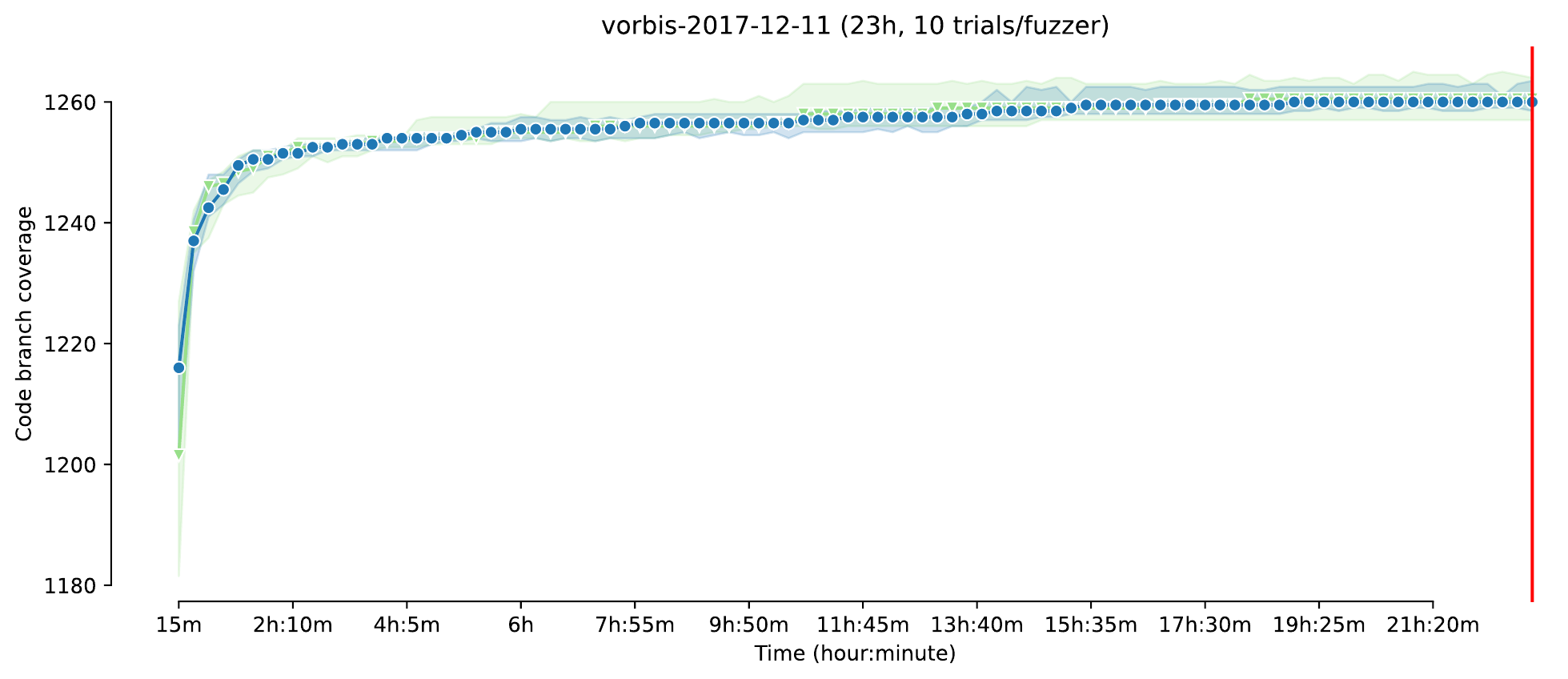} }}%
    \quad
    \subfloat[\centering zlib\_uncompress]{{\includegraphics[width=4cm]{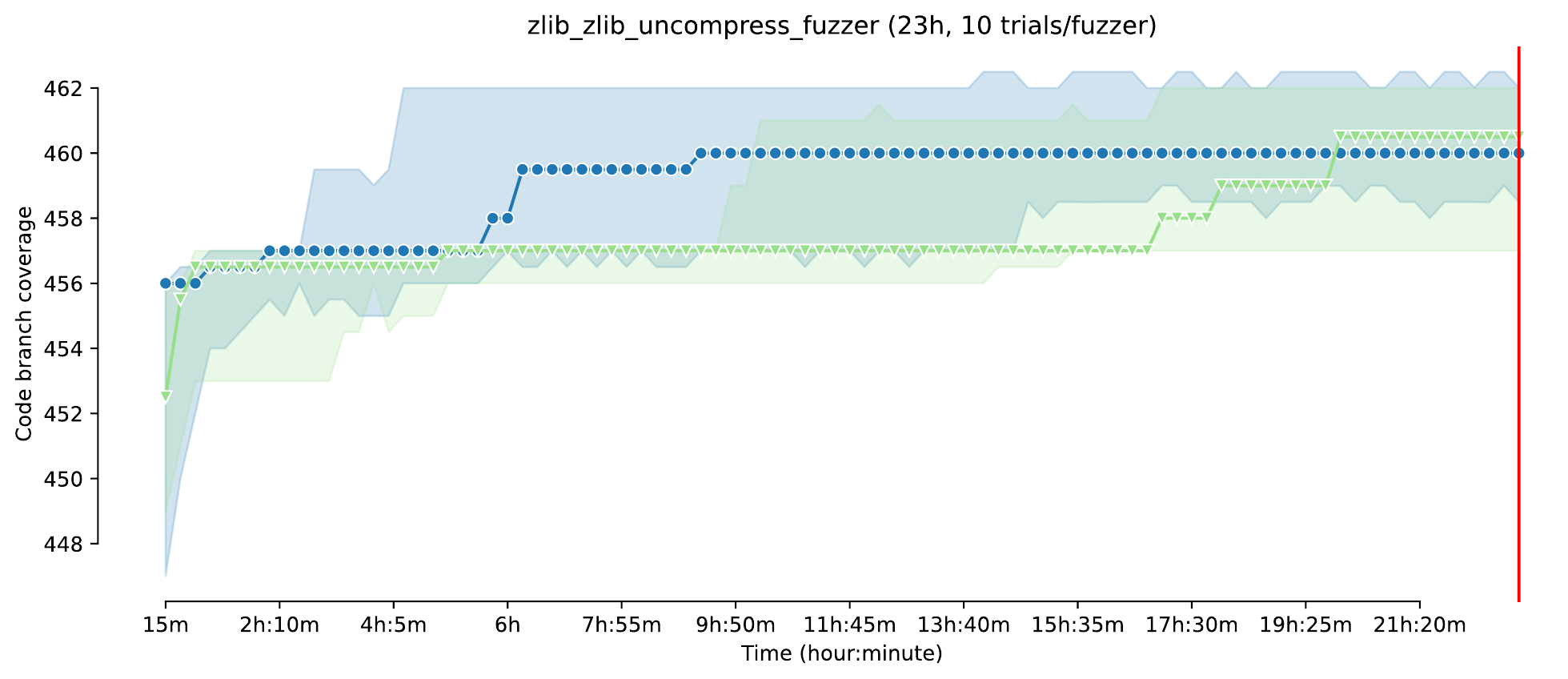} }}%
    \quad
    \subfloat{{\includegraphics[width=4cm]{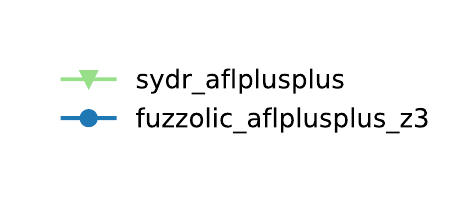} }}%
    \caption{Sydr-Fuzz vs FUZZOLIC (23h).}%
    \label{fig:fuzzolic_res}%
\end{figure}

\end{document}